\documentclass[prodmode,acmtosn]{acmsmall}
\usepackage[ruled]{algorithm2e}
\renewcommand{\algorithmcfname}{ALGORITHM}
\SetAlFnt{\small}
\SetAlCapFnt{\small}
\SetAlCapNameFnt{\small}
\SetAlCapHSkip{0pt}
\IncMargin{-\parindent}

\usepackage{graphicx}

\usepackage{booktabs}
\usepackage{multirow}

\usepackage{amssymb}

\usepackage[usenames,dvipsnames]{xcolor}

\usepackage{framed}
\usepackage{url}
\usepackage{picture}
\usepackage{subfigure}
\usepackage{picins}
\usepackage{balance}
\usepackage{amsmath}

\usepackage{cespvect}

\def\scale2fig{0.85}

\graphicspath{{./figures/}{./figures_final/}}

\newcommand{\negativeskip}{\vspace{-4mm}}

\newcommand{\R}{\mathbb{R}}

\DeclareMathOperator{\E}{\mbox{\large E}}

\def\N{\mathcal{N}}

\newcommand{\bea}{\begin{eqnarray}}
\newcommand{\eea}{\end{eqnarray}}
\newcommand{\be}{\begin{equation}}
\newcommand{\ee}{\end{equation}}
\newcommand{\bi}{\begin{itemize}}
\newcommand{\ei}{\end{itemize}}
\newcommand{\ben}{\begin{enumerate}}
\newcommand{\een}{\end{enumerate}}
\newcommand{\bc}{\begin{cases}}
\newcommand{\ec}{\end{cases}}
\newcommand{\eq}[1]{(\ref{#1})}
\newcommand{\fig}[1]{Fig.~\ref{#1}}

\newcommand{\OO}[1]{\operatorname{O}\bigl(#1\bigr)}

\newcommand{\Cth}{C_{\rm th}}
\newcommand{\dd}{\mathrm{d}}
\newcommand{\bth}{b_{\rm th}}
\newcommand{\mS}{\mathcal{S}}
\newcommand{\U}{\mathcal{U}}
\newcommand{\X}{\mathcal{X}}
\newcommand{\XN}{\mathcal{X_N}}
\newcommand{\B}{\mathcal{B}}

\newcommand{\toff}{t_{\rm off}}
\newcommand{\toffopt}{t^*_{\rm off}}
\newcommand{\ton}{t_{\rm on}}
\newcommand{\tdc}{t_{\rm dc}}
\newcommand{\tdclim}{t^{\rm lim}_{\rm dc}}
\newcommand{\tdcmin}{t^{\min}_{\rm dc}}
\newcommand{\tdcopt}{t^*_{\rm dc}}
\newcommand{\tU}{t_{\rm U}}
\newcommand{\tUopt}{t^*_{\rm U}}

\newcommand{\tUlim}{t^{\rm lim}_{\rm U}}
\newcommand{\tdata}{t_{\rm data}}
\newcommand{\tcpu}{t_{\rm cpu}}
\newcommand{\tint}{t_{\rm int}}
\newcommand{\trpl}{t_{\rm rpl}}
\newcommand{\tout}{t_{\rm out}}
\newcommand{\ttx}{t_{\rm TX}}
\newcommand{\trx}{t_{\rm RX}}
\newcommand{\tcts}{t_{\rm cts}}

\newcommand{\tack}{t_{\rm ack}}
\newcommand{\tv}{t_{\rm v}}

\newcommand{\itx}{I_{\rm TX}}
\newcommand{\irx}{I_{\rm RX}}
\newcommand{\iinte}{I_{\rm INT}}
\newcommand{\icpu}{I_{\rm CPU}}
\newcommand{\iidle}{I_{\rm IDLE}}
\newcommand{\icca}{I_{\rm CCA}}
\newcommand{\ioff}{I_{\rm OFF}}
\newcommand{\io}{I_{\rm out}}
\newcommand{\iolim}{I^{\rm lim}_{\rm out}}
\newcommand{\iomin}{I^{\min}_{\rm out}}
\newcommand{\ic}{i_{\rm c}}
\newcommand{\ir}{i_{\rm r}}
\newcommand{\itr}{i_{\rm t}}
\newcommand{\is}{i_{\rm s}}

\newcommand{\fu}{f_{\rm U}}

\newcommand{\ftxdg}{f_{\rm TX,DG}}
\newcommand{\frxdg}{f_{\rm RX,DG}}
\newcommand{\ftxrpl}{f_{\rm TX,RPL}}
\newcommand{\frxrpl}{f_{\rm RX,RPL}}
\newcommand{\finte}{f_{\rm INT}}

\newcommand{\rtx}{r_{\rm TX}}
\newcommand{\rrx}{r_{\rm RX}}
\newcommand{\rinte}{r_{\rm INT}}
\newcommand{\rcpu}{r_{\rm CPU}}
\newcommand{\ridle}{r_{\rm IDLE}}

\newcommand{\ku}{k_{\rm U}}

\newcommand{\nc}{n_{\rm c}}
\newcommand{\nin}{n_{\rm i}}
\newcommand{\nint}{n_{\rm int}}

\newcommand{\cc}[1]{\textcolor{blue}{#1}}
\renewcommand{\cc}[1]{{#1}}

\newcommand{\mr}[1]{\textcolor{blue}{#1}}
\renewcommand{\mr}[1]{{#1}}

\newcommand{\bn}[1]{\textcolor{blue}{#1}}
\renewcommand{\bn}[1]{{#1}}

\newcommand{\nb}[1]{\textcolor{orange}{#1}}
\renewcommand{\nb}[1]{{#1}}

\newcommand{\etx}{e_{\rm t}}
\newcommand{\ecoll}{e_{\rm c}}
\newcommand{\ecollm}{e_{\rm c,\max}}
\newcommand{\pcoll}{p_{\rm c}}
\newcommand{\ep}{e_{\rm p}}
\newcommand{\fup}{f_{\rm U}^\prime}

\renewcommand{\negativeskip}{}
\newcommand{\figarch}{
\begin{figure}[t]
    \centering
    \includegraphics[width=0.8\textwidth]{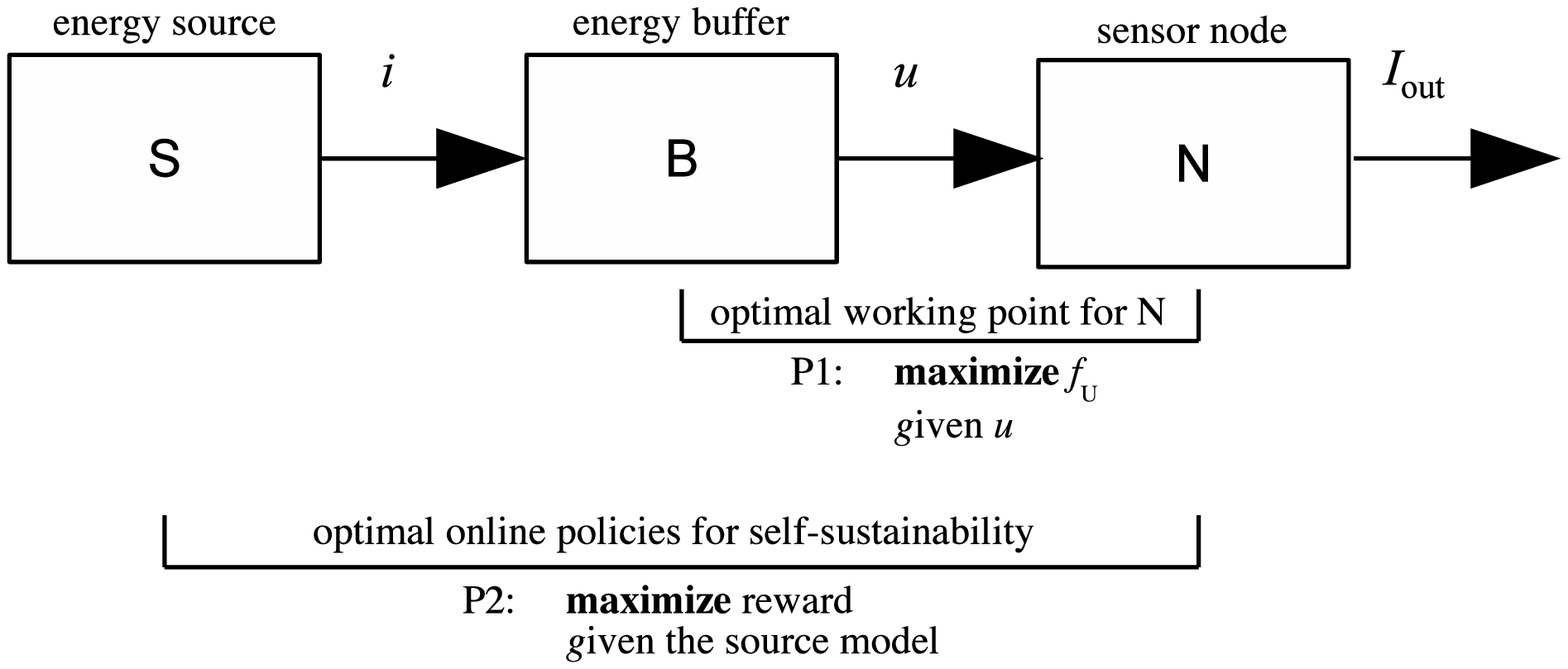}
    \caption{Sensor node diagram.}
    \label{fig:sensor_diagram}
\end{figure}
}

\newcommand{\figmac}{
\begin{figure*}[t]
\begin{center}
	\setlength{\unitlength}{1mm}
	\begin{picture}(80,60)(0,0)
	\put(0,0){\includegraphics[width=80mm]{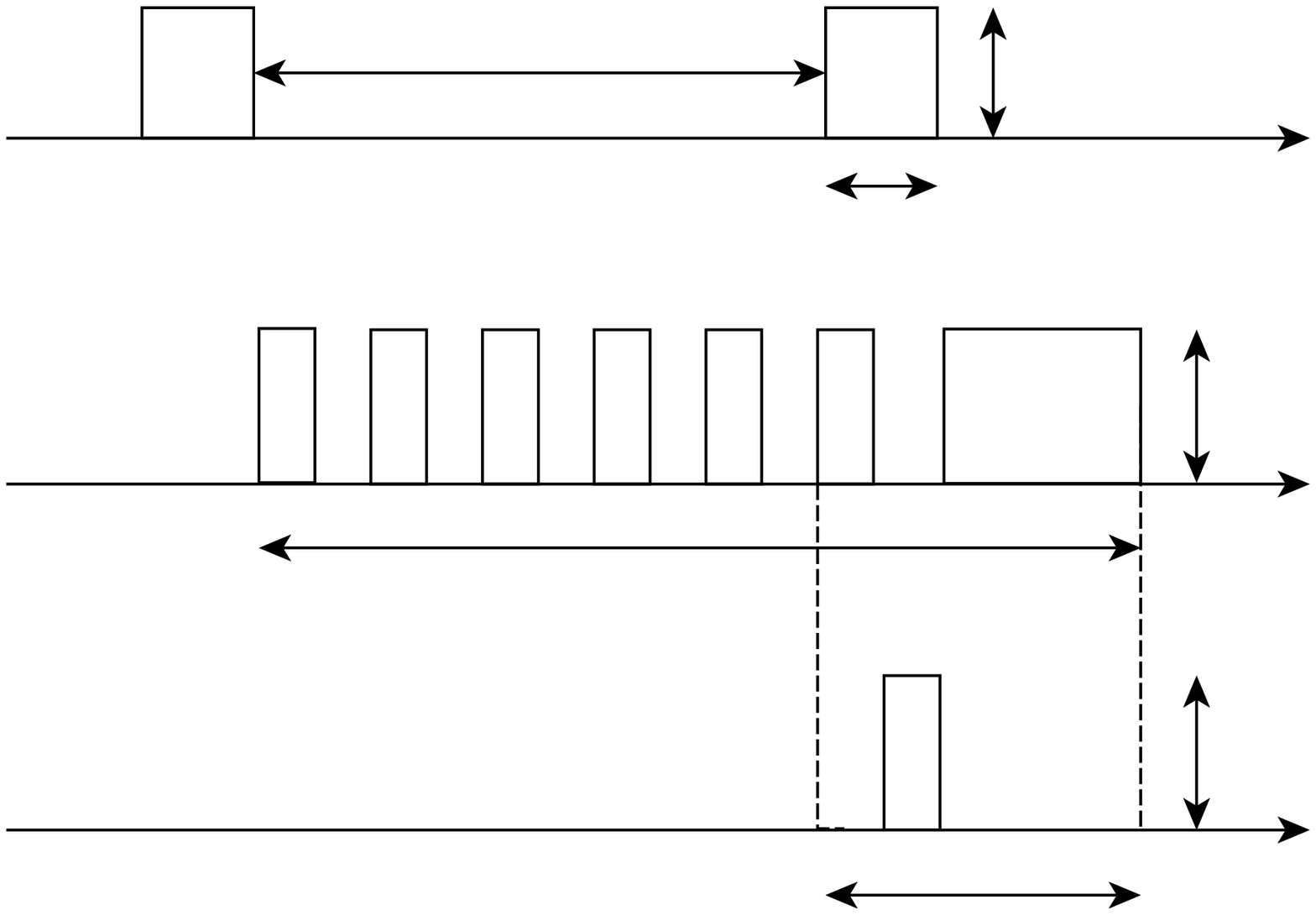}}
	\w(53,42)[13]{$\ton$}
	\w(35,54)[13]{$\toff$}
	\w(70,51.5)[13]{$\irx+\icpu$}
	\w(79,47)[13]{$t$}
	\w(45,22)[13]{$t_{\rm TX}$}
	\w(80,33)[13]{$\itx+\icpu$}
	\w(79,27)[13]{$t$}
	\w(58,3)[13]{$t_{\rm RX}$}
	\w(80,13)[13]{$\irx+\icpu$}
	\w(79,7)[13]{$t$}
	\w(14,58)[13]{CCA}
	\w(18.7,40)[13]{RTS}
	\w(53.9,20.2)[13]{CTS}
	\w(61.5,40)[13]{DATA}
	\w(5,46)[13]{Idle}
	\w(5,26.5)[13]{Sender}
	\w(5,7)[13]{Receiver}
	\end{picture}
\end{center}
	\caption{MAC timings for CCA (top), TX (middle) and RX phases (bottom).}
	\label{fig:mac_timings}
\end{figure*}
}

\newcommand{\figcoll}{
\begin{figure*}[t]
\begin{center}
	\setlength{\unitlength}{1mm}
	\begin{picture}(80,50)(0,0)
	\put(0,0){\includegraphics[width=80mm]{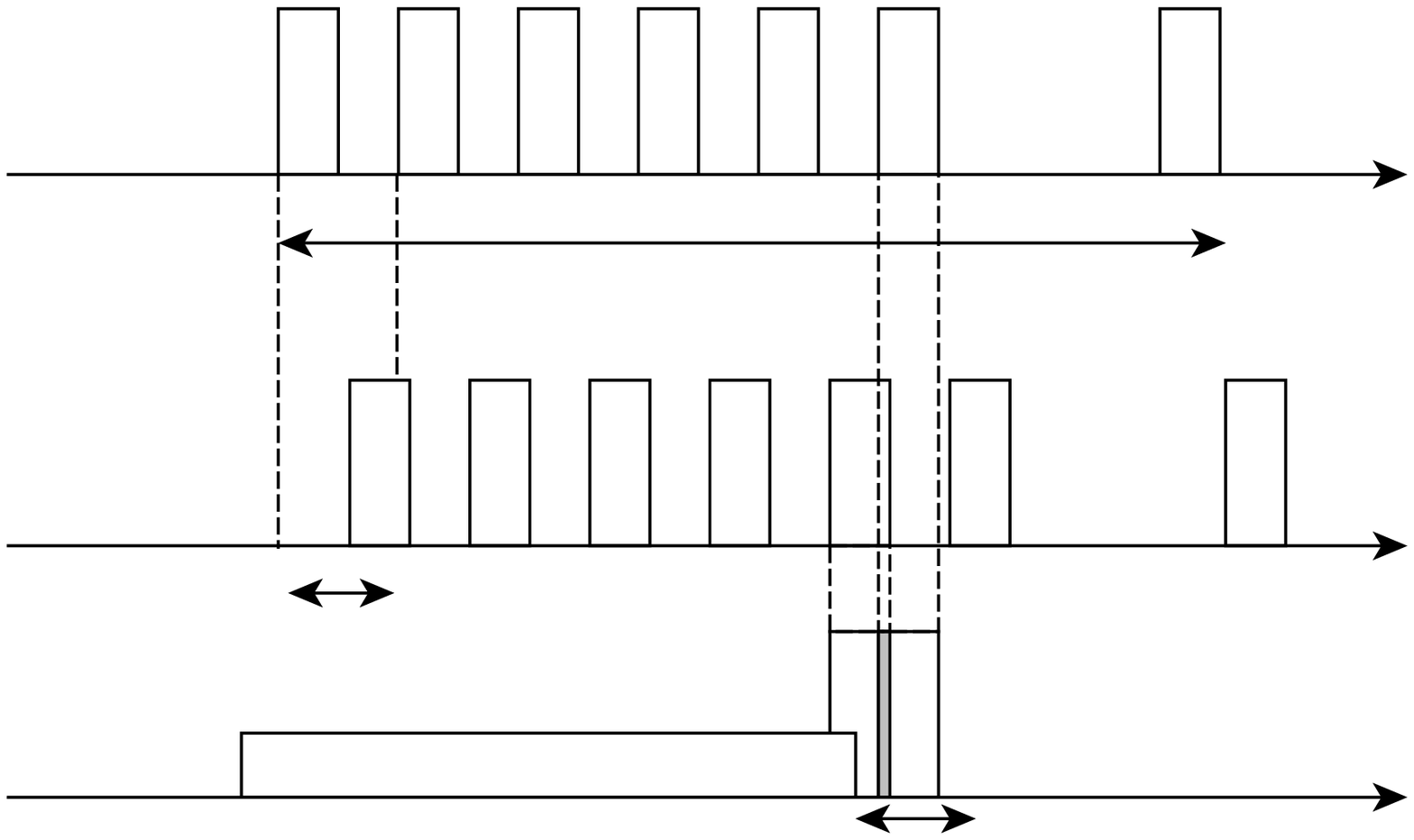}}
	\w(45,36.5)[13]{$\tdc$}
	\w(79,40)[13]{$t$}
	\w(20,18)[13]{$\tv$}
	\w(79,20)[13]{$t$}
	\w(79,7)[13]{$t$}
	\w(18.5,52.5)[13]{RTS}
	\w(58,46)[13]{\large$\cdots$}
	\w(61,26.3)[13]{\large$\cdots$}
	\w(50,6)[13]{$\ton$}
	\w(59,13)[13]{Collision}
	\w(5,39.5)[13]{Sender 1}
	\w(5,20)[13]{Sender 2}
	\w(5,7)[13]{Receiver}
	\w(30,11.3)[13]{RX off}
	\end{picture}
\end{center}
	\caption{Graphical example of a collision: the first sender (top) starts sending periodical RTSs; before the vulnerability time, $\tv$, has elapsed, the second sender starts sending RTSs too; since none of them is aware of the other. They keep on transmitting RTSs for $\tdc$ seconds (the duration of the RTS burst). When one of the intended destinations wakes up, it will receive a corrupted RTS (collision).}
	\label{fig:coll}
\end{figure*}
}

\newcommand{\fignrganal}{
\begin{figure}[t]
\begin{center}
\setlength{\unitlength}{1mm}
	\begin{picture}(80,60)(0,0)
	\put(0,0){\includegraphics[width=80mm]{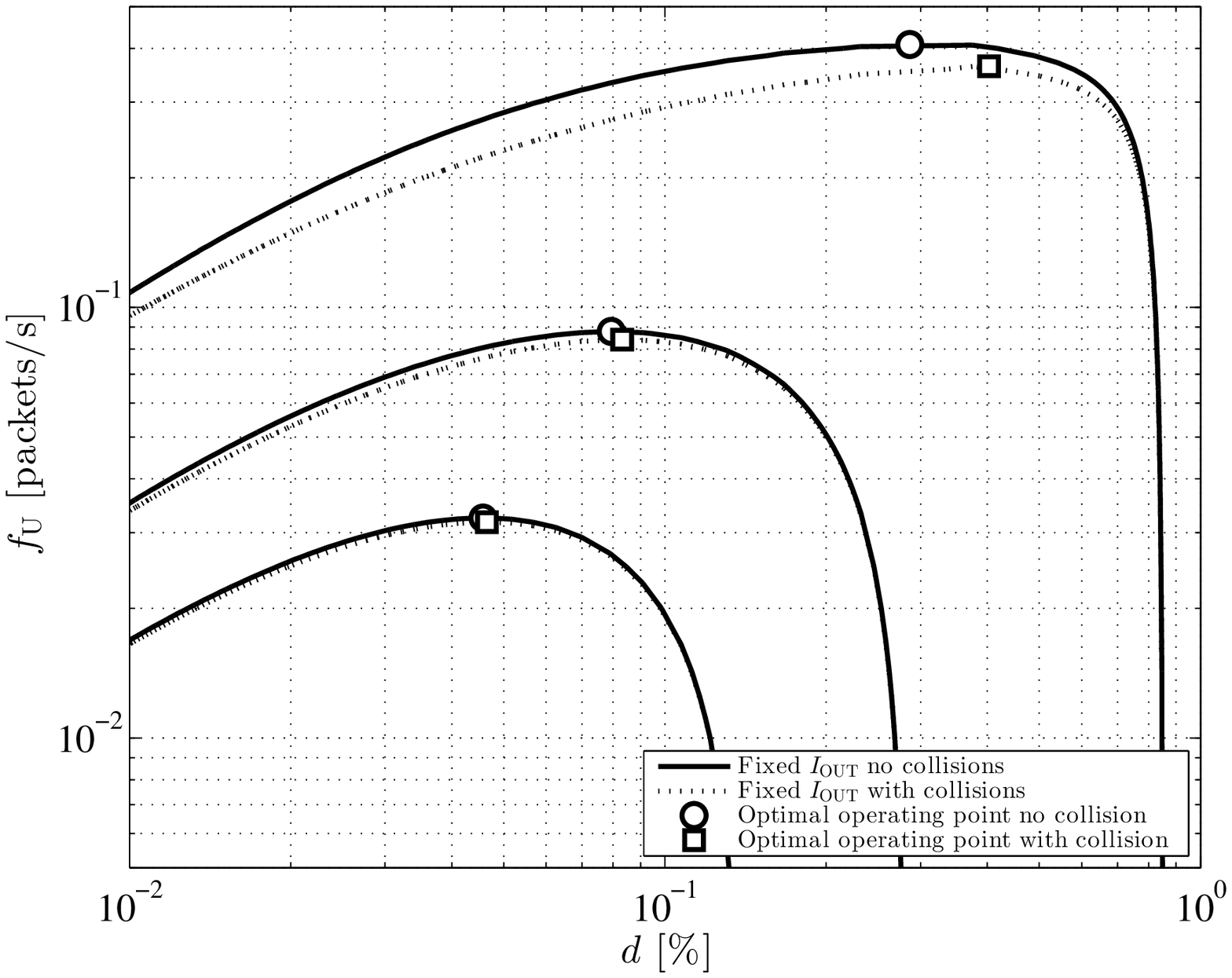}}
	\w(60,53)[13]{$\io = 30$ mA}
	\w(40,35)[13]{$\io = 10$ mA}
	\w(28,23)[13]{$\io = 5$ mA}
	\end{picture}
	\end{center}
\caption{Contour lines in the $(d,\fu)$ plane for different output current levels ($\io \in \{5, 10, 30\}$ mA): dotted lines represent the numerical solution to the complete problem \eq{eq:OOP}, while dash-dotted lines show the solution for a collision-free channel for the same $\io$ levels. The optimal working points are also plotted for both problems (using white squares for the complete problem and white circles to indicate the solution for a collision-free channel).}
\label{fig:nrganal}
\end{figure}
}

\newcommand{\figopttoff}{
\begin{figure}[t]
\begin{center}
\setlength{\unitlength}{1mm}
	\begin{picture}(80,65)(0,0)
	\put(0,0){\includegraphics[width=80mm]{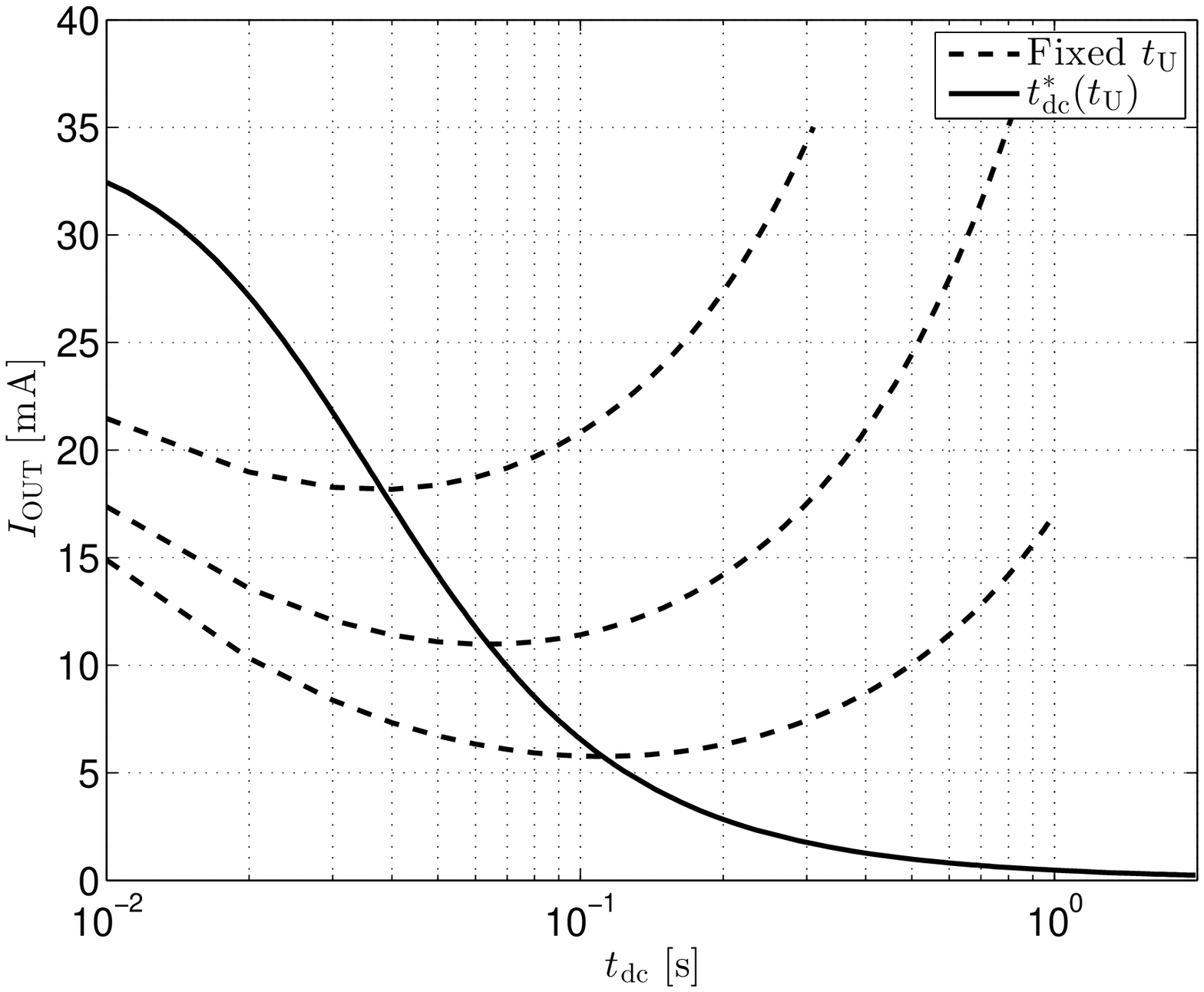}}
	\w(29,39)[13]{$\tU = 5$ s}
	\w(37,29)[13]{$\tU = 10$ s}
	\w(45,19)[13]{$\tU = 25$ s}
	\end{picture}
	\end{center}
\caption{Dashed lines represent $\io(\tU,\tdc)$ as a function of $\tdc$, considering a fixed inter-packet transmission time $\tU \in \{5, 10, 25\}$ seconds. The locus of the optimal solutions $\tdcopt$, obtained through \eq{eq:opttoff}, is plotted as a solid line.}
\label{fig:opttoff}
\end{figure}
}

\newcommand{\figoptimal}{
\begin{figure}[t]
\begin{center}
\setlength{\unitlength}{1mm}
	\begin{picture}(80,65)(0,0)
	\put(0,0){\includegraphics[width=80mm]{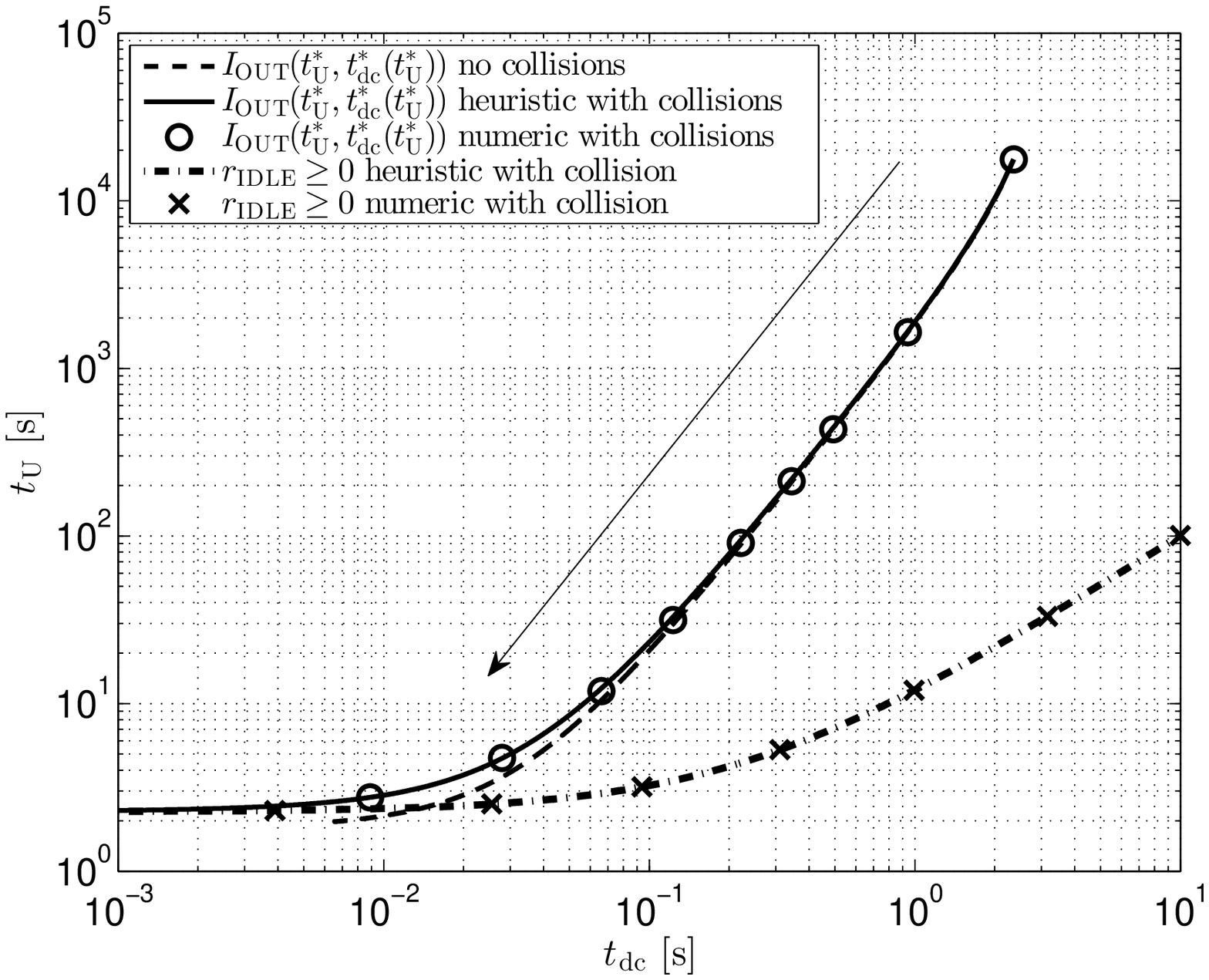}}
	\w(34,42)[13]{Increasing $u$}
	\end{picture}
	\end{center}
\caption{Comparison between closed form and exact solution of \eq{eq:OOP}. The dashed line shows the results obtained with the closed form solution considering a collision-free channel, the dots represent the numerical solution of the problem with collisions, and the solid line corresponds to the closed form solution, heuristically adapted to keep collisions into account. In addition, the constraint $\ridle(\tU,\tdc) = 0$ is also shown (crosses indicate the exact bound obtained numerically, the dash-dotted line is obtained using the heuristically adapted closed form).}
\label{fig:optimal}
\end{figure}
}

\newcommand{\figreward}{
\begin{figure}[t]
\begin{center}
\setlength{\unitlength}{1mm}
	\begin{picture}(80,68)(0,0)
	\put(0,0){\includegraphics[width=80mm]{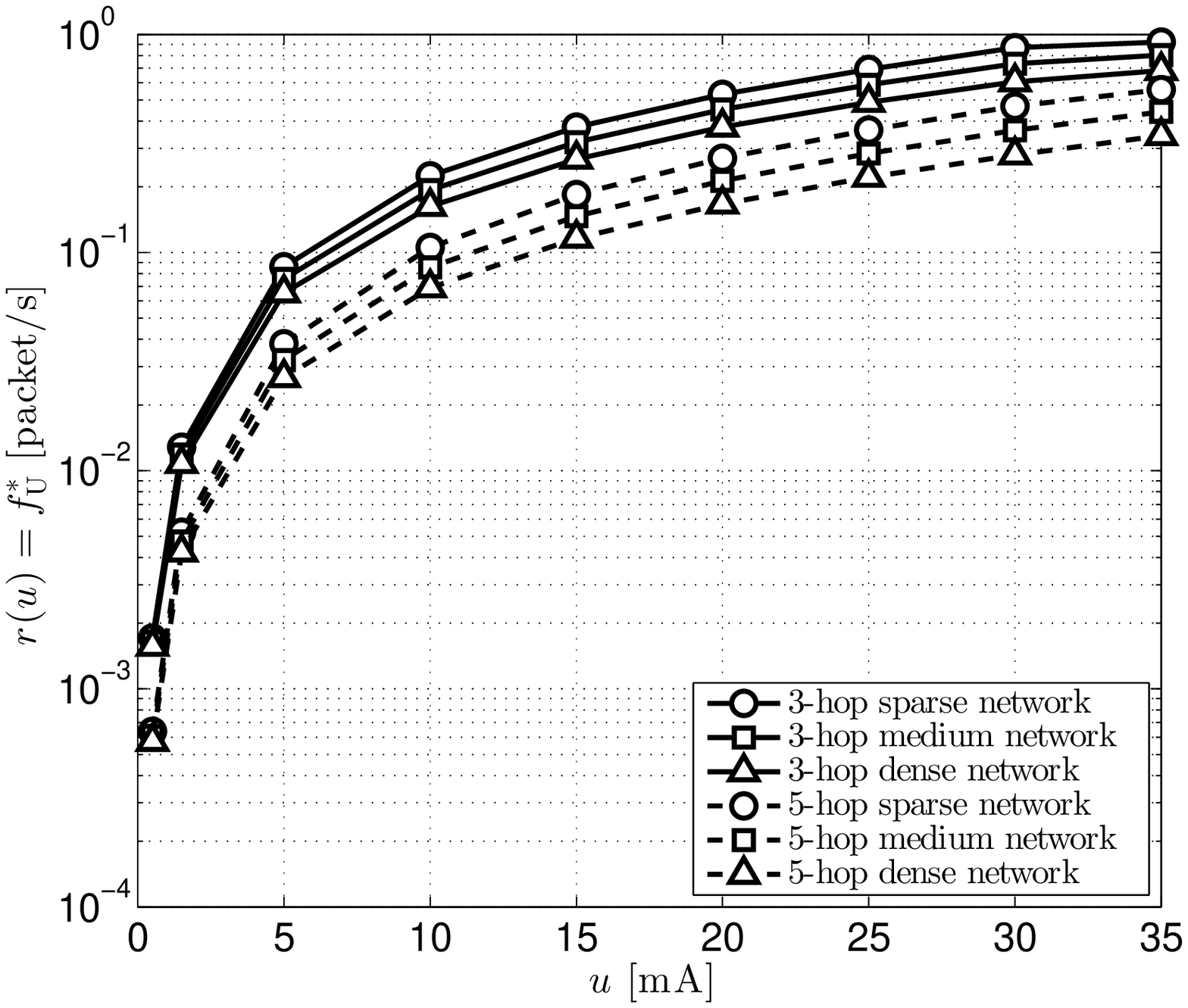}}
	\end{picture}
	\end{center}
\caption{Reward function $r(u)$ for different network topologies.}
\label{fig:reward}
\end{figure}
}

\newcommand{\figtopo}{
\begin{figure}[t]
\begin{center}
\setlength{\unitlength}{1mm}
	\begin{picture}(80,70)(0,0)
	\put(0,0){\includegraphics[width=80mm]{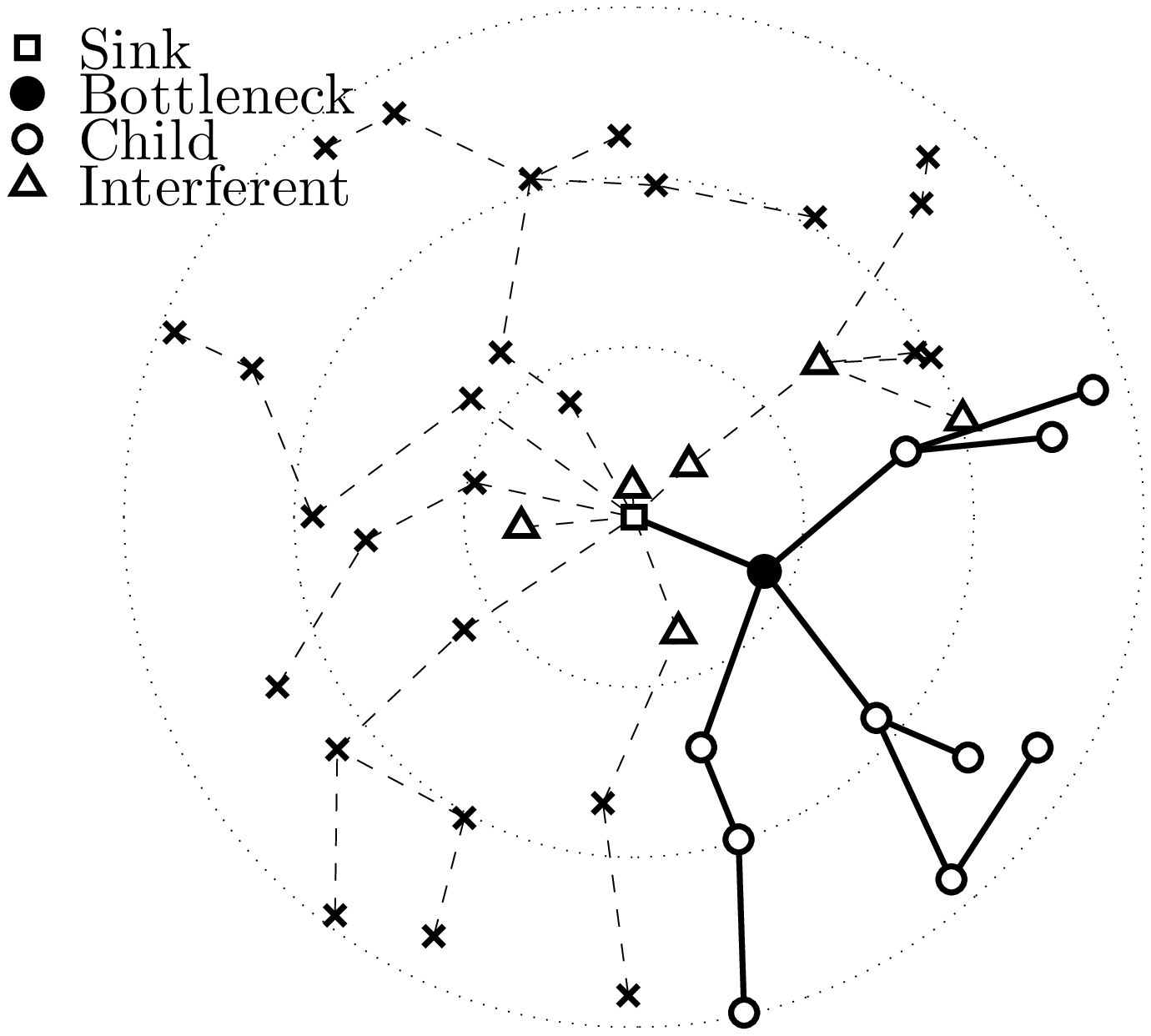}}
	\end{picture}
	\end{center}
\caption{Topology.}
\label{fig:topo}
\end{figure}
}

\newcommand{\tabcoeffabc}{
\begin{table}[b]
    \tbl{Coefficients $a$, $b$ and $c$.}{
    \centering
    \begin{tabular}{ l | l }
    \toprule
		$a_1=(1+\nc)(1/2 + e_t/(1-e_t))$ 	& $b_1 = c_1a_1$ \\
		$a_2=(1+\nc)(\tdata+\ton/2)$		& $b_2 = c_1a_2$ \\
		$a_3=(2+\nc)(1/2 + e_t/(1-e_t))/\trpl$ & $b_3 = c_1a_3$ \\
		$a_4=(2+\nc)(\tdata+\ton/2)/\trpl$ 	& $b_4 = c_1a_4$ \\
		$a_5=\nc\tdata$ 			& $b_5 = c_2a_5$ \\
		$a_6=(1+\nc+\nin)\tdata/\trpl$ 	& $b_6 = c_2a_6$ \\
		$a_7=\tint \nint$ 			& $b_7 = c_2a_7$ \\
		$a_8=\tint \nint/\trpl$ 		& $b_8 = c_2a_8$ \\
		$a_9=\tcpu \ku$ 			& $b_9 = c_3a_9$ \\
		$a_{10}=1-a_4-a_6-a_8$ 		& $b_{10}=-a_1c_4$ \\
		$a_{11}=a_2+a_5+a_7+a_9$ 	& $b_{11}=-a_1\ton c_5 - a_{11}c_4$ \\
		\cline{1-1} $c_1 = \ic + \itr$	& $b_{12}=-a_3c_4$\\
		$c_{2}=\ic+\ir$		 		& $b_{13}=-a_3\ton c_5 + a_{10}c_4$ \\
		$c_{3}=\ic$		 		& $b_{14}=a_{10}c_5\ton$ \\
		$c_{4}=\is$		 		& $b_{15}=-a_{11}c_5\ton$\\
		$c_{5}=\ic + \ir - \is$ 			& \\
    \bottomrule
    \end{tabular}}
    \label{tab:coeff1}
\end{table}
}

\newcommand{\tabcoeffdef}{
\begin{table}[htb]
    \tbl{Coefficients $d$, $e$ and $f$.}{
    \centering
    \begin{tabular}{ l | l }
    \toprule
    $d_1 = b_1 + b_{10}$ & $e_0 = 4d_1d_6 - d^2_2$\\
    $d_2 = b_2 + b_5 + b_7 + b_9 + b_{11}$ & $e_1 = 4 d_1d_5+4d_3d_6-2d_7d_2$\\
    $d_3 = b_3 + b_{12}$ & $e_2= 4 d_5 d_3 - d^2_7$ \\
    \cline{2-2} $d_4 = b_4 + b_6 + b_8 + b_{13}$ & $f_0=-a_{10} d_6-a_{11} d_5$\\
    $d_5 = b_{14}$ & $f_1= a_3 d_6 - a_1 d_5$\\
    $d_6 = b_{15}$ & $f_2= d_1 a_{10}+d_3 a_{11}$\\
    $d_7 = d_4 - u$ & $f_3= -d_1 a_3 + a_1 d_3$\\
    \bottomrule
    \end{tabular}}
    \label{tab:coeff2}
\end{table}
}

\newcommand{\tabnets}{
\begin{table}[htb]
	\tbl{Network parameters. $R$ is the radio coverage range.}{
	\centering
	\begin{tabular}{ l | c | c | c | c | c }
	\toprule
	& $N$ & $\rho$ [nodes/$R^2$] & $\nc$ [nodes] & $\nin$ [nodes] & $\nint$ [packets] \\
	\midrule
	$3$-hop sparse		& $15$ & $0.53$ & $5$ & $4$ & $16$ \\
	$3$-hop medium	& $25$ & $0.88$ & $5$ & $8$ & $32$ \\
	$3$-hop dense		& $38$ & $1.35$ & $5$ & $13$ & $54$ \\
	$5$-hop sparse		& $42$ & $0.53$ & $15$ & $4$ & $48$ \\
	$5$-hop medium	& $68$ & $0.86$ & $15$ & $8$ & $96$ \\
	$5$-hop dense		& $106$ & $3.23$ & $15$ & $13$ & $160$ \\
	\bottomrule
	\end{tabular}}
	\label{tab:nets}
\end{table}
}

\newcommand{\tabpar}{
\begin{table}[t]
\tbl{System parameters.}{
\centering
\begin{tabular}{c|c|c|c|c|c|c|c|c}
\toprule
$\ton$&$\tdata$&$\tint$&$\tcpu$&$\trpl$&$\itr$&$\ir$&$\ic$&$\is$\\
\midrule
$6$ ms & $14$ ms & $10$ ms & $40$ ms & $6$ h & $14$ mA & $12.3$ mA & $42$ mA & $31$ $\mu$A\\
\bottomrule
\end{tabular}}
\label{tab:nets2}       
\end{table}
}

\newcommand{\figresone}{
\begin{figure}[t]
\begin{center}
	\subfigure[Optimal policy $u(x)=\mu(x)$ ($x_s=0$)]{%
		\includegraphics[width=0.48\columnwidth]{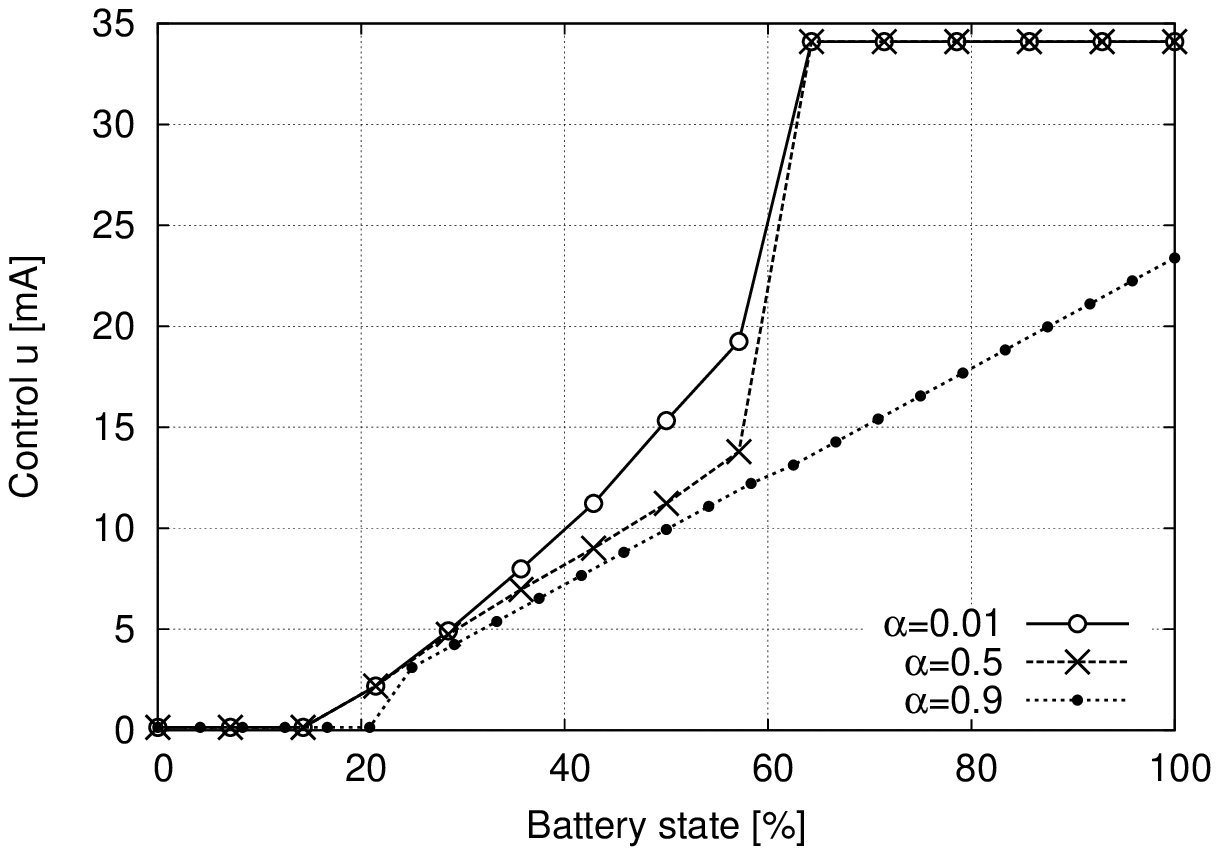}
    		\label{subfig:op_alpha_xs0}
	}
    	\subfigure[Steady-state distribution $P(x)$ ($x_s=0$)]{%
		\includegraphics[width=0.48\columnwidth]{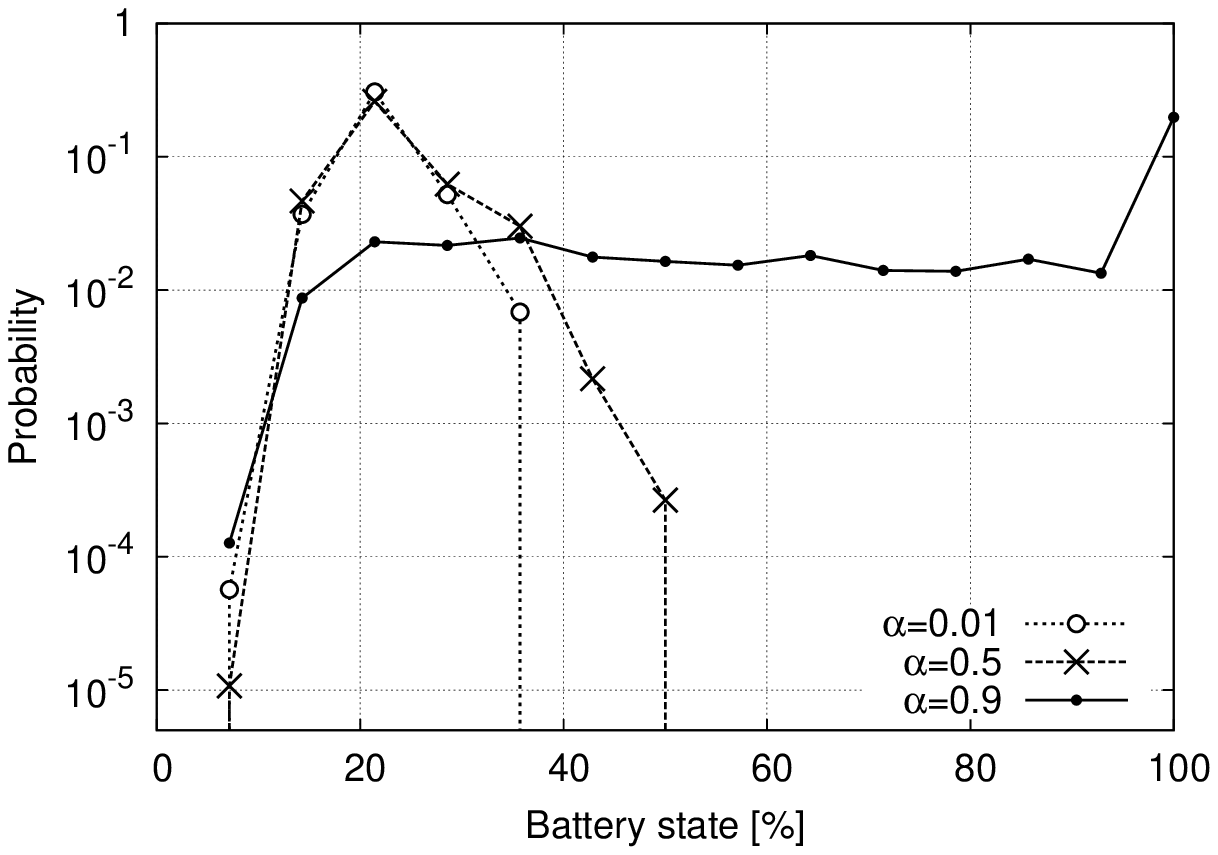}
    		\label{subfig:ss_alpha_xs0}
	}
\caption{Optimal policy $u(x) = \mu(x)$ and associated steady-state distribution $P(x)$ for the energy state $x_s=0$, $\alpha \in \{0.01,0.5,0.9\}$, $b_{\max}=250$~mAh, $b_{\rm th}=50$~mAh.}
\label{fig:op_alpha_xs0}
\end{center}
\end{figure}
}

\newcommand{\figrestwo}{
\begin{figure}[t]
\begin{center}
	\subfigure[Optimal policy $u(x)=\mu(x)$ ($x_s=1$)]{%
		\includegraphics[width=0.48\columnwidth]{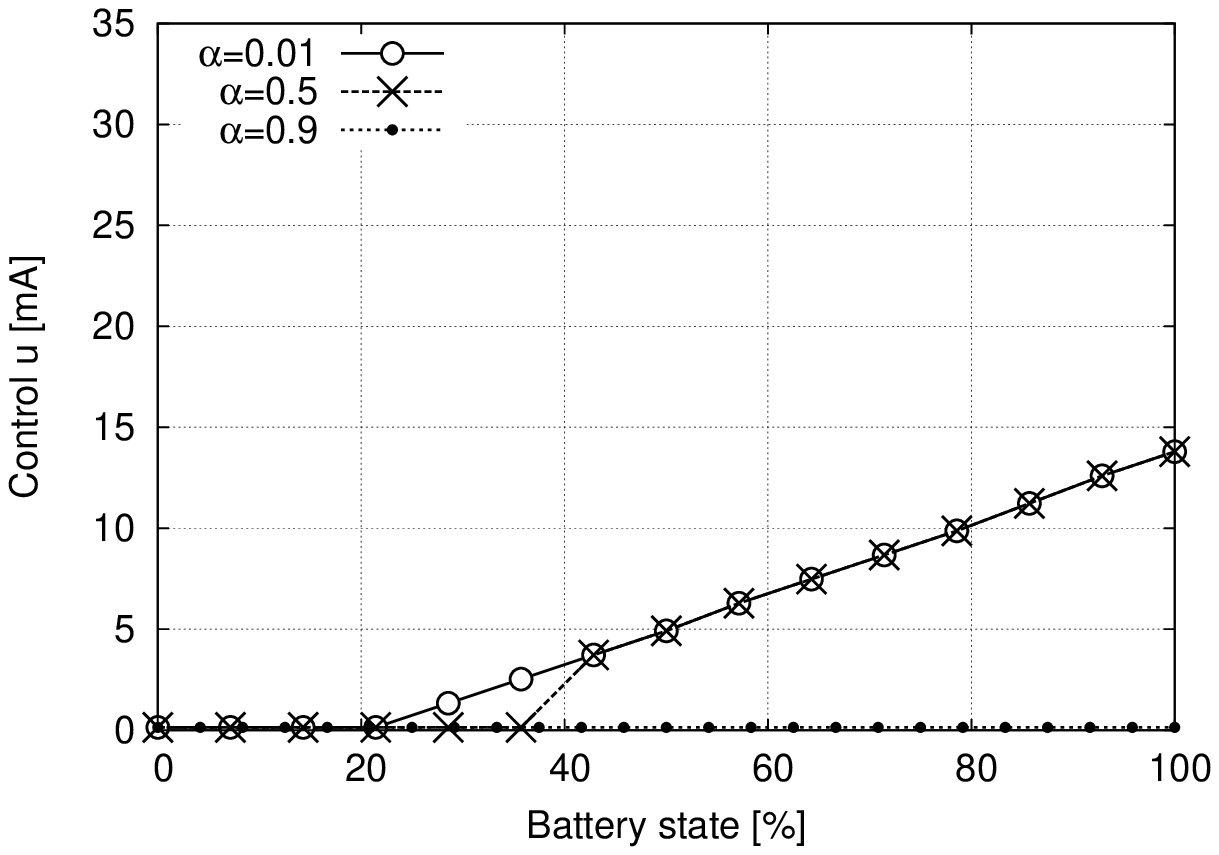}
    		\label{subfig:op_alpha_xs1}
	}
    	\subfigure[Steady-state distribution $P(x)$ ($x_s=1$)]{%
		\includegraphics[width=0.48\columnwidth]{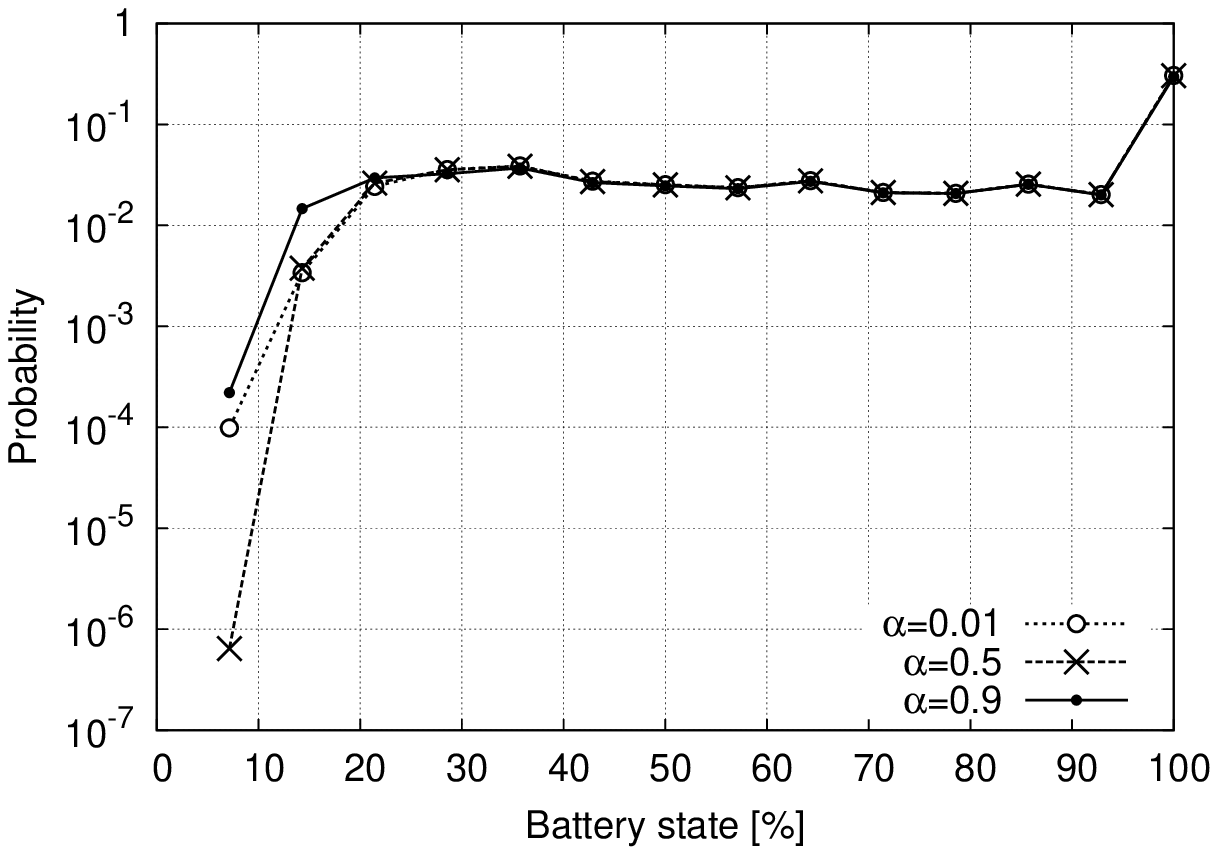}
    		\label{subfig:ss_alpha_xs1}
	}
\caption{Optimal policy $u(x) = \mu(x)$ and associated steady-state distribution $P(x)$ for the energy state $x_s=1$, $\alpha \in \{0.01,0.5,0.9\}$, $b_{\max}=250$~mAh, $b_{\rm th}=50$~mAh.}
\label{fig:op_alpha_xs1}
\end{center}
\end{figure}
}

\newcommand{\figresthree}{
\begin{figure}[t]
\begin{center}
	\subfigure[Optimal policy $u(x)=\mu(x)$ ($x_s=0$)]{%
		\includegraphics[width=0.48\columnwidth]{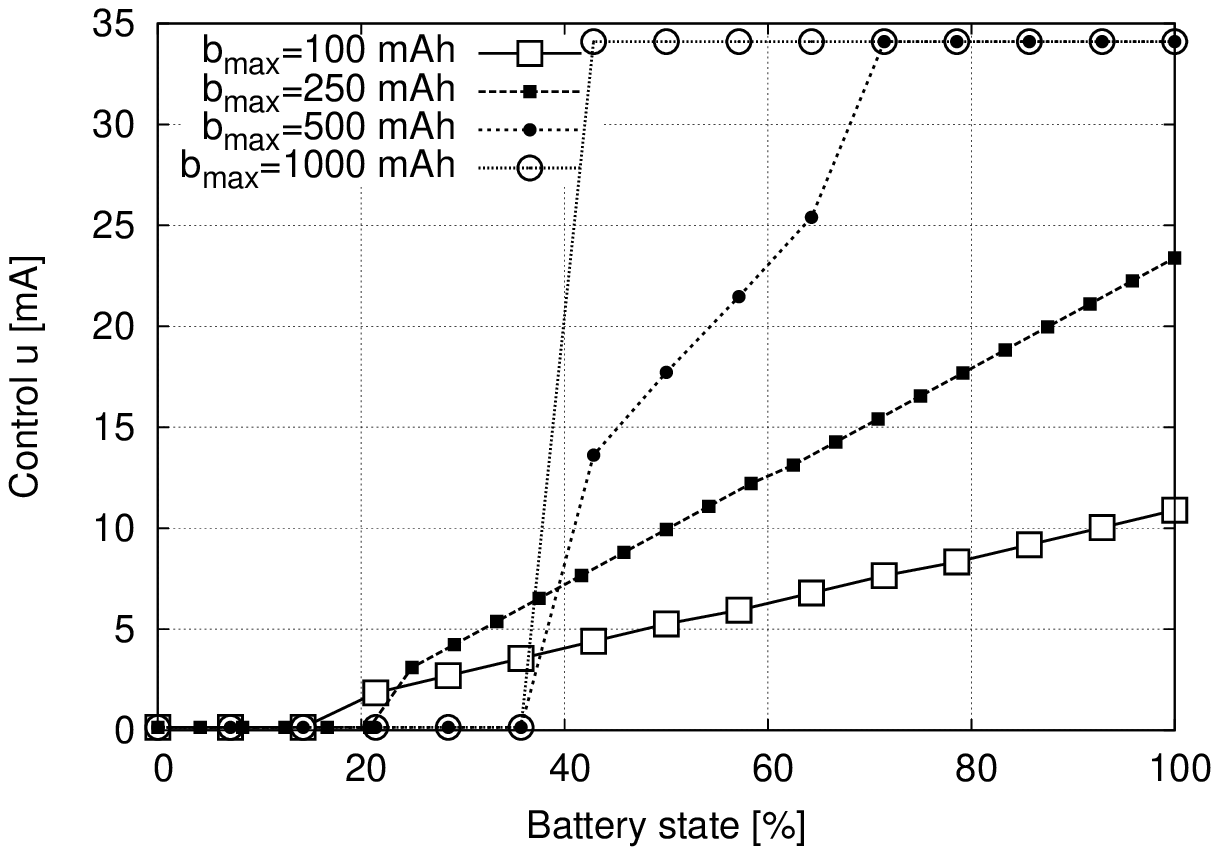}
    		\label{subfig:op_Bsize_xs0}
	}
    	\subfigure[Steady-state distribution $P(x)$ ($x_s=0$)]{%
		\includegraphics[width=0.48\columnwidth]{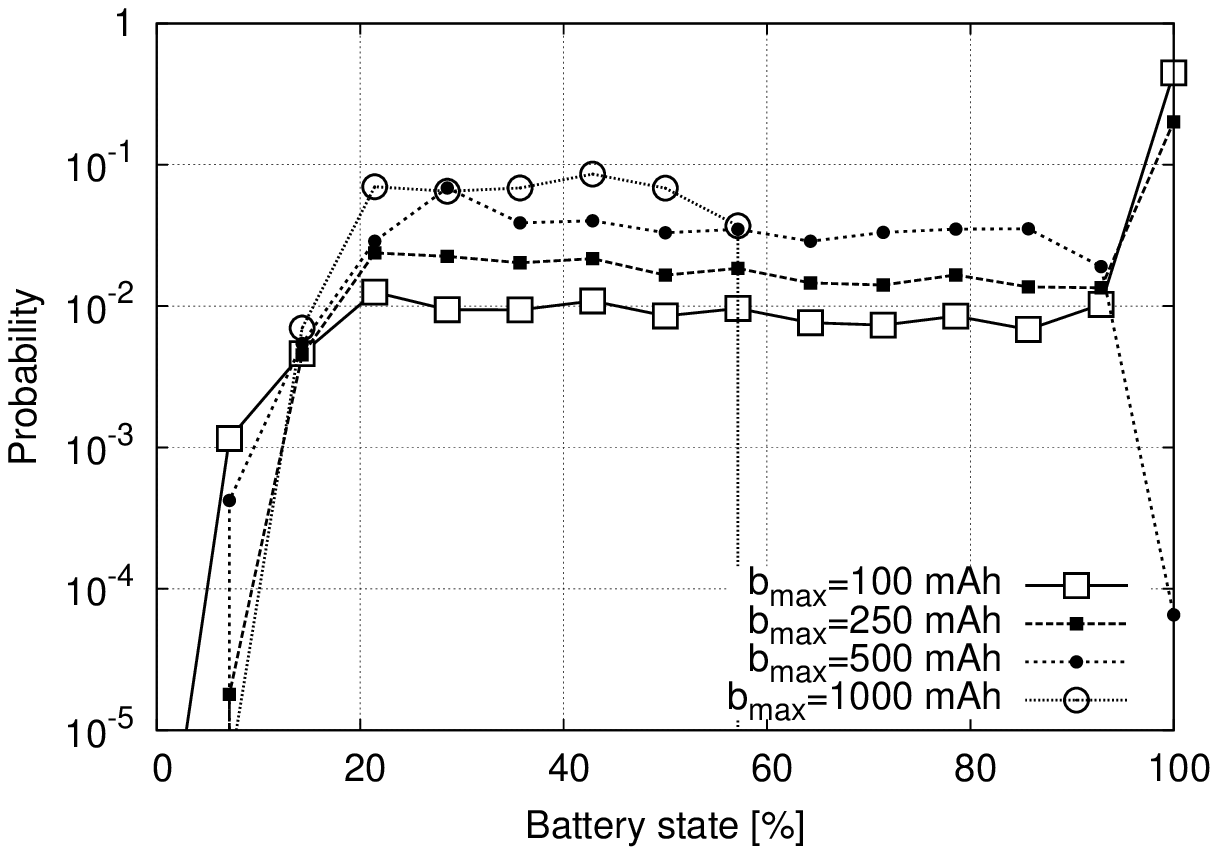}
    		\label{subfig:ss_Bsize_xs0}
	}
\caption{Optimal policy $u(x) = \mu(x)$ and associated steady-state distribution $P(x)$ for the energy state $x_s=0$, $\alpha =0.9$, $b_{\max} \in \{100,250,500,1000\}$~mAh, $b_{\rm th}=0.2 b_{\max}$.}
\label{fig:op_Bsize_xs0}
\end{center}
\end{figure}
}

\newcommand{\figresfour}{
\begin{figure}[t]
\begin{center}
	\subfigure[Optimal policy $u(x)=\mu(x)$ ($x_s=1$)]{%
		\includegraphics[width=0.48\columnwidth]{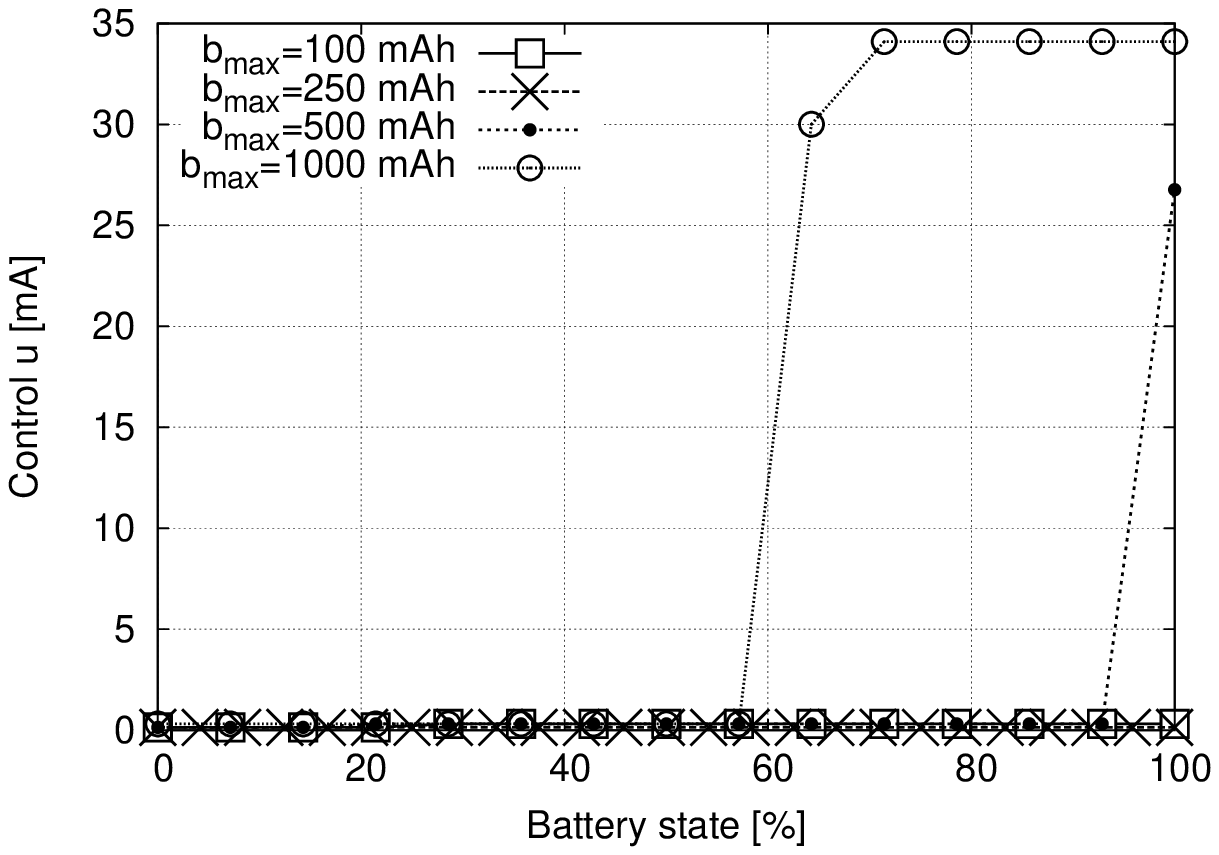}
    		\label{subfig:op_Bsize_xs1}
	}
    	\subfigure[Steady-state distribution $P(x)$ ($x_s=1$)]{%
		\includegraphics[width=0.48\columnwidth]{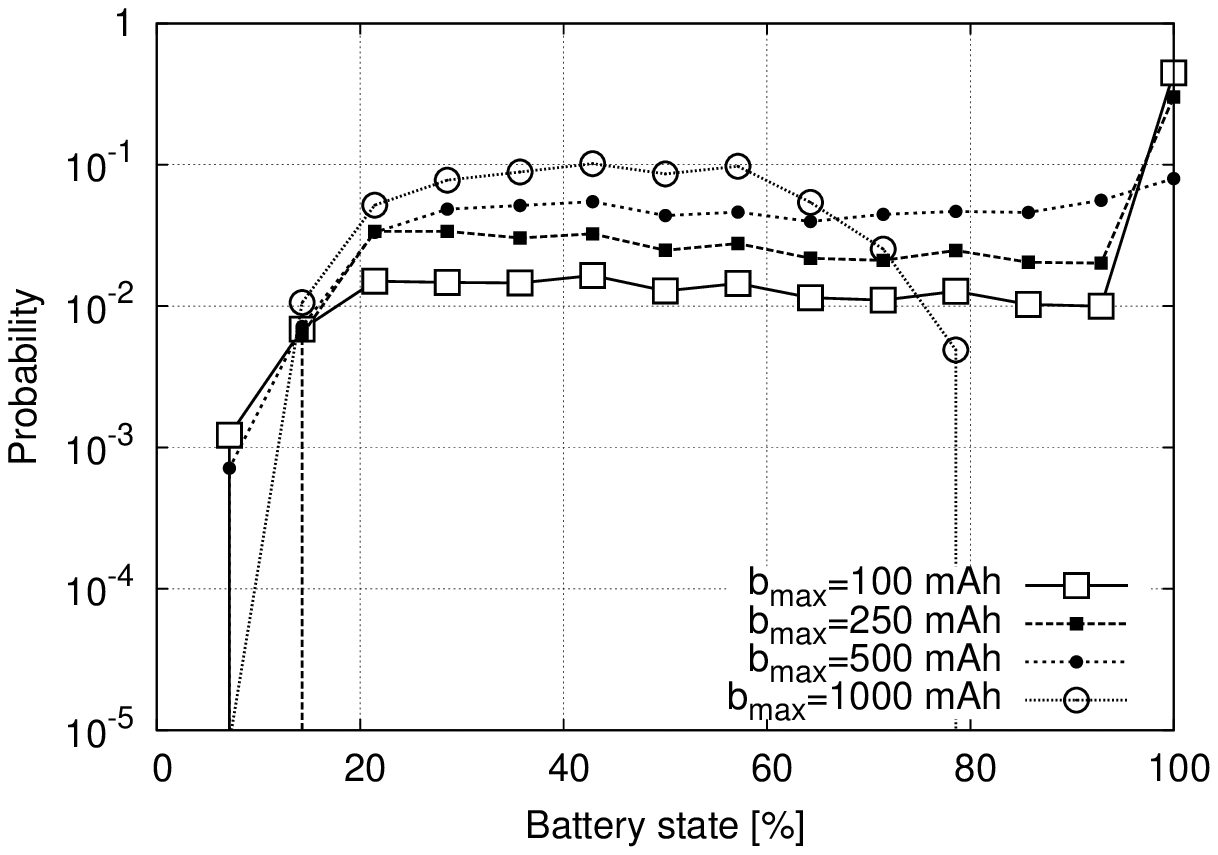}
    		\label{subfig:ss_Bsize_xs1}
	}
\caption{Optimal policy $u(x) = \mu(x)$ and associated steady-state distribution $P(x)$ for the energy state $x_s=1$, $\alpha =0.9$, $b_{\max} \in \{100,250,500,1000\}$~mAh, $b_{\rm th}=0.2 b_{\max}$.}
\label{fig:op_Bsize_xs1}
\end{center}
\end{figure}
}

\newcommand{\figresfivebis}{
\begin{figure}[t]
\begin{center}
\includegraphics[width=0.8\columnwidth]{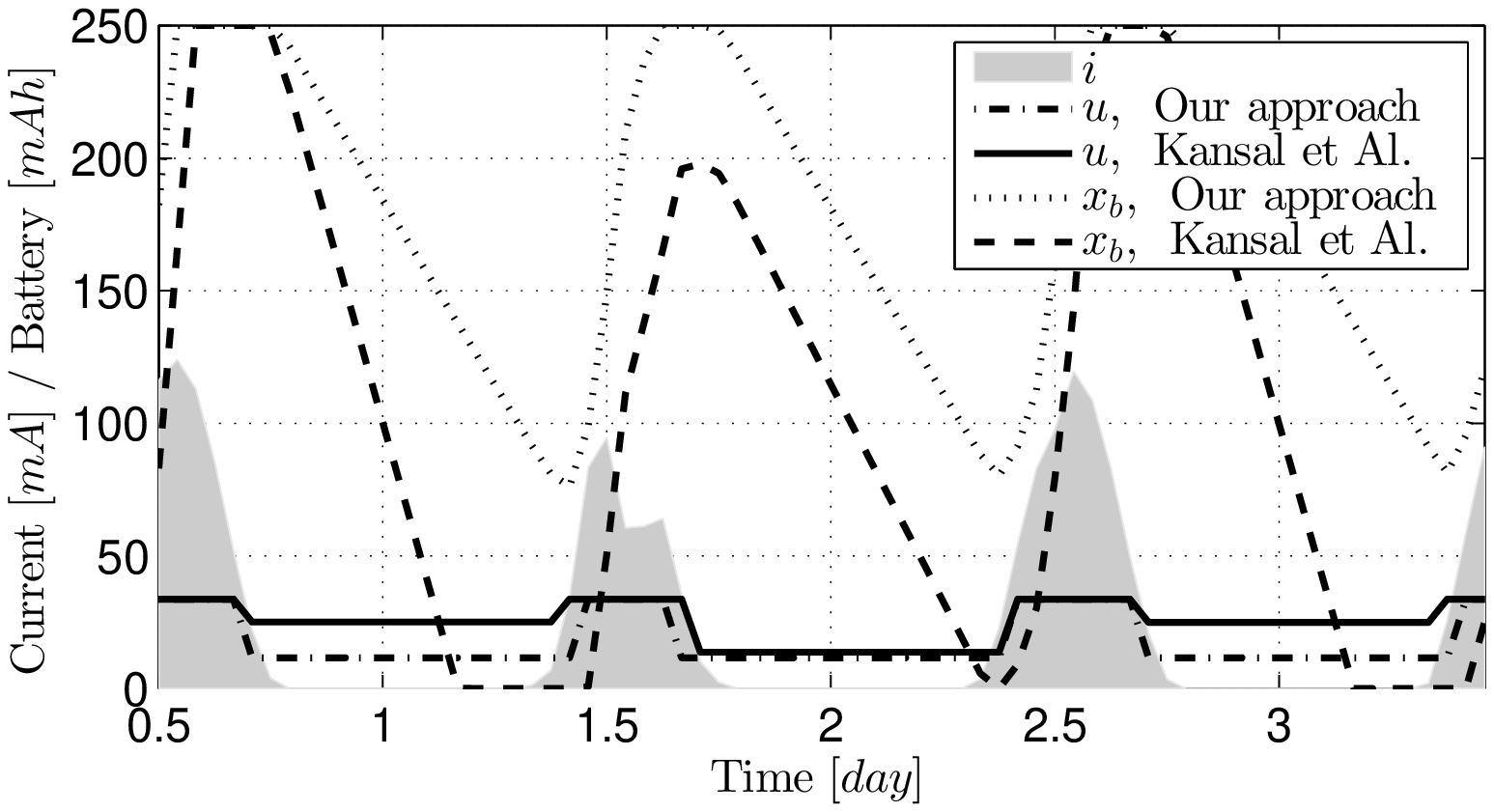}
\caption{\cc{Performance comparison between our solution and Kansal's during three simulated days for the month of August, considering a panel side of $10$~cm. The $x$-axis shows the simulation time, whereas the $y$-axis is used to visualize the amount of current harvested, drained, and the battery state.}}
\label{fig:fluct_comp}
\end{center}
\end{figure}
}

\newcommand{\figresfiveter}{
\begin{figure}[t]
\begin{center}
	\subfigure[Throughput]{%
		\includegraphics[width=0.4775\columnwidth]{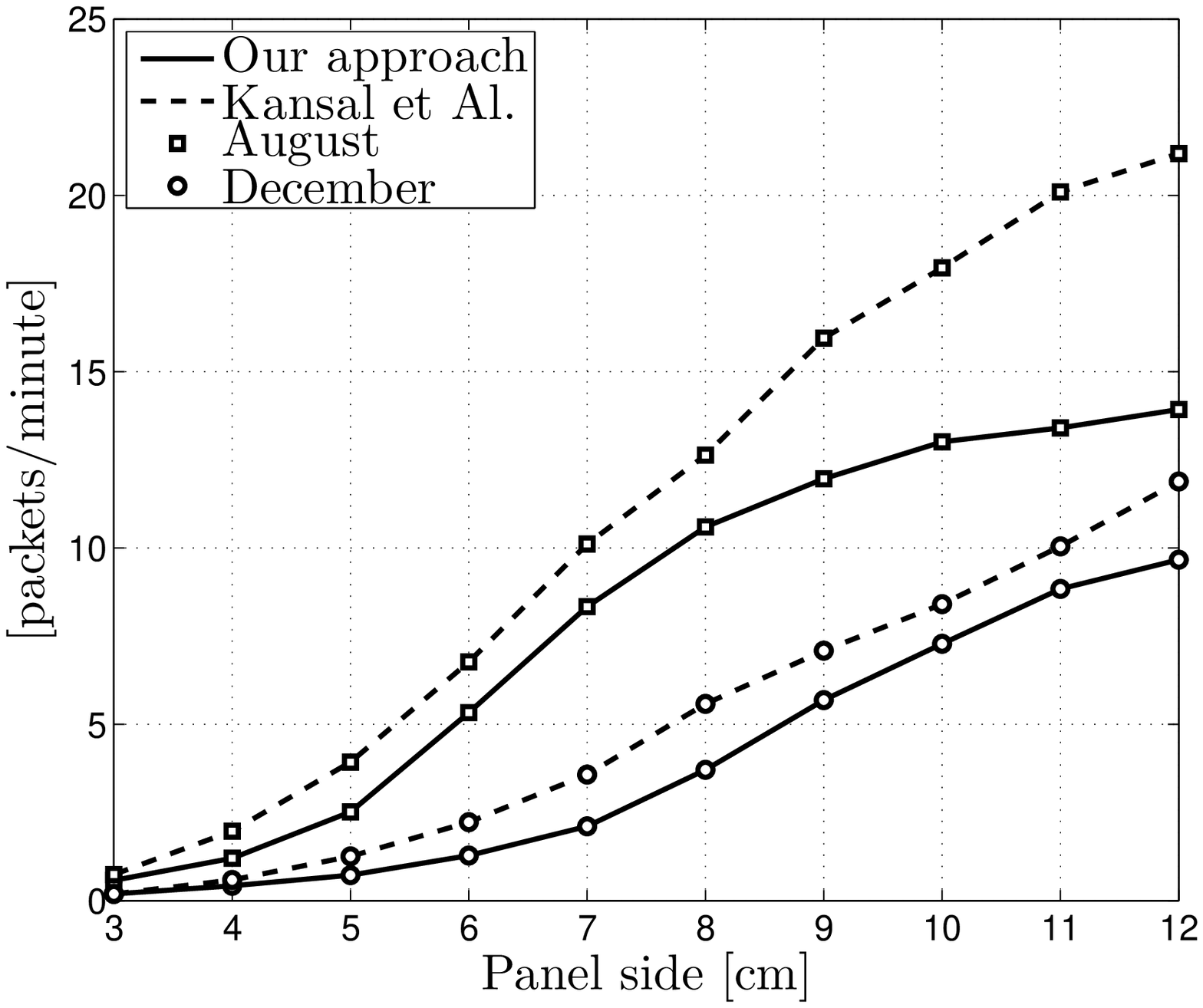}
    		\label{subfig:thr_comp}
	}
    	\subfigure[Outage]{%
		\includegraphics[width=0.495\columnwidth]{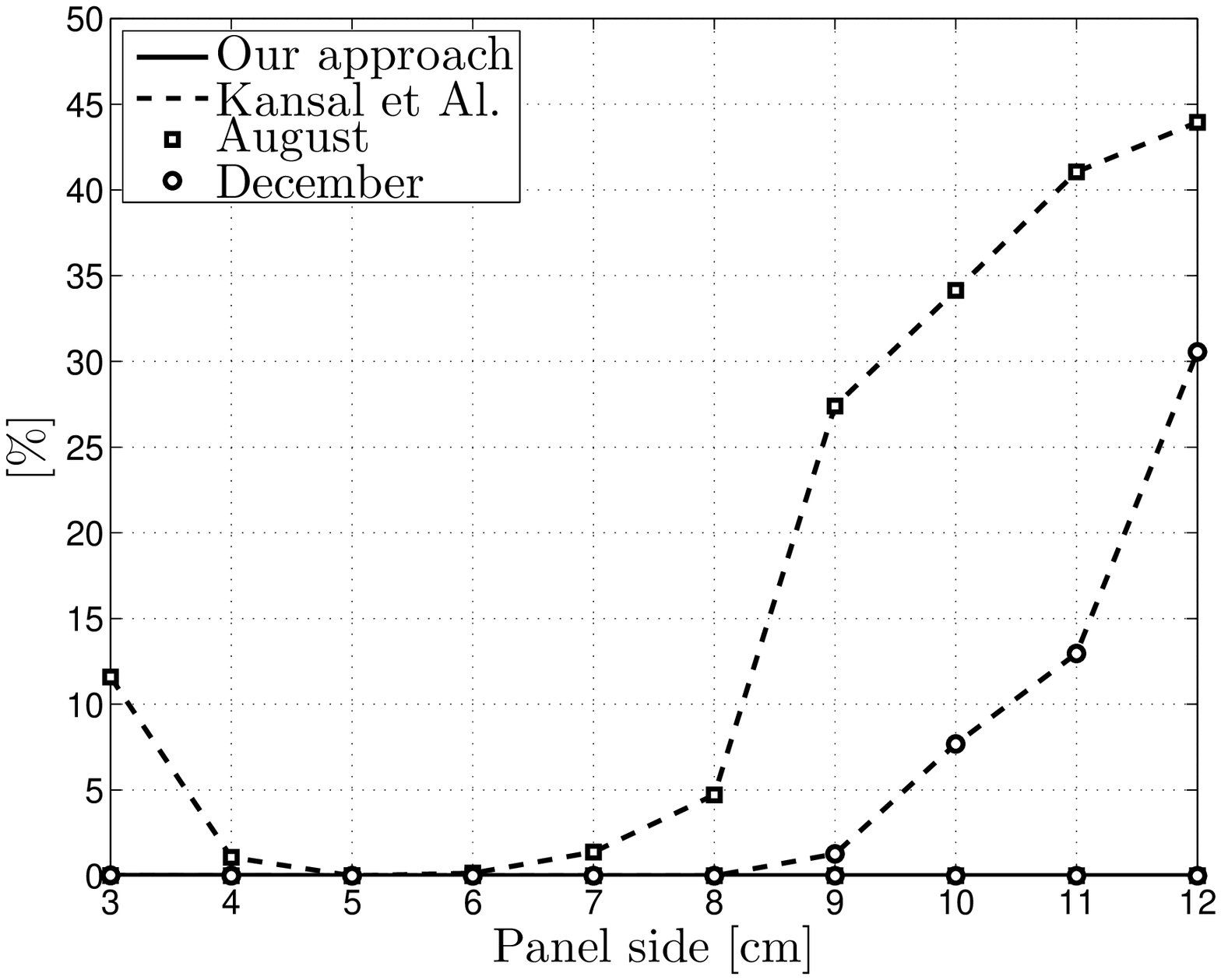}
    		\label{subfig:cost_comp}
	}
\caption{\cc{Performance comparison between our proposed technique (solid line) and that proposed by Kansal et al. (dashed line) for the months of August and December and varying the panel size.}}
\label{fig:comparisons}
\end{center}
\end{figure}
}

\newcommand{\figressix}{
\begin{figure}[t]
\begin{center}
	\subfigure[$f_\iota(i |x_s,p)$ (harvested current pdf)]{%
		\includegraphics[width=0.48\columnwidth]{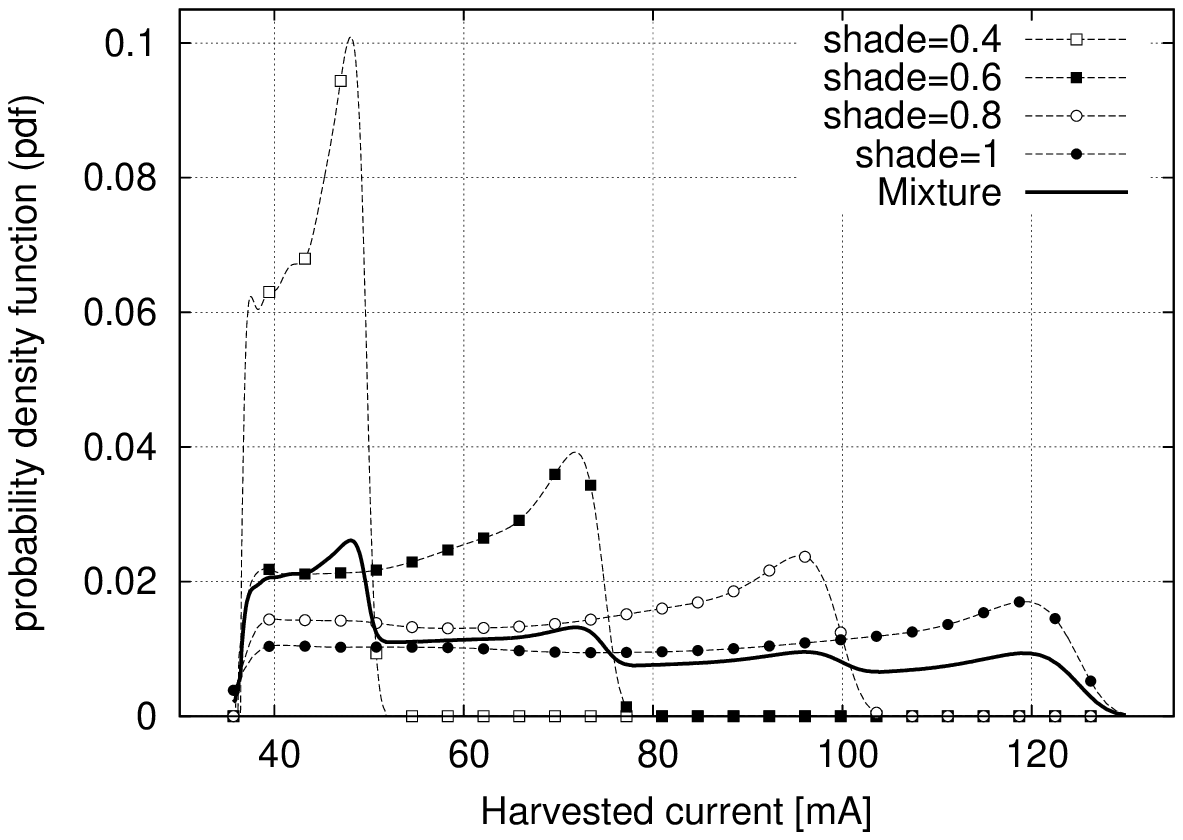}
    		\label{subfig:pdf_mixture_energy}
	}
    	\subfigure[$f_\tau(t |x_s,p)$ (permanence time pdf)]{%
		\includegraphics[width=0.48\columnwidth]{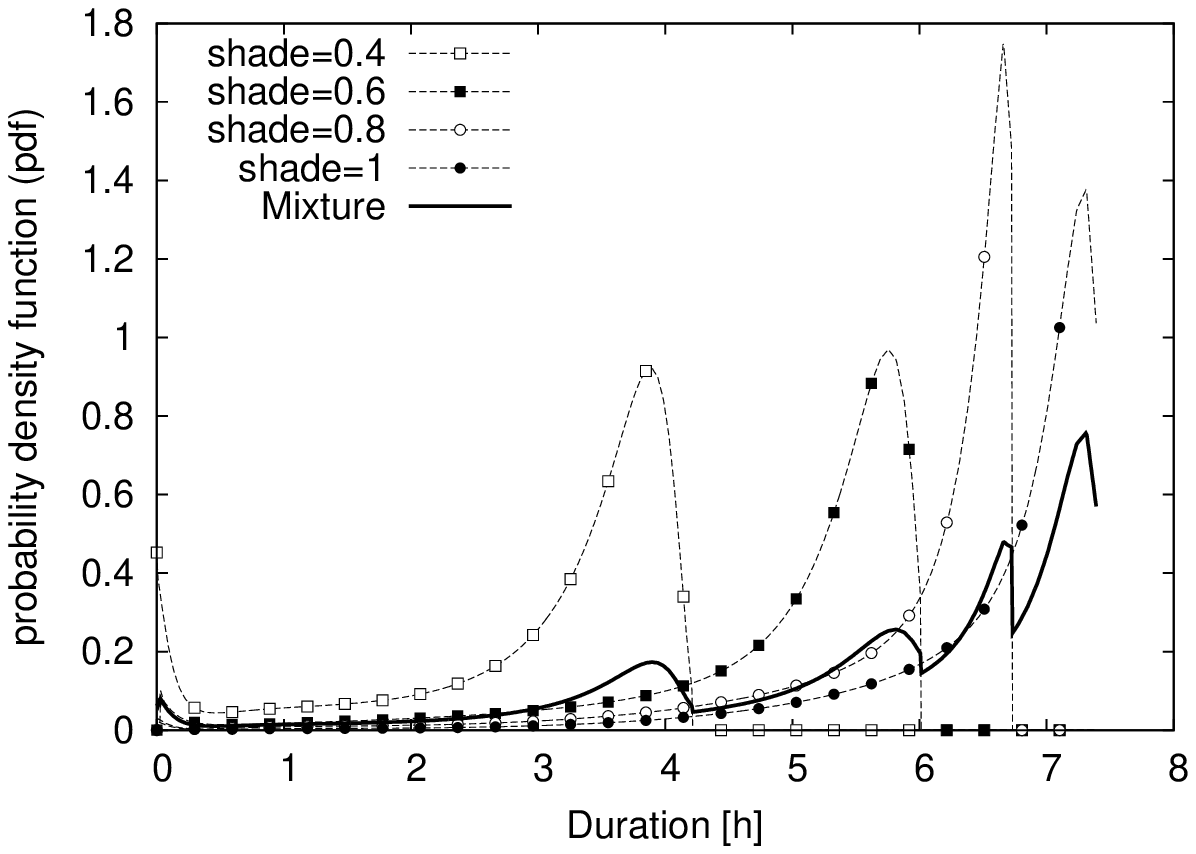}
    		\label{subfig:pdf_mixture_tau}
	}
\caption{Pdfs $f_\iota(i|x_s,p)$ and $f_\tau(t|x_s,p)$ obtained for $p \in \mathcal{D}$, state $x_s=0$ (daytime) for Los Angeles in the month of August. A  thick solid line is used to indicate the mixture densities.}
\label{fig:pdf_mixture}
\end{center}
\end{figure}
}

\newcommand{\figresheterofluct}{
\begin{figure}[t]
\begin{center}
\includegraphics[width=0.8\columnwidth]{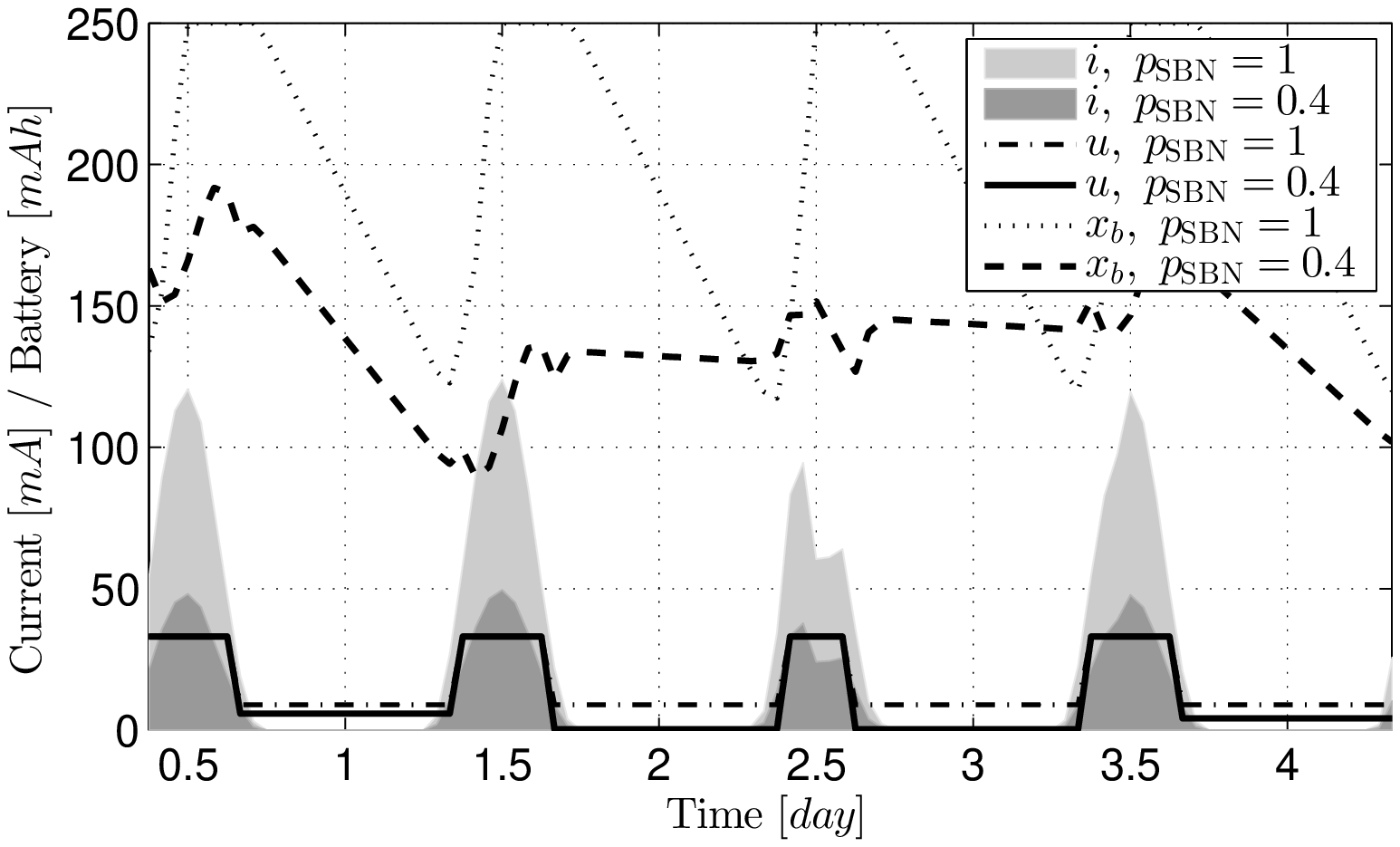}
\caption{\cc{Microscopic behavior of the heuristic policy in three simulated days. The $x$-axis shows the simulation time, whereas the $y$-axis represents, for the SBN node, the amount of current harvested, drained, and the battery state. The comparison is between the best ($p_{\rm SBN}=1$) and the worst ($p_{\rm SBN}=0.4$) case for the SBN node, considering $p=1$ (no shading) for the bottleneck node.}}
\label{fig:hetero_fluct}
\end{center}
\end{figure}
}

\newcommand{\figresheterocomp}{
\begin{figure}[t]
\begin{center}
\includegraphics[width=80mm]{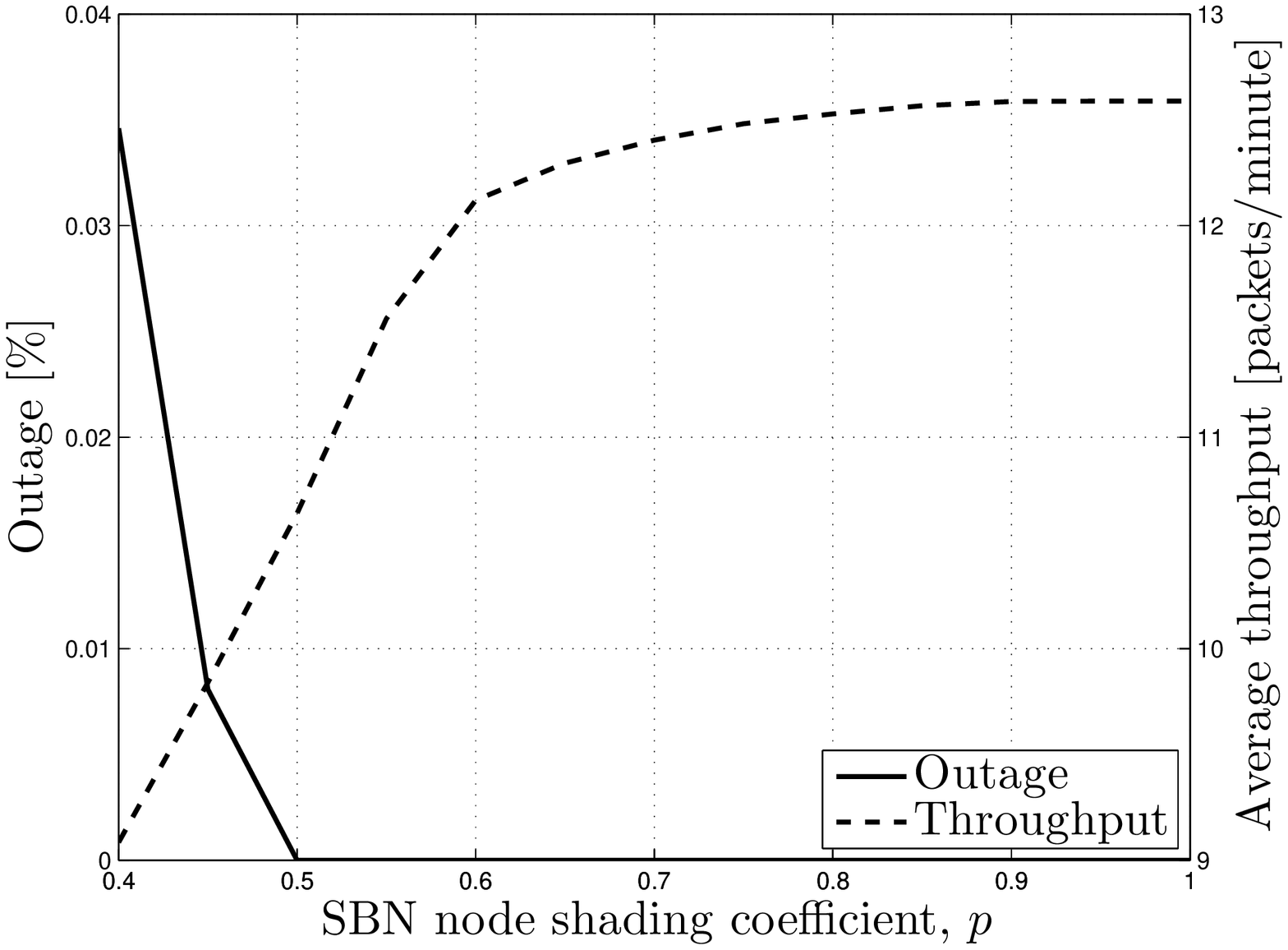}
\caption{\cc{Performance of the heuristic policy by varying the shade parameter $p \in [0.4,1]$ ($x$-axis). On the left $y$-axis we show the outage probability for the SBN node (solid line), while in the right $y$-axis we show the throughput.}}
\label{fig:hetero_comp}
\end{center}
\end{figure}
}

\newcommand{\tabsymone}{
\begin{table}[htb]
	\cc{
	\tbl{Notation.}{
	\centering
	\begin{tabular}{ l | l }
	\toprule
	Capital letters: N, S, etc. & denote system states and functional blocks. \\
	Capital letters, italic: $\io$, $\itx$, etc. & denote average quantities. \\	
	Lower letters, italic: $\toff$, $\tdc$, etc. & denote variables. \\
	Calligraphic font: $\mS$, $\U$, etc. & denotes sets. \\
	Greek letters: $\tau$, $\iota$, etc. & denote random variables. \\
	Bold letters: $\bm{p}$, $\bm{\rho}$, etc. & denote vectors. \\
	\end{tabular}}
	\label{tab:symb}}
\end{table}
}

\newcommand{\tabsymbtwo}{
\begin{table}[htb]
	\cc{
	\tbl{Symbol definitions.}{
	\centering
	\begin{tabular}{ l | l }
	\toprule
	S & energy source block in the system model. \\
	B & energy buffer (battery) block in the system model. \\
	N & energy consumer (sensor node) block in the system model. \\
	$\mathcal{N}$ & sensor nodes set.\\
	$i$ & harvested current.\\
	$u$ & \mr{control policy (drained current)}.\\
	$d_c$ & duty cycle.\\
	$\io$ & \mr{average current consumption for a given network configuration}.\\
	$\fu$ & packet transmission rate for endogenous traffic (reward).\\	
	\end{tabular}}
	\label{tab:symbtwo}}
\end{table}
}

\newcommand{\tabsymbthree}{
\begin{table}[htb]
	\cc{
	\tbl{Symbol definitions.}{
	\centering
	\begin{tabular}{ l | l }
	\toprule
	$x \in \XN$ & node operational state $x$ and state set $\XN$.\\
	$\fup$ & modified reward function accounting for retransmissions.\\
	$\ton$, $\toff$, $\tdata$, $\tdc$, $\tU$, $\tv$, $\trpl$ & sensor node timings.\\
	$i_x$, $I_x$ & instantaneous ($i_x$) and average ($I_x$) currents drained in state $x$.\\
	$\ic$, $\itr$, $\ir$, $\is$ & currents drained by the cpu ($\ic$), radio ($\ir, \itr$) and sensing unit ($\is$).\\
	$t_x$, $r_x$, $f_x$ & average duration, frequency and fraction of time spent in state $x$.\\
	$\ku$ & constant accounting for energy drained due to sensing and computation.\\
	$\nc$, $\nin$, $\nint$ & network topology parameters.\\
	$\etx$, $\ecoll$, $\ep$ & channel error ($\etx$), collision ($\ecoll$) and total error ($\ep$) probabilities.\\
	\end{tabular}}
	\label{tab:symbthree}}
\end{table}
}

\newcommand{\tabsymbfour}{
\begin{table}[htb]
	\cc{
	\tbl{Symbol definitions.}{
	\centering
	\begin{tabular}{ l | l }
	\toprule
	$t_x^*$ & optimal values for the variable $t_x$.\\
	$x^{\rm lim}$ & optimal values for the variable $x$, computed assuming no energy constraint.\\
	$x^{\rm min}$ & optimal values for the variable $x$, computed assuming zero reward (zero throughput).\\
	$a_i$, $b_i$, $c_i$, $d_i$, $e_i$, $f_i$ & coefficients. See Appendix~\ref{app:closed} and Table~\ref{tab:coeff2} for their complete definition.\\
	$u_{\rm max}$ & maximum allowed energy consumption for a sensor node.\\
	$u_{\rm min}$ & minimum required energy consumption so that the system remains operational.\\
	$\rho$ & node density.\\
	\end{tabular}}
	\label{tab:symbfour}}
\end{table}
}

\newcommand{\tabsymbfive}{
\begin{table}[htb]
	\cc{
	\tbl{Symbols used in the energy source model.}{
	\centering
	\begin{tabular}{ l | l }
	\toprule
	$x_s \in \mathcal{S}$ & energy source state $x_s$ and the set of all energy states, $\mathcal{S}$.\\
	$t_k$, $\Delta_k$ & transition time $t_k$ and epoch duration $\Delta_k$.\\
	$\tau_{x_s}$, $f_{\tau}(t | x_s)$ & r.v. and pdf describing the permanence time in state $x_s$.\\
	$\iota_{x_s}$, $f_{\iota}(i | x_s)$ & r.v. and pdf describing the current harvested in state $x_s$.\\
	$p_{ij}$ & transition probabilities of the source model's embedded Markov chain.\\
	$\delta = \delta_{\rm in} + \delta_{\rm out}$ & r.v.s. describing the total variation ($\delta$), the harvested ($\delta_{\rm in}$) \\
	& and the consumed ($\delta_{\rm out}$) charge in a decision epoch.\\
	$f_{\delta}(d | u,x_s)$ & pdf of the variation of charge in state $x_s$ when the control is $u$.\\ 
	\end{tabular}}
	\label{tab:symbfive}}
\end{table}
}

\newcommand{\tabsymbsix}{
\begin{table}[htb]
	\cc{
	\tbl{Symbols used in the MDP analysis.}{
	\centering
	\begin{tabular}{ l | l }
	\toprule
	$x_b \in \mathcal{B}=[0,b_{\rm max}]$ & buffer state $x_b$, buffer state set $\mathcal{B}$ and buffer size $b_{\rm max}$.\\
	$x=(x_s,x_b) \in \mathcal{X}=\mathcal{S} \times \mathcal{B}$ & system state $x$ in the current decision epoch, system state set $\mathcal{X}$, \\
	& source state set $\mathcal{S}$ and buffer state set $\mathcal{B}$.\\
	$y=(y_s,y_b) \in \mathcal{X}$ & system state in the next decision epoch.\\
	$u \in \mathcal{U} = [u_{\rm min},u_{\rm max}]$ & action (control) $u$ and action set $\mathcal{U}$.\\
	$\pi$, $\mu$ & policy $\pi$ and mapping $\mu$ between states $x$ and actions $u$. \\
	$r(u)$ & reward associated with action $u$.\\
	$R(x,u)$, $C(x,u)$ & single-stage expected reward $R(x,u)$ and cost $C(x,u)$.\\
	$J_R(x)$, $J_C(x)$ & optimal expected reward $J_R(x)$ and cost $J_C(x)$.\\
	$C_{\rm th}$ & threshold on the cost for the admissibility of the solution.\\
	$\alpha$ & discount factor.\\
	$\lambda$, $L_{\lambda}(x,u)$ & Lagrangian multiplier $\lambda$ and Lagrangian reward $L_{\lambda}(x,u)$.\\
	\end{tabular}}
	\label{tab:symbsix}}
\end{table}
}

\newcommand{\papertitle}{Staying Alive: System Design for Self-Sufficient Sensor Networks}
\newcommand{\papertitleshort}{Staying alive: system design for self-sufficient sensor networks}

\title{\papertitle}
\author{NICOLA BUI \affil{IMDEA Networks Institute}
MICHELE ROSSI \affil{University of Padova}}

\begin{abstract}
Self-sustainability is a crucial step for modern sensor networks. Here, we offer an original and comprehensive framework for autonomous sensor networks powered by renewable energy sources. We decompose our design into two nested optimization steps: the inner step characterizes the optimal network operating point subject to an average energy consumption constraint, while the outer step provides online energy management policies making the system energetically self-sufficient \cc{in the presence of} \mr{unpredictable} \cc{and intermittent energy sources}.
\mr{Our framework sheds new light into the design of pragmatic schemes for the control of energy harvesting sensor networks} and permits to gauge the impact of key sensor network parameters, such as the battery capacity, the harvester size, the information transmission rate and the radio duty cycle. 
\mr{We analyze the robustness of the obtained energy management policies in the cases where the nodes have differing energy inflow statistics and where topology changes may occur, devising effective heuristics.}
\mr{Our energy management policies are} \cc{finally evaluated considering real solar radiation traces, validating them against state of the art solutions and describing the impact of relevant design choices in terms of achievable network throughput and battery level dynamics.}
\end{abstract}




\acmformat{Nicola Bui, and Michele Rossi. 2014.
\papertitle.}

\begin{document}
\begin{bottomstuff}
 The research leading to these results has received funding from the Seventh Framework Programme (FP7/2007-2013) under grant agreement no. 251557 (Project SWAP).

Author's addresses: Nicola Bui, IMDEA Networks Institute, Av. Mar del Mediterraneo, 22, 28918, Madrid, Spain. Email: {\tt nicola.bui@imdea.org}. Michele Rossi, Department of Information Engineering (DEI), University of Padova, Via Gradenigo 6/B, 35131 Padova, Italy. Email: {\tt rossi@dei.unipd.it}.
\end{bottomstuff}

\maketitle

\section{Introduction}
\label{sec:intro}

The operation of wireless sensor networks powered by renewable sources is a very lively area of research, both theoretical and applied. 
This is due to the increasing inclination toward green systems and to the need for Wireless Sensor Networks (WSN) that can last unattended indefinitely. 
In fact, despite the advances in microprocessor fabrication and protocol design, batteries are expected to last for less than ten years for many applications and their replacement is in some cases prohibitively expensive. 
This problem is particularly severe for urban sensing applications, e.g., sensors placed below the street level, where the installation of new power cables is impractical. Other examples include body sensor networks or WSNs deployed in remote geographic areas~\cite{Wang-review-11}. 
In contrast, WSNs powered by energy scavenging devices provide potentially maintenance-free perpetual networks, which are particularly appealing, especially for highly pervasive Internet of Things~\cite{Morabito-IoT-10}.

In the past few years, a vast literature has emerged on energy harvesting WSNs. These networks are made of tiny sensor devices with communication capabilities, that also have an onboard rechargeable battery (also referred to as energy buffer) and are capable of scavenging energy from the surrounding physical environment. 
Most of the research papers that have been published so far deal with the {\it energy neutral design} of transmission policies, where the concept of energy neutrality accounts for the fact that the energy used, in the long term, should be equal to that harvested. 
%
Within this body of work, two well established approaches have been adopted to find energy neutral policies, namely, {\it offline} and {\it online}. Offline solutions are concerned with finding optimal packet transmission schedules, assuming that the nodes have full knowledge of the harvesting and information generation processes.
Although this is unrealistic, it provides useful insights into the design of online strategies. 
On the other hand, online approaches only assume some prior statistical knowledge about the energy arrival and the input data processes. 

\textbf{Offline approaches:} 
\cite{Ozel-offline-2011} considers a single sensor node transmitting data over a wireless fading channel with additive Gaussian noise and causal channel state information at the transmitter.
The authors of this paper obtain optimal policies considering two objectives: maximize the throughput by a deadline and minimize the transmission completion time. 
\cite{Yang-offline-2012} generalizes the results of~\cite{Ozel-offline-2011} by relaxing the assumption on packet arrivals, which can now arrive during transmissions. 
Also, this paper derives fast search algorithms leveraging structural properties of the solution.
Another recent work~\cite{Gregori-offline-2013} relaxes the assumption that the battery is infinite, obtaining optimal transmission policies for given Quality of Service (QoS) constraints, while fulfilling data and energy causality constraints. To the best of our knowledge, no papers in this category studied energy management policies for network of devices. 

\textbf{Online approaches:} these approaches differ in the stochastic model considered for the energy arrival process and in the optimization objective. Notably, only a few contributions addressed aspects related to multiple access and routing in distributed networks. \cite{Vigorito-SECON-07} presents a decentralized strategy for the control of an energy buffer with stochastic replenishment, through the adaptation of the transmission duty-cycle. This paper models the optimal buffer management as an online optimization problem, estimating the system dynamics using a gradient descent update rule and implementing energy-centric policies. 
%
\cc{Similarly, \cite{hsu2006adaptive} presents an adaptive duty cycling algorithm for energy harvesting sensor nodes.} 

The authors of~\cite{kansal2007power} study fundamental properties of energy harvesting processes and utilize them to devise an algorithm which maximizes the throughput based on energy prediction. 
\cite{fan2008steady} proposes a solution for high throughput with fairness guarantees, devising centralized and distributed algorithms that compute the optimal lexicographic rate assignment for all nodes.
\cite{Lei-online-09} develops a Markov decision analysis for a sensor node with i.i.d. stochastic replenishments (i.e., fixed energy arrival rate) and a finite energy buffer. 
The authors of this paper devise optimal online policies that depend on the importance of packets, which is modeled through a generic probability distribution function (pdf). 
The authors of~\cite{Vinod-online-10} propose throughput as well as delay optimal online policies for a sensor node with infinite data and energy queues. This paper considers stationary and ergodic arrival processes for data and energy and transmission over fading channels.
\cite{Michelusi-online-13} generalizes the results of~\cite{Lei-online-09}: 
it models energy replenishment through a two-state Markov model and associates a cost with data transmission. 
Optimal and heuristic policies are characterized considering the long-term data importance of transmitted data through a dynamic programming formulation. 
The focus of~\cite{Luo-online-13} is instead on practical circuits for energy harvesting wireless transmitters and on their impact on the design of optimal transmission policies for TDMA channel access.
The paper optimizes the time spent in storing energy and transmitting, while accounting for QoS constraints and a TDMA access scheme. 

Other approaches dealing with multiple access channels and, in turn, considering the simultaneous interaction of multiple sensor nodes are \cite{Gatzianas-2010,Longbo-2013},~\cite{Michelusi-ICC-13} and \cite{Tapparello-2014}. To our knowledge,~\cite{Gatzianas-2010} is the first contribution that has dealt with the distributed control of energy harvesting WSNs. There, the authors present an online and adaptive policy for the stabilization and optimal control of these networks using tools from Lyapunov optimization. 
This line of work has been continued by~\cite{Longbo-2013}, which tackles the distributed routing problem using the Lyapunov optimization theory combined with the idea of weight perturbation, see, e.g.,~\cite{Neely-TNET-08}. 
The authors of~\cite{Michelusi-ICC-13} consider a single hop WSN where each node harvests energy from the environment and randomly accesses the channel to transmit packets of random importance to a sink node. Thus, optimal distributed policies, based on a Game theoretic formulation of the random access problem are proposed. 
\cite{Tapparello-2014} presents a theoretical framework which extends~\cite{Gatzianas-2010,Longbo-2013} by proposing joint transmission, data compression (distributed source coding, DSC) and routing policies that minimize the long-term expected distortion of the signal reconstructed at the sink, while assuring the energetic stability of the network. 

Other research directions deal with energy sharing networks~\cite{zhu2010eshare} and laser-power beaming~\cite{bhatti2014sensors}. However, in the present contribution we neither look at the possibility of exchanging energy among nodes nor at performing wireless energy transfer. 
Further extensions may involve the adoption of energy aware programming languages~\cite{sorber2007eon}.

\textbf{Our contribution:} our present work belongs to the online category and \mr{considers} networks of energy harvesting devices. \mr{Specifically, we propose a framework based on the dynamic adaptation of two key protocol parameters, namely, the radio {\it duty cycle} $d_c$  and the {\it transmission frequency} for the own generated traffic, $\fu$. This framework permits to assess the performance of energy harvesting sensor networks, while shedding new light into the pragmatic design of energy management solutions.}

Toward this end, we account for: 1) the network topology, 2) the transmission of endogenous (own packets) data, 3) the relaying of exogenous (forwarded) data, 4) the amount of energy consumed for transmission, reception, idling, processing, etc., 5) the channel access mechanism and 6) the harvested energy inflow dynamics. 
For the channel access, we consider the Low Power Listening (LPL) MAC~\cite{buettner2006x,bonetto2012mcmac}, whereas routing dynamics are modeled through the IETF Routing for low Power Lossy networks (RPL)~\cite{ko2011evaluating,bui2012}. 

Technically, our first contribution is a model that, for any pair $(d_c,\fu)$, returns the associated average energy consumption of a sensor node, taking 1)--5) as input. We obtain (in closed form) the pair $(d_c^*,\fu^*)$ that maximizes the node throughput subject to a given energy constraint. 
We subsequently locate the bottleneck node in the network (the one suffering the highest amount of interference) and we carry out a further optimization step based on 6) keeping this worst case into account. 
The resulting policies dynamically select the pair $(d_c,\fu)$
considering the state of the bottleneck node along with the stochastic model of the harvested energy. 
Being dimensioned for the worst case, the obtained policies can be applied at all nodes, leading to the self-sufficient operation of the entire WSN. \mr{Hence, we comment the behavior of the obtained energy management policies and we compare their performance against that of competing solutions from the state of the art}. \cc{\nb{Finally, we relax each of the model assumptions, showing that the solutions so obtained are still robust.}}


In summary, the main contributions of the present paper are:
\begin{enumerate}
\item a model for the energy consumption of a network of embedded wireless devices;
\item a closed form formula for the optimal operating point of the network;
\item a mathematical framework to 
maximize the throughput performance, 
while allowing the perpetual operation of the entire sensor network;
\item a performance evaluation of the proposed energy management policies; 
\item a validation of the proposed solution when the model assumptions are relaxed.
\end{enumerate}


\cc{In Table~\ref{tab:symb}, we introduce the notation used in the rest of the paper. Additional definitions will be given at the beginning of each section.}

\tabsymone

The remainder of this paper is organized as follows. In Section~\ref{sec:architecture}, we describe the workflow of the paper, detailing the objectives of our design and how these are accomplished by the analyses that follow. 
In Section~\ref{sec:node} and Section~\ref{sec:nodeanal}, we characterize the energy consumption of a sensor node according to the network properties and we derive the optimal operating point for the network subject to input energy constraints. 
In Section~\ref{sec:source} we present a stochastic semi-Markov model for the harvested energy and in Section~\ref{sec:smdp} we obtain energy management policies for self-sufficient networks of embedded devices. In Section~\ref{sec:results} and Section~\ref{sec:relax}, we \mr{evaluate the proposed policies} and, in Section~\ref{sec:conclusions}, we present our closing remarks.  

\section{Problem Formulation}
\label{sec:architecture}

\cc{In this section we describe the problem formulation as two nested optimization problems. The list of used symbols is given in Table~\ref{tab:symbtwo}}.

\tabsymbtwo

\figarch

We consider a wireless sensor network $\mathcal N$ composed of $N=|\N|$ {\it homogeneous} embedded devices, where sensor nodes transmit their readings to a data collector node (referred to as {\it sink}). The nodes are deployed according to a certain multi-hop topology, and the data packets are routed toward the sink through a pre-determined collection tree, as detailed in Section~\ref{sec:node}. Each sensor node is described through the diagram in Fig.~\ref{fig:sensor_diagram}. Specifically:
\bi
\item \textbf{Energy source (S):} this block accounts for the presence of some energy scavenging circuitry that feeds a storage unit. The amount of harvested current is described by the variable $i$. A detailed description of a stochastic semi-Markov model of S is provided in Section~\ref{sec:source}. Note that, while the energy scavenged is stochastic across time, we initially assume that it is described by the same Markov source for all nodes. The extension to heterogeneous energy sources is provided in Section~\ref{sec:hetero}.
\item \textbf{Battery (B):} the storage unit (e.g., either a rechargeable battery or a super-capacitor) provides an average current $u$ to the following block N, see Section~\ref{sec:smdp}.
\item \textbf{Sensor node (N):} this block models the aggregate energy consumption of a sensor node, which is referred to as $\io$. This accounts for the energy drained by the sensor node hardware, including the network protocol stack (e.g., routing,  channel access and physical layer), the onboard sensors and the CPU. The energy consumption of block N is characterized in Section~\ref{sec:node}. 
\ei

The overall objective of our analysis is providing dynamic and energy-dependent (i.e., depending on the state of S and B) configurations for the sensor nodes in $\N$ so that the entire network will be energetically self-sufficient. 

To accomplish this, for a given network setup, we first identify the so called {\it bottleneck} node, which is the node experiencing the highest traffic load.
This node is by definition the one subject to the highest energy consumption (more precise details will be given in Section~\ref{sec:node} and in Appendix~\ref{app:stability}).

Our analysis develops along the following two optimization steps:
\bi
\item[1)] We first characterize the energy consumption of the bottleneck node, for the given routing topology and channel access technology. In detail, we relate its average energy consumption, $\io$ (assumed constant for this first analysis), to two key parameters: the {\it radio duty-cycle}, $d_c$, and the {\it transmission frequency} for the endogeneous traffic, $\fu$. Given this, we solve a first optimization problem P1 (the inner problem in Fig.~\ref{fig:sensor_diagram}), where we seek the operational point (i.e., the pair $(d_c,\fu)$) for which $\fu$ is maximized considering $u$ as the the energy consumption constraint. To solve P1, we model the interaction of the bottleneck node with respect to the other sensors in $\N$, accounting for the transmission behavior of all nodes within range (e.g., the amount of relay traffic from the children nodes, the total traffic that these forward on behalf of their children, the number of interferers and their transmission rate, etc.). 
Subsequently, we derive in closed form the optimal protocol configuration $(d_c,\fu)$ for a given average energy consumption constraint $u$.

\item[2)] In the second optimization step (problem P2), we additionally account for the presence of blocks S and B, where S is modeled through a stochastic time-correlated Markov model, where the harvested current $i$ is assumed to be a time-varying, correlated stochastic process and $u$ is now the control variable. Problem P2 consists of dynamically selecting the control $u$ (or, equivalently, the pair $(d_c,\fu)$, where the relation $u \to (d_c,\fu)$ follows from the solution of P1), for the given energy source model, so that the bottleneck will maximize its own throughput, while being energetically self-sufficient.
\ei

At this point, we combine the results of P1 and P2: 
P1 decides the optimal operating point for the bottleneck as a function of $u$, whereas P2 dictates how $u$ should vary as a function of the battery state and on some statistical knowledge of the energy harvesting process. 
This combined optimization amounts to a dynamic selection of the current level $u$ that has to be drained by the node, depending on the state of S and B, so that the throughput is maximized (P1) and the node is energetically self-sufficient (P2). 

After solving this combined problem, the self-sufficiency of all network nodes can be assured by the following scheme. The time is divided into a number of slots, which depend on the temporal characterization of the energy scavenging process, see~Section~\ref{sec:source}. 
A {\it decision epoch} occurs at the beginning of each slot, i.e., when the source model transitions to a new state. Thus, at each epoch the sink collects the information about the state of the battery of the bottleneck node, computes the optimal actions (using P1 and P2) for the next time slot for this node, and sends back a description of the computed optimal policy to all network nodes. Thus, all nodes will implement, in the next time slot, the policy that is optimal for the bottleneck. Consequently, the energetic stability at all nodes is assured. \cc{This can be conveniently implemented through a practical network management and routing protocol such as RPL~\cite{winter2010rpl}.}

In this paper we look at a course-grained control of the protocol behavior of the nodes. In fact, one control command has to be sent out to the nodes at the beginning of every time slot, whose duration depends on the number of states that are used to model the energy inflow during a typical day. While our mathematical analysis holds for any number of energy states, practical considerations related to the network overhead incurred in sending control actions to the nodes, and to the number of states that is sufficient to accurately model, e.g., typical solar sources, lead to slot durations of the order of hours. 

In Section~\ref{sec:node}, for a given network scenario (i.e., transmission model, topology and data collection tree), we characterize the energy consumption of the bottleneck node. Thus, in the next Sections~\ref{sec:nodeanal} and~\ref{sec:smdp} we respectively solve problems P1 and P2 for this node, assuming that all the remaining nodes in the network behave in the same exact manner as the bottleneck does. 

In Section~\ref{sec:hetero} we extend our analysis to the case where the sensor nodes harvest different amounts of energy. 

\section{Node Consumption Model}
\label{sec:node}

The symbols used in this section are listed in the following Table~\ref{tab:symbthree}.

\tabsymbthree

In this section, we discuss the sensor node block of our architecture: this entails the definition of a tractable framework to model the interactions among nodes, including routing and channel access (MAC). 
We require the model to track network characteristics such as the topology, the adopted MAC protocol, channel errors and internal processing (assembling data packets, etc.). 
Although our framework develops along the lines of~\cite{fischione-13}, we aim at obtaining simple and meaningful relationships, that will make it possible to compute the optimal throughput in closed-form.

For tractability, we make the following assumptions: 
\begin{itemize}
\item[1)] there exists a node that consumes more energy than any other sensor. This node is referred to as the \emph{bottleneck} node; 
\item[2)] every sensor operates as the bottleneck node in terms of information generation rate, $\fu$ (expressed in packets per second), and duty cycle, $d_c=\ton/\tdc = \ton/(\ton+\toff)$, where $\tdc = \ton+\toff$, whereas $\ton$ and $\toff$ are the durations of the active and sleeping portions of the duty cycle, respectively;
\item[3)] the sink at each decision epoch (see Section~\ref{sec:smdp}) collects the status of the bottleneck, in terms of energy reserve, and broadcasts a feedback message to adapt the protocol behavior of all nodes. 
We provide practical considerations on how to deal with dissemination delays in Section~\ref{sec:relax}; 
\item[4)] the sensor nodes maintain the same behavior for long enough to justify the use of average energy consumption figures. Specifically, the time scale at which the sink takes control actions is much coarser than that related to the radio duty cycling. 
\end{itemize}

To start with, we identify the operational states of a sensor node and, for each of them, the associated energy expenditure (expressed here in terms of the current $i_x$ drained in each state $x$):
\begin{itemize}
\item {\bf TX}: this is the transmission state. Here, both the microprocessor and the radio transceiver are active and the current drained in these states is $\ic$ and $\itr$, respectively. 
\item {\bf RX}: in this state a node receives and decodes a radio frame. As for the TX state, both the microprocessor and the radio transceiver are on and, in this case, their energy drainage is $\ic$ and $\ir$, respectively. 
\item {\bf INT}: in this state the node receives a frame that is neither intended for it nor it has to be forwarded by it. Here, the node drains exactly the same current as in state RX. In the following analysis, we track this state separately from RX as the rate of interfering and successful transmissions may differ.
\item {\bf CPU}: the node is busy with operations that do not require any radio activity (e.g., sensing, data processing, encoding, etc.). In this state, the radio transceiver is off or in a power saving state, thus the consumption is just $\ic$.
\item {\bf IDLE}: the node is idle and can switch to some low-power state. However, since preamble-sampling MAC protocols, such as X-MAC~\cite{buettner2006x} or Low-Power Listening (LPL)~\cite{moss2007low}, need to periodically sample the radio channel while idling, it is convenient to split this state into two sub-states:
\begin{itemize}
\item {\bf CCA}: in this state, the node samples the channel (Clear Channel Assessment). Hence, it drains the same current as in RX. 
\item {\bf OFF}: this is the state with the lowest energy consumption. Here, the microprocessor and the radio transceiver are in power saving mode and the total current drained by the device is $\is$, which is much smaller than all the other energy consumption figures ($\is \ll i_x, x \in \{{\rm t},{\rm r},{\rm c}\}$).
\end{itemize}
\end{itemize}

We now formally introduce the system state set
\be
\label{eq:state_set}
\XN = \{{\rm TX, RX, INT, CPU, CCA, OFF}\} ,
\ee
where for the IDLE state it holds ${\rm IDLE} = {\rm CCA} \cup {\rm OFF}$. The main idea behind our model consists of computing the average current $I_x = \E[i_x]$ drained by the bottleneck node for each state $x \in \XN$, for the given protocol and network parameters. Note that, in our model computing average currents is equivalent to computing powers, as we assume that the sensors operate according to a fixed supply voltage. For each $x \in \XN$, we have that: $I_x = i_x t_x f_x$, where $i_x$, $t_x$ and $f_x$ correspond to the drained current, the average permanence time (duration) in state $x$ and the average rate (frequency) at which state $x$ is entered, respectively. In addition, we  use the quantity $r_x = t_x f_x$ to indicate the average fraction of time the node spends in state $x$. Hence, the average output current $\io$ is obtained by the sum of the average currents:
\be
\label{eq:iot_tot}
\io = \sum_{x \in \XN} I_x .
\ee
To find $f_x$ and $t_x$, we make the following choices: 
\begin{enumerate}
\item[1)] the main function of the nodes is that of sensing environmental data and sending them to the sink \cc{\nb{(Section~\ref{sec:relax} describes how to account for event-driven WSNs}});
\item[2)] at the channel access, we adopt a preamble-based transmitter-initiated MAC protocol, such as X-MAC (that exploits a Low Power Listening strategy)~\cite{buettner2006x};
\item[3)] network configuration and maintenance is managed via a distributed protocol, such as RPL (IPv6 Routing Protocol for Low power and Lossy Networks)~\cite{winter2010rpl}.
\end{enumerate}

From the first choice, we assume that the nodes periodically sense the environment and generate their data at a constant rate of $\fu$ packets per second, where $\tU=1/\fu$ is the average inter-packet generation time \cc{\nb{(practical details on how to deal with non periodic traffic are provided in Section~\ref{sec:relax})}}. Also, each data packet is assembled considering the data from $\ku \geq 1$ sensor readings; $\ku$ can be used to account for additional processing of data and other operations that do not involve radio activity. Note that $\fu$ is the nominal transmission rate, that is only obtained for a collision and error-free channel. In practice, given that multiple nodes share the same transmission medium, packets can be lost due to, e.g., collisions or transmission errors. Taking some error recovery into account (retransmissions), the actual transmission rate will be $\fu^\prime \geq \fu$. 

For the routing, each node forwards its data packets either to the sink or to its next-hop node (referred to as its parent node). Also, each node sends its own information packets (this is referred to as {\it endogenous} traffic), as well as the packets generated by other nodes ({\it exogenous} traffic, in case the node acts as a relay for its children nodes). 

To illustrate our network setting we refer to the topology example of \fig{fig:topo}, where the bottleneck node is represented as a black dot, while the sink is placed in the center of the network. In this figure, a possible realization of the routing tree is also shown. In particular, the links represented with solid lines belong to the sub-tree rooted at the bottleneck. White filled dots indicate the nodes that use the bottleneck to forward their data to the sink (these are referred to as children nodes), while white triangles indicate the nodes whose traffic can interfere with that of the bottleneck (interfering nodes). Crosses indicate the position of all the other nodes.

\figtopo

For our model, we consider the topology, the data gathering tree and the coverage range as given. Also, we only track the number of children and interfering nodes, disregarding their actual position. Given this, next we refer to the following quantities as the input parameters for our analysis: 
\bi
\item[1)] $\nc$: is the number of children nodes, i.e., the total number of nodes in the sub-tree rooted at the bottleneck. $\nc$ governs the total traffic that has to be relayed by the bottleneck node.
\item[2)] $\nin$: is the number of interfering nodes (white triangles of \fig{fig:topo}). These are within the transmission range of the bottleneck (i.e., within one hop from it), but the latter is not their intended next-hop. Any transmission from one of these $\nin$ nodes can either be a spurious reception or a collision for the bottleneck.
\item[3)] $\nint$: this corresponds to the total number of packets the bottleneck may be interfered from, \cc{i.e., the sum of the traffic load (endogenous and exogenous) from all the interfering nodes.}
Note that in general $\nint > \nin$. 
\ei
Note that $\nc$ especially depends on the size of the network in terms of number of communication hops, while $\nin$ and $\nint$ increase with the node density. Finally, the analysis that follows, we assume that no node in the network has larger $\nc$, $\nin$, and $\nint$ than the bottleneck node and, for each node but the bottleneck, at least one of the three parameters is strictly smaller than that of the bottleneck.\\

We are now ready to compute the various quantities needed to calculate \eq{eq:iot_tot} for the bottleneck node. We start with states TX and RX. Note that packet transmissions and receptions depend on $\nc$. In fact, given that all the nodes generate a packet every $\tU$ seconds (homogeneous network behavior), on average, the bottleneck will receive $\nc$ packets from its children nodes and will transmit $\nc +1$ packets (the exogenous traffic plus its own endogenous) every $\tU$ seconds. This leads to:
\bea
\ftxdg & = & (1+\nc)/\tU,\\
\frxdg & = & \nc/\tU,
\eea
where $\frxdg$ and $\frxdg$ are the data gathering components of the transmission and reception frequencies, disregarding for the moment the traffic due to RPL.

To account for the impact of the MAC protocol, we summarize here its basic functionalities. The X-MAC LPL protocol specifies that each idling node periodically wakes up to perform a clear channel assessment (CCA) operation. The duty-cycle period lasts $\tdc$ seconds and is composed of a sleeping phase of $\toff$ seconds and a wake-up phase lasting $\ton$ seconds, during which CCA is performed. A node wanting to send a unicast packet transmits a burst of short request to send (RTS) preambles, for long enough so that the intended receiver will detect at least one of these RTSs in its next wake-up period. Since the nodes are in general not synchronized, to be sure of hitting the intended receiver, a node will be sending preambles for the duration of an entire duty-cycle $\tdc = \ton+\toff$. Due to the lack of synchronization, the receiver can detect an RTS at any time within this period. Whenever a node detects an incoming RTS destined to itself, it sends a clear to send (CTS) message back to the sender and waits for the transmission of the actual data packet. After the complete reception of the data, the receiver sends an acknowledgment (ACK) to the sender. This channel access mechanism is illustrated in Fig.~\ref{fig:mac_timings} (where we omit the transmission of the ACK for simplicity). In this figure, the sixth RTS from the sender is detected by the intended receiver, which immediately replies with a CTS. The node at the top of the diagram also detects the RTS, but it does not take any action as it is not the intended destination. 

\figmac

For this channel access scheme, the average time needed to carry out a successful transmission is $\ttx = \ton + \toff/2 + \tcts + \tdata + \tack$, where the term $\toff/2$ follows from the fact that the time needed for the receiver to detect an incoming RTS is assumed to be uniformly distributed in $[0, \toff]$. The terms $\tdata$, $\tcts$, and $\tack$ correspond to the durations associated with the transmission of a data packet, a CTS and an ACK, respectively. The reception time is $\trx=\tcts+\tdata+\tack$. Note that the RTS time is not considered in $\ttx$ nor in $\trx$, because it is accounted for by the CCA state. Also, to simplify the notation in the following analysis we include $\tcts$ and $\tack$ in $\tdata$. 

\cc{Now, if $\fu=1/\tU$ is the transmission rate (packets/second) for an error-free channel, in the presence of packet collisions and transmission errors the actual transmission rate becomes $\fup \geq \fu$. For the sake of clarity, the complete characterization of the channel access problem in this case is provided in the Appendices~\ref{app:MAC_modeling} and \ref{app:collision}.}

Thus, the average transmission time can be expressed as:
\be
\label{eq:ttxcoll}
\ttx = \ton + \toff/2+ \tdata + (\fup/\fu-1) \tdc,
\ee
where the factor $\fup/\fu-1$ represents the average number of retransmissions. Note that \eq{eq:ttxcoll} implies a stop-and-wait retransmission policy, where an infinite number of retransmissions is allowed for each data packet. Instead, we assume that the impact of channel errors and collisions on spurious receptions and interfering packets is negligible as, in these cases, the intended receiver does not stay awake to receive the data packet and, thus, its energy expenditure is already accounted for by the CCA state.

We now model the energy expenditure associated with the maintenance of the routing topology. The selected routing algorithm, RPL, consists of a proactive technique that periodically disseminates network information through DODAG\footnote{Destination oriented directed acyclic graph (DODAG).} information objects (DIO) and, subsequently, builds a routing tree by sending destination advertisement objects (DAO) toward the sink. RPL timing is governed by the trickle timer, which exponentially increases up to a maximum value for a static topology. In this paper, we analyze the steady state phase of RPL, considering static networks. This implies the following operations: for every trickle timer epoch, which lasts $\trpl$ seconds, the bottleneck node must send its own DIO message, its own DAO and has to forward $\nc$ DAOs for its children. This leads to a transmission frequency for RPL messages of:
\be
\ftxrpl = (2+\nc)/\trpl.
\ee
In addition, the bottleneck node will receive $\nc$ DAOs from its children and $\nin$ DIOs from its interfering nodes (note that DIOs  are not treated as interference, as they are broadcast). Thus the reception frequency for RPL messages is:
\be
\frxrpl = (1+\nin + \nc)/\trpl,
\ee
where $\ftxrpl$ and $\ftxrpl$ are the contributions of RPL to the transmission and reception frequencies, respectively.

Finally, our model accounts for the energy expenditure due to the reception of messages that are detected during CCA but are not destined to the receiver. In this case, the receiver behaves as during a reception, but, as soon as it decodes the packet header, it recognizes that the message is not intended for itself. At this point, the node drops the message and goes back to sleep. Interfering messages can be either due to data gathering or to networking traffic and occur at a rate proportional to $\nint$. Thus, we have:
\be
\finte = \nint (1/\tU + 1/\trpl).
\ee
Also, we refer to $\tint < \trx$ as the time needed to decode the packet header and therefore detect whether a node is the intended destination for that message.

From the above reasonings, we are able to express the average current consumption for each state:
\bea
\label{eq:currtx}
\itx & = & (\ic + \itr)[\tdc/2+\ton/2+\tdata + (\fu^{\prime}/\fu-1) \tdc ]\times \nonumber \\ 
  & & \times[(1+\nc)/\tU+(2+\nc)/\trpl]\\
\label{eq:currrx}
\irx & = & (\ic + \ir)\tdata [\nc/\tU + (1+\nc+\nin)/\trpl] \\
\label{eq:currint}
\iinte & = & (\ic + \ir) \tint \nint (1/\tU+1/\trpl)\\
\label{eq:currcpu}
\icpu & = & \ic \tcpu \ku/\tU \\
\label{eq:currcca}
\icca & = & (\ic + \ir)d_c \, \ridle \\
\label{eq:curroff}
\ioff & = & \is(1-d_c)\ridle,
\eea
where $\tcpu$ is the average time spent in operations that do not involve the radio and $\ridle$ is the fraction of time that the node spends in the IDLE state, which is computed as one minus the fraction of time spent in the remaining states:
\be
\label{eq:curridle}
\ridle = 1 - \rtx - \rrx - \rinte - \rcpu.
\ee
The total energy consumption is finally given by:
\be
\label{eq:currout}
\io = \itx+\irx+\iinte+\icpu+\icca+\ioff.
\ee

\section{Node Consumption Analysis}
\label{sec:nodeanal}

In this section, we present the solution of problem P1: identifying the optimal network's operating point given a target consumption $\io = u$. The symbols used in this section are listed in Table~\ref{tab:symbfour}.

\tabsymbfour

\noindent \cc{Problem P1 can be formally written as:\\}

\noindent \textbf{Problem P1:}
\bea
\label{eq:OOP}
\underset{\tU, \tdc}{\textrm{\cc{maximize} }} & & \fu \nonumber \\
\textrm{subject to:} && \io \leq u, \nonumber\\
	&& r_{x} \geq 0,\; \forall x \in \XN, \nonumber\\
	&& \tU \geq 0,\; \tdc \geq \ton .
\eea

P1 \eq{eq:OOP} amounts to finding the optimal pair $(\tUopt, \tdcopt)$ that maximizes the node throughput, \mbox{$\fu = 1/\tU$}, subject to the maximum allowed consumption $u$ and to time and frequency constraints. The problem can be numerically solved through two nested dichotomic searches (as shown in~\cite{bui2013dimensioning}): the inner search looks for the optimal $\toffopt$ given $\tU$,\footnote{Note that in this paper we consider $\ton$ as a constant that depends on the considered sensor architecture, whereas the nodes can adapt the duration of their off phase, $\toff$, of the duty cycle. Hence, optimizing over $d_c = \ton / (\ton + \toff)$, $\tdc = \ton + \toff$ or $\toff$ is equivalent.} while the outer search looks for the optimal $\tUopt$. Instead, our objective here is to obtain the solution in closed form. This will permit to solve problem P2 in a reasonable amount of time, while also facilitating the implementation of optimal energy management policies on constrained sensor devices.

Despite the simple problem formulation, \eq{eq:ttxcoll} introduces a polynomial of $\nin$-th degree on the independent variable $\tU$, which makes it difficult to express the solution through tractable and still meaningful equations. Thus, we solve the problem for a \mbox{collision-free} channel and we subsequently adapt the results to keep collisions into account through a heuristic. 

In fact, removing collisions allows for a simpler expression for $\fu^\prime$, i.e., $\fu^{\prime} = \fu/(1-e_t)$, which removes the $\nin$-th degree polynomial on $\tU$. In order to illustrate that this approach is reasonable within the solution space, in~\fig{fig:nrganal}, we show some preliminary results. 

\fig{fig:nrganal} shows contour lines in the $(d_c,\fu)$ plane for different output current levels ($\io \in \{5, 10, 30\}$~mA): dotted lines represent the numerical solution for the complete problem, while dash-dotted lines represent the solution for a collision-free channel for the same $\io$ levels. The locations of the optimal operating points in these two cases are also plotted for comparison (white squares and white circles for the complete problem and that without collisions, respectively). For a given $\io$ the maximum throughput is achieved for a unique value of the duty cycle $d_c$. 
Hence, 
it is not possible to find a feasible solution with higher throughput nor one with the same throughput and a different duty cycle.

From~\fig{fig:nrganal} we deduce the following facts:
\begin{itemize}
 \item the impact of collisions increases with $\io$ which implies that the difference between the optimal working points with and without collisions is an increasing function of the energy consumption $\io$.
 \item the maximum allowed $\fu$ increases with $\io$, which is expected and means that the transmission rate for the endogenous data is an increasing function of the energy consumption $\io$.
 \item the duty-cycle $d_c$ has a critical point, beyond which the throughput $\fu$ suddenly drops; which implies that $\tdc$ has a critical point too.
 \item \cc{the search for the optimal operating point involves the joint optimization of the transmission rate $\fu$ ($\tU$) and the duty-cycle period ($\tdc$) as these two quantities are intertwined.}  
\end{itemize}

\fignrganal

For the sake of readability, the full derivation of the closed form solution in the collision-free case is given in Appendix~\ref{app:closed}. In what follows, we confine ourselves to the discussion of the adopted approach and of the main results.
First, $\io$ has been rewritten as a function of $\tdc$ and $\tU$, which makes it possible to find the mathematical expression of $\tdc^*$ (as a function of $\tU$, which is still a free parameter). This is achieved by taking the partial derivative of $\io$ with respect to $\tdc$, equating it to zero and solving for $\tdc$. In doing so, we observe that $\partial \irx/\partial \tdc = 0$, $\partial \iinte/\partial \tdc = 0$, and $\partial \icpu/\partial \tdc = 0$, as they do not depend on $\tdc$. This leads to: 
\bea
\label{eq:opttoff}
\frac{\partial \io(\tU,\tdc)}{\partial \tdc} & = & \frac{\partial}{\partial \tdc} (\itx(\tU,\tdc)+\icca(\tU,\tdc)+\ioff(\tU,\tdc)) = 0 \nonumber\\
& \Rightarrow & \tdcopt(\tU)=\sqrt{\frac{d_{6}/\tU + d_{5}}{d_{1}/\tU + d_{3}}},
\eea
where coefficients $d_1$, $d_3$, $d_5$ and $d_6$ are given in Table~\ref{tab:coeff2}.

To illustrate the behavior of \eq{eq:opttoff}, in \fig{fig:opttoff} we show $\io$ by varying $\tdc$ and keeping $\tU$ fixed in the set $\tU \in \{5, 10, 25\}$~seconds (see dashed lines). The locus of the optimal solutions $\tdcopt$, obtained through \eq{eq:opttoff}, is plotted as a solid line. The closed form for the optimal $\tdc$ crosses $\io$ (without collisions) where the latter is minimized, as requested.

At this point, it is possible to replace $\tdc$ with $\tdcopt(\tU)$ in $\io(\tU,\tdc)$ (see \eq{eq:currout}) expressing the output current as $\io(\tU,\tdcopt(\tU))$, which becomes a function of the single independent variable $\tU$. Since $\fu$ increases with $\io$, the maximum achievable $\fu$ for a given target current $u$ is obtained at the equality point $\io(\tU,\tdcopt(\tU)) = u$. 

\figopttoff

Also, $u$ cannot be increased indefinitely, because, beyond a given threshold $\tU \leq \tUlim$ the problem becomes bound by the frequency constraint $\ridle \geq 0$. In this region, the system drains the maximum current $u_{\max}$, which cannot be further increased as the channel is saturated.
$\tUlim$ is the smallest feasible inter-packet transmission time for the considered system and can be analytically derived by observing that the optimality condition, see \eq{eq:opttoff}, and the frequency constraint $\ridle(\tU,\tdc)=0$ must concurrently hold for $\tU = \tUlim$. Thus, from $\ridle(\tUlim,\tdc)=0$ we obtain the relationship between $\tUlim$ and $\tdc$, i.e., $\tUlim(\tdc)=(a_1\tdc+a_{11})/(a_{10}-a_3\tdc)$. Whereas replacing $\tU$ with $\tUlim$ in \eq{eq:opttoff} leads to $\tdclim=\tdcopt(\tUlim)$. Using $\tUlim(\tdclim)$ in place of $\tUlim$ in the latter equation returns a third order polynomial in the only variable $\tdclim$, which allows the calculation of $\tdclim$ and, in turn, of $\tUlim$. The coefficients $\{a_1,a_3,a_{10},a_{11}\}$ are given in Table~\ref{tab:coeff1}, whereas the involved mathematical derivations are detailed in Appendix~\ref{app:closed}.
Computing $\io(\tU,\tdc)$ for $(\tUlim,\tdclim)$ returns the maximum current that can be consumed by the bottleneck node using an optimal configuration, i.e., $\iolim = \io(\tUlim,\tdclim)$. The maximum control is therefore given by $u_{\max} = \iolim$.

Conversely, there is a minimum current $\iomin$ that has to be drained in order to keep the system running and operational. 
$\iomin$ is found as $\iomin = \lim_{\tU \to +\infty} \io(\tU,\tdcopt)$, which amounts to solely considering the energy consumption due to the periodic transmission of control traffic (taken into account through $\trpl$). The minimum energy consumption, also corresponds to the smallest control action $u_{\min} = \iomin$.

Finally, the optimal working point, $\tUopt$, is found as the solution of $\io(\tU,\tdcopt(\tU))=u$ with $u \in [u_{\min},u_{\max}]$, which can be expressed as:
\be
(\tUopt,\tdcopt) = 
\begin{cases}
 (+\infty,\sqrt{d_5/d_3}) & {\rm if } \, u < u_{\min} \\
 (\tUopt, \tdcopt(\tUopt)) & {\rm if } \, u_{\min} \leq u \leq u_{\max}\\
 (\tUlim, \tdclim) & {\rm if } \, u > u_{\max},
\end{cases}
\ee
where $\tUopt$ is the positive solution of the quadratic equation $e_2 \tU^2 + e_1 \tU + e_0 = 0$ and $\tdclim$ is the largest solution of the cubic equation $f_3 \tdc^3 + f_2 \tdc^2 + f_1 \tdc + f_0 = 0$. The reader is referred again to Appendix~\ref{app:closed} for mathematical insights and the definition of the coefficients (see Table~\ref{tab:coeff2}).

\figoptimal

\fig{fig:optimal} shows the optimal operating point $(\tUopt,\tdcopt)$ by varying the control $u$ as the independent parameter. The dashed line corresponds to the result of \eq{eq:OOP} for a \mbox{collision-free} channel, the white filled circles represent the numerical results of the complete problem with collisions and the solid line shows the results achieved from the closed form solution, which has been adapted through a heuristic to keep collisions into account. In addition, the crosses and the dash-dotted line illustrate the solution of $\ridle(\tU,\tdc) = 0$ obtained for the complete problem and using the closed form heuristically modified, respectively.

The adopted heuristic is a rigid translation of the closed form for a collision-free channel so that the latter equals the numerical solution with collisions for the maximum allowed control $u_{\max}$. The error introduced through this approach is very small for high values of $u$ and increases for decreasing $u$. However, this error is negligible throughout most of the solution space, as it grows slower than $\tU$ does and it always provides a feasible solution for the system. 

\figreward

\tabnets
\tabpar

Finally, in \fig{fig:reward} we plot the reward function: 
\be
\label{eq:reward}
r(u) = 1/\tUopt(u) .
\ee
$r(u)$ corresponds to the maximum achievable throughput for the given multi-hop network. In Fig.~\ref{fig:reward}, we show results for dense, medium and sparse networks (represented with squares, circles and triangles, respectively) of $3$ and $5$ hops (solid and dashed lines, respectively). The parameters of these networks are given in Table~\ref{tab:nets}, where $N$ is the total number of nodes and $\rho$ is the network density. Increasing the number of hops has a much larger impact on the reward function than increasing the node density. All the graphs of this paper have been obtained considering a sensor platform characterized by the energy consumption and timing parameters of Table~\ref{tab:nets2}. The optimal throughput of \eq{eq:reward} will be used in Section~\ref{sec:smdp} as the reward function for problem P2, which considers a stochastic energy source.

\section{Optimization Framework}
\label{sec:source}

The objective of the following sections is to solve problem P2, which translates into finding {\it optimal} and {\it online} energy consumption strategies for the sensor nodes, given the energy consumption model (see problem P1), their current energy reserve and a statistical characterization of future energy arrivals (i.e., of the energy source S). This requires to link the energy consumed to that harvested and to the instantaneous energy buffer state. In the analysis that follows, we assume that the amount of charge in the energy buffer is a known quantity or, equivalently, that it can be reliably estimated at the sensor nodes. Based on this, we formulate our optimal control as a Markov Decision Process (MDP). We observe that heuristic approaches, which base their energy consumption policies on energy estimates, are also possible but are not considered here and are left as a future work. Nevertheless, in Section~\ref{sec:performance} the performance of the obtained policies is compared against that of heuristic solutions from the literature.

Here, we present the stochastic model that will be used to describe the source S, as per our sensor diagram of Fig.~\ref{fig:sensor_diagram}. This will be used in Section~\ref{sec:smdp} to solve problem P2.
The resulting energy management policies are validated in Section~\ref{sec:results}.

\cc{In Table~\ref{tab:symbfive} we define the symbols used in this section.}

\tabsymbfive

\noindent \textbf{Energy source:} the energy source dynamics are captured by a continuous-time Markov chain with $N_S$ states $x_s \in \mS = \{0,1,\dots,N_S-1\}$. We refer to $t_k$, with $k \geq 0$, as the time instant where the source transitions between states and to $\Delta_k = t_k - t_{k-1}$ as the time elapsed between two subsequent transitions. Also, the system between $t_{k-1}$ and $t_{k}$ is said to be in {\it stage} $k$, and its duration $\Delta_k$ is described by a r.v. $\tau_{x_s} \in [t_{\min}(x_s),t_{\max}(x_s)]$, depending on the source state $x_s$ in the stage. $\tau_{x_s}$ has an associated probability distribution function (pdf) $f_\tau(t | x_s)$. Moreover, during stage $k$, the source provides a constant current $i_k$ that is fed into the battery and is assumed to remain constant until the next transition, occuring at time $t_{k}$. This input current is described by the r.v. $\iota_{x_s} \in [i_{\min}(x_s),i_{\max}(x_s)]$ with pdf $f_\iota(i | x_s)$. We assume that $\tau_{x_s}$ and $\iota_{x_s}$ have bounded support. $p_{ij} = \textrm{Prob}\{x_s(k) = j | x_s(k-1) = i\}$ with $i,j \in \mS$ are the transition probabilities of the associated embedded Markov chain, which are invariant with respect to $k$.\\
%

\noindent \textbf{Discrete-Time Formulation:} we describe the energy source model through an equivalent discrete-time Markov process. This will make it possible to conveniently characterize the optimal policies through a Discrete-Time Constrained Markov Decision Process (DT-CMDP), in Section~\ref{sec:smdp}. For improved clarity of exposition and conciseness, in the remainder of this paper we omit the time index $k$ from the symbols, unless explicitly stated otherwise.

To describe the energy source through a discrete time model, for any given $k$, we map the random nature of the stage duration into the corresponding variation of charge during the stage. To do this, we define the two r.v.s $\delta_{\rm in}$ and $\delta_{\rm out}$ that respectively describe the amount of charge that enters the system during the stage (stored into the energy buffer) and the amount of charge consumed by the sensor node. $\delta = \delta_{\rm in} - \delta_{\rm out}$ is the r.v. describing the overall variation of charge during the stage. We recall that $u$ is our control variable, corresponding to the current drained by the sensor node during the stage. 
$u$ for a given policy is a known quantity and it will be considered as a constant in the following derivations. 
We have that:
\cc{\be
\label{eq:delta}
\delta_{\rm in} = \tau \iota \, , \, \delta_{\rm out} = \tau u \, , \, \delta = \delta_{\rm in} - \delta_{\rm out} = \tau (\iota - u) . 
\ee}
Hence, the r.v. $\delta$ is obtained as the product of the two r.v.s $\tau$ and $\iota-u$. From the theory in~\cite{Papoulis-2002}, the pdf of $\delta$ when the source is in state $x_s$ and the control is $u$, $f_\delta(d | u, x_s)$, is obtained as:
\be
\label{eq:pdf_joint_charge}
f_\delta(d | u, x_s) = \int_{t_{\min}(x_s)}^{t_{\max}(x_s)} f_{\tau}(t | x_s) f_{\iota}(d/t + u | x_s) |t|^{-1} \dd t \, , \, d \in \R .
\ee
Henceforth, the energy source is equivalently characterized by a discrete-time Markov chain with $N_S$ states and transition probabilities $p_{ij}$, $i,j \in \mS$. Moreover, when the current state is $x_s \in \mS$ and the control is $u$, the corresponding variation of charge during a stage is accounted by the r.v. $\delta$ with pdf given by~\eq{eq:pdf_joint_charge}.

\section{Markov Decision Process Analysis}
\label{sec:smdp}

This section presents our analysis of the outer optimization problem P2, which is framed as a Markov Decision Process. For improved clarity, this analysis is split into four subsections: in Section~\ref{sec:def}, we define the basic ingredients of the MDP, in Section~\ref{sec:polform} we formulate the optimal policy, discussing its properties and detailing an algorithm for its computation (see Section~\ref{sec:polcomp}). Finally, in Section~\ref{sec:polcomplex} we report our considerations on computational complexity and on the usage model for the computed policies. The list of symbols used in the MDP analysis is given in Table~\ref{tab:symbsix}.

\tabsymbsix

\subsection{Definitions}
\label{sec:def}

We consider the sensor system of Fig.~\ref{fig:sensor_diagram}
and we assume without loss of generality that the system evolves in discrete time. Hereafter, at time $k \geq 0$, the system is said to be in stage $k$ and the terms ``time'' and ``stage'' will be used interchangeably in the following analysis. The source S feeds energy into the energy buffer B and is modeled according to the discrete-time Markov chain presented in the previous section. At any time $k$, the source S is in a certain state $x_s$, whereas the energy buffer hosts an amount of charge $x_b \in \B = [0,b_{\max}]$, where $b_{\max}$ is the buffer capacity. At the generic time $k$, we define the system state as $x=(x_s,x_b) \in \X$, where $\X = \mS \times \B$. The system state at the following time $k+1$, defined as $y=(y_s,y_b) \in \X$, depends on the dynamics of $S$, on the control $u$ for the current stage $k$ and on the total variation of charge $\delta$ during stage $k$. For the battery at the beginning of the next stage $k+1$, $y_b$, we have:
\be
\label{eq:buffer_dynamics}
y_b = \min \{ \max \{ x_b + \delta, 0\}, b_{\max} \} = [ x_b + \delta ]^+ ,
\ee      
where $\delta$ is expressed in \eq{eq:delta} and depends on the control $u$ for the current stage $k$, whereas $[a]^+$ is defined as $[a]^+ =  \min \{ \max \{ a, 0\}, b_{\max} \}$, with $a \in \R$.

We model the sensor system through a discrete time MDP. At every stage $k$ a decision $u$ has to be made based on the current state $x \in \X$. 
In addition to the system state and its dynamics, a Markov decision process is characterized by a control set $\U = [u_{\min},u_{\max}]$, where $u_{\min}=\iomin$ and $u_{\max}=\iolim$. $\U$ contains all the feasible current consumption levels for the sensor (see Section~\ref{sec:nodeanal}). 
In this paper, we consider mixed and stationary Markov (i.e., history independent) policies. The term mixed means that there exists a mapping $\mu$ that, for any possible state $x \in \X$, returns a vector of pairs $(u(i),p(i))$, of size $M \geq 1$, with $\sum_{i=1}^M p(i) = 1$. 
This vector represents the decision to be made when the system state is $x$ and indicates that control $u(i)$ must be implemented with the associated probability $p(i)$. 
A mixed {\it policy} $\pi$ is a collection of such mappings $\pi = \{\mu_0, \mu_1,\mu_2, \dots\}$ for all stages. 
Our problem belongs to the class of MDPs with unichain structure, bounded costs and rewards. For these, it is sufficient to consider the set of admissible Markov policies as the optimal policy can always be found within this class, see~\cite{Derman1966}, \cite{Altman-1999} or Theorem 13.2 of~\cite{Feinberg1995}. 
The boundedness of rewards and costs follows from the finite support of $\tau$, $\iota$ and from the fact that the instantaneous reward function is also bounded. Thus, for the problem addressed in this paper it is sufficient to restrict our attention to Markov stationary policies, which means that $\mu_k$ only depends on the system state at time $k$ (past stages $0,\dots,k-1$ are not considered) and that the mapping functions do not depend on $k$, i.e., $\pi = \{\mu, \mu,\mu, \dots\}$. \\

\noindent {\textbf{Reward:}} the reward function takes into account the throughput of the system. Specifically, from the derivations in Section~\ref{sec:nodeanal}, we know that for a given control $u$ the optimal instantaneous throughput of a sensor node is given by $r(u)$, as defined in \eq{eq:reward}. Now, let $x=(x_b,x_s)$, with $x \in \X$,  be the system state at the beginning of a generic decision stage $k$. Moreover, let $t$ and $i$ respectively represent the realization of the r.v. $\tau_{x_s}$, describing the duration of the stage, and the realization of the r.v. $\iota_{x_s}$, quantifying the input current from the source. Taking~\eq{eq:delta} into account and recalling that the input current $i$ and the control $u$ are both constant during the stage, we have that the amount of charge varies linearly within a stage until it either hits the buffer capacity $b_{\max}$ or drops to $0$, depending on the sign of $i-u$. Hence, during the stage, the total variation of charge is $d = t (i-u)$ (see~\eq{eq:delta}) and the amount of time the level of charge in the energy buffer is greater than zero is given by the following function: 
\be
g_{>0}(d,t,u,x_b) = 
\bc
t & d \geq 0 \\
\displaystyle \min \left \{\frac{-x_b t}{d},t \right \} & d < 0 .
\ec
\ee
Furthermore, as long as the buffer level is above zero, the throughput remains constant and equal to $r(u)$, whereas it drops to zero in case the energy buffer gets empty. Given this, the single-stage expected reward, when the system state at the beginning of the stage is $x=(x_b,x_s)$ and the control is $u$, is computed as:
\bea
\label{eq:reward_smdp}
R(x,u) & = & \E [r(u) g_{>0}(\xi,t,u,x_b) | x, u]  \nonumber \\
& = & \int_{-\infty}^{+\infty} \int_{t_{\min}(x_s)}^{t_{\max}(x_s)}  r(u) g_{>0}(\xi,t,u,x_b) f_{\tau}(t | x_s) f_{\iota}(\xi / t + u | x_s) |t|^{-1} \dd t \dd \xi \nonumber \\
& = & r(u) \E [g_{>0}(d,t,u,x_b) | x, u] ,
\eea
where $\E [g_{>0}(d,t,u,x_b)]$ represents the average amount of time the energy buffer contains a positive amount of charge during the stage. 
In the previous equation $r(u)$ 
remains constant during a stage 
when $u$ is given. The actual average throughput is then modulated through the average amount of time the energy buffer state is greater than zero in the stage, i.e., $\E [g_{>0}(d,t,u,x_b) | x, u]$.\\


\noindent {\textbf{Cost:}} for the cost, we account for a penalty whenever the energy buffer drops below a given threshold $\bth \in (0, b_{\max}]$. This threshold is a design parameter that may be related to the minimum energy reserve that is required to keep the system operational and responsive. 
Also, $\bth$ is in general implementation dependent and besides depending on application requirements, it depends on hardware constraints. In fact, too low a charge may not be sufficient to guarantee the correct operation of the sensor nodes.

The cost is obtained as the average time spent with the energy buffer level below $\bth$. The amount of time the energy buffer level is below $\bth$ is given by the following function: 
\be
\label{eq:time_below_th}
g_{<\bth}(d,t,u,x_b) = 
\bc
\displaystyle \max \left \{ 0, \min \left \{ \frac{(\bth-x_b)t}{d}, t \right \} \right \} & d \geq 0 \\
\displaystyle \min \left \{ \max \left \{ 0, \left ( 1 - \frac{(\bth-x_b)}{d} \right ) t \right \},t \right \} & d < 0.
\ec
\ee
Hence, the single-stage expected cost when the system state at the beginning of the stage is $x=(x_b,x_s)$ and the control is $u$, is obtained as:
\bea
\label{eq:cost}
C(x,u) & = & \E [g_{<\bth}(\xi,t,u,x_b) | x,u ] \nonumber \\
& = & \int_{-\infty}^{+\infty} \int_{t_{\min}(x_s)}^{t_{\max}(x_s)} g_{<\bth}(\xi,t,u,x_b) f_{\tau}(t | x_s) f_{\iota}(\xi / t + u | x_s) |t|^{-1} \dd t \dd \xi .
\eea \\

\subsection{Optimal Policy - Formulation} 
\label{sec:polform}

We now formulate our optimal control problem as a DT-CMDP. The total expected reward that is earned over an infinite horizon by a feasible policy $\pi$ is expressed as:
\be
J_R(x_o) = \lim_{N \to +\infty} \E \left [ \sum_{k=0}^{N-1} \alpha^k R(x(k),u(k)) \right | x(0) = x_o , \pi \Bigg ] ,
\ee
where $\alpha \in [0,1)$ is the discount factor, $x(k)$ and $u(k)$ are respectively the system state and the control at stage $k$ and $x_o$ is the initial state. If we disregard the cost, having the sole objective of maximizing the throughput (reward), the optimal policy is the one that solves the following Bellman optimality equation:
\bea
\label{eq:Bellman_reward}
J_R(x) & = & \max_{u \in \U} \left \{ R(x,u) + \alpha \sum_{y_s \in \mS} p_{x_s y_s} \int_{-\infty}^{+\infty} f_{\delta}(\xi | u, x_s) J_R(y) \dd \xi  \right \} , \nonumber \\
\textrm{with: } y & = & (y_b,y_s), \, y_b = [x_b + \xi]^+ ,
\eea
where if the current state is $x$, $J_R(x)$ represents the optimal expected reward from the current stage onwards and is obtained, maximizing over the admissible controls, the sum of the single-stage expected reward (the immediate reward, accrued in the present stage) and the expected optimal reward from the next stage onwards (where future rewards $J_R(y)$ are weighted accounting for the system dynamics, i.e., $f_{\delta}(\cdot)$ and $p_{x_s y_s}$). \eq{eq:Bellman_reward} can be solved through Value Iteration (VI), as detailed in Section 1.3.1 of~\cite{Bertsekas-2012}. 
\cc{In short, VI amounts to using \eq{eq:Bellman_reward} as an update rule, which is iterated for all states starting from an initial estimate of $J_R(x)$}.\footnote{Setting $J_R(x)=0$, $\forall \, x$ in the first iteration of the algorithm also assures convergence.} It can be shown that the optimality equation $J_R(x)$ is a {\it contraction mapping}. This property assures that the VI iterations converge, at which point the optimal estimates $J_R(x)$ computed in the previous step equal the new ones, that are obtained using the right-hand side (RHS) of \eq{eq:Bellman_reward}. Hence, the optimal policy, for any given $x \in X$, is given by the control $u$ that maximizes the RHS of \eq{eq:Bellman_reward}. Note that the optimal control corresponding to \eq{eq:Bellman_reward} is a {\it pure policy} whereby a single control $u$ is  associated with each state $x \in \X$, i.e., there exists a mapping function $\mu(x)$ such that, $u(x) = \mu(x)$ for each state $x \in \X$ and $u(x)$ is unique for each $x$.

Analogously, solely taking the cost into account, the total expected and discounted cost of a given policy $\pi$ for an initial state $x$ is obtained as the solution of the following Bellman equation:
\bea
\label{eq:Bellman_cost}
J_C(x) & = & \max_{u \in \U} \left \{ C(x,u) + \alpha \sum_{y_s \in \mS} p_{x_s y_s} \int_{-\infty}^{+\infty} f_{\delta}(\xi | u, x_s) J_C(y) \dd \xi  \right \} , \nonumber \\
\textrm{with: } y & = & (y_b,y_s), \, y_b = [x_b + \xi]^+ .
\eea
The DT-CMDP problem for our controlled sensor node is thus written as:\\

\noindent \textbf{Problem P2:}
\bea
\label{eq:CMDP}
\underset{\pi}{\textrm{maximize }} && \E_{x}[J_R(x) | \pi] \nonumber \\
\textrm{subject to:} && \E_{x} [J_C(x) | \pi] \leq \Cth ,
\eea

where the maximization is taken over the set of all feasible policies and $\E_{x}[\cdot]$ represents the expectation taken with respect to the steady-state distribution of $x \in \X$ induced by policy $\pi$. $\Cth$ is a positive constant and a policy is termed {\it feasible} if its average cost satisfies the constraint of \eq{eq:CMDP}. For the selection of $\Cth$, note that, as shown in~\cite{Altman-1999}, the average cost per stage corresponding to a total expected cost $\Cth$ and a discount factor $\alpha$, is obtained as $\Cth^\prime = \Cth (1-\alpha)$. Moreover, from the definition of the cost (see \eq{eq:time_below_th}), this quantity corresponds to the average amount of time in a stage where the amount of charge in the energy buffer is below $\bth$. Thus, dividing $\Cth^\prime$ by the average stage duration, $T = \E[\tau_{x_s}]$, returns the maximum tolerable fraction of time in a stage during which the amount of charge in the energy buffer can be smaller than $\bth$, i.e., a buffer outage occurs. Thus, the average fraction of time in a stage that the buffer is in outage is found as:
\be
\label{eq:outage}
\tout = \frac{\Cth (1 - \alpha)}{T} .
\ee
This relation facilitates the tuning of $\Cth$, associating it to a tangible concept.

The inequality constraint in \eq{eq:CMDP} limits the maximum energy consumption by imposing a maximum expected cost $\Cth$. The optimal policy is thus tunable through $\alpha$ and $\Cth$. The former determines how much we look ahead in the optimization; for instance $\alpha=0$ represents a {\it myopic} decision maker where the control is uniquely chosen based on the current stage and the future system evolution is disregarded. Higher values of $\alpha$ generate optimal policies with better look-ahead capabilities. In particular, as $\alpha \to 1$, the associated optimal policies converge to the policy that maximizes the average reward over an infinite time horizon, see~\cite{White1993}. Instead, decreasing $\Cth$ generates less aggressive policies, which will be more parsimonious in the consumption of the energy stored in the buffer.\\

\subsection{Optimal Policy - Computation} 
\label{sec:polcomp}

From the analysis in~\cite{Ross-1985} (Theorem 4.3) and \cite{Altman-1999} (Theorem 12.7) we know that \eq{eq:CMDP} can be solved through the definition of a Lagrangian reward $L_{\lambda}(x,u)$ \cc{(referred to as {\it Lagrangian relaxation}):}
\be
\label{eq:Lagrangian_reward_00}
L_{\lambda}(x,u) = R(x,u) - \lambda C(x,u) ,
\ee
where $\lambda \geq 0$ is the Lagrangian, whereas $R(x,u)$ and $C(x,u)$ are respectively defined in \eq{eq:reward_smdp} and \eq{eq:cost}. Thus, we define an unconstrained discounted problem depending on $\lambda$ and having the following Bellman optimality equation: 
\bea
\label{eq:Bellman_lagrangian}
J_{\lambda}(x) & = & \max_{u \in \U} \left \{  Q(x,u,\lambda) \right \} , \nonumber \\
\textrm{with: }  & & Q(x,u,\lambda) \stackrel{\rm def}{=} L_{\lambda}(x,u) + \alpha \sum_{y_s \in \mS} p_{x_s y_s} \int_{-\infty}^{+\infty} f_{\delta}(\xi | u, x_s) J_{\lambda}(y) \dd \xi , \nonumber \\
\textrm{and:} & & y = (y_b,y_s), \, y_b = [x_b + \xi]^+.
\eea
For a fixed $\lambda$, \eq{eq:Bellman_lagrangian} represents a standard discrete-time Markov Decision problem and can be solved through VI obtaining the corresponding pure optimal policy $\pi_\lambda$. For a given $\lambda$, the function $J_{\lambda}(x)$
returns the optimal Lagrangian reward associated with the optimal policy $\pi_\lambda$. We denote the expected log-term Lagrangian reward of this optimal policy by $J_{\lambda}=\E_x[J_{\lambda}(x) | \pi_\lambda]$, where the expectation is taken over the steady-state distribution of $x$ induced by the optimal policy $\pi_\lambda$.

\cc{Intuitively, considering \eq{eq:Lagrangian_reward_00} one can easily see that an increasing $\lambda$ puts more weight on the cost $C(x,u)$ making the policy more conservative.
While a smaller $\lambda$ will instead put more weight on the reward $R(x,u)$
giving a higher priority to the throughput. 
These facts are used in the algorithm below to exploit $\lambda$ to search within the solution space. The optimal $\lambda$ is the one that achieves the maximum reward, whilst leading to an average cost smaller than or equal to $\Cth$, see~\eq{eq:CMDP}.}

Next, we propose an efficient algorithm that exploits a dichotomic search over $\lambda$. \cc{Note that this search strategy is possible because, as proven in Lemmas 3.1 and 3.2 of~\cite{Ross-1985}, for our discounted MDP, the optimal Lagrangian reward $J_{\lambda}(x)$ is a uniformly absolutely continuous, monotone and non-increasing function of $\lambda$. This means that the reward $J_{\lambda}(x)$ is well-behaved as a function of $\lambda$, i.e., it does not have local minima or maxima.}

Moreover, \cc{the results in~\cite{Ross-1985} (see Theorems 4.3 and 4.4) guide us in the search for the optimal $\lambda$}. In fact, for the optimal policy there can only be the following two possibilities: 1) an optimal $\lambda$, termed $\lambda^*$, exists such that the average cost of $\pi_{\lambda*}$ is equal to $\Cth$; in this case $\pi_{\lambda*}$ is the optimal policy that we are looking for and belongs to the class of pure policies, or 2) there exist two values of $\lambda$,  say $\lambda^-$ and $\lambda^+$ with $\lambda^- < \lambda^+$, for which the cost of $\pi_{\lambda^-}$ is larger than $\Cth$, whereas that of $\pi_{\lambda^+}$ is smaller than $\Cth$, the two policies differ in at most one state and the optimal policy we are looking for is a mixed policy that consists of using, at every decision epoch, $\pi_{\lambda^-}$ with a certain probability $p$ and $\pi_{\lambda^+}$ with probability $1-p$. Case 2 is always verified, even when a pure policy exists, whereas a pure policy may or may not exist, depending on the structure of the MDP. 

Given this, our algorithm seeks a mixed policy that maximizes the total expected Lagrangian reward $J_{\lambda} = \E_x [J_{\lambda}(x) | \pi_{\lambda}]$, while satisfying the constraint $\E_x [J_C(x) | \pi_{\lambda}] \leq \Cth$, where we define $C_\lambda = \E_x [J_C(x) | \pi_{\lambda}] $. The algorithm is described next:

\begin{enumerate}

\item Pick the initial values for $\lambda^-$ and $\lambda^+$, where $\lambda^+$ is a small value for which $C_{\lambda^-} > \Cth$ and $C_{\lambda^+}$ is such that $C_{\lambda^+} < \Cth$.

\item Compute $\lambda = (\lambda^+ + \lambda^-)/2$ and apply VI to \eq{eq:Bellman_lagrangian} for this $\lambda$. This returns the optimal Lagrangian reward function $J_{\lambda}(x)$ ($\forall \, x \in \X$), which is the unique solution of \eq{eq:Bellman_lagrangian}. Once $J_{\lambda}(x)$ is known, the associated optimal policy $\pi_\lambda$ is described by the mapping $u(x) = \mu_{\lambda}(x)$, where:
\be
\label{eq:optimal_mapping}
\mu_{\lambda}(x) = \arg \! \max_{u \in \U} \left \{ Q(x,u,\lambda) \right \} , 
\ee
\cc{where $Q(x,u,\lambda)$ is defined in the second line of \eq{eq:Bellman_lagrangian}.}

\item Obtain the stationary distribution of $x$ induced by $\pi_\lambda$, referred to as $P(x)$, which is computed by numerically solving the recursion:
\be
\label{eq:recursion_steady_01}
P(y) = \int_{x \in \X} P(x) f(y | x, u(x)) \dd x ,
\ee
under the constraint $\int_{x \in \X} P(x) \dd x = 1$, where $P(x)$ represents the steady-state distribution evaluated in state $x=(x_b,x_s) \in \X$, whereas $f(y | x, u(x))$ is the conditional probability distribution function that the system moves to $y=(y_b,y_s) \in \X$ at the end of a given stage, given that the initial state is $x$ and the action taken is $u(x) = \mu_{\lambda}(x)$. For our problem, \eq{eq:recursion_steady_01} specializes to:
\be
P(y) = \sum_{x_s \in \mS} p_{x_s y_s} \int_{x_b \in \B} P(x) \int_{I(x_b,y_b)} f_{\delta}(\xi | \mu_{\lambda}(x), x_s) \dd \xi \dd x_b ,
\ee
where $x=(x_b,x_s)$, $y=(y_b,y_s)$ and $I(x_b,y_b) = \{y_b-x_b\}$ if $y_b > 0$ and $b < b_{\max}$, whereas $I(x_b,y_b)=[y_b-x_b,+\infty)$ if $y_b=b_{\max}$ and $I(x_b,y_b) = [-\infty,y_b-x_b]$ if $y_b=0$.

\item At this point, the average long-term cost performance $J_C(x)$ associated with policy $\pi_\lambda$ is obtained by solving \eq{eq:Bellman_cost} through VI, where $\max_{u \in \U} $ is replaced with $\max_{u \in \{\mu_{\lambda}(x)\}}$, which means that the single optimal action $\mu_{\lambda}(x)$ is used in placed of set $\U$; so the maximization reduces to the evaluation of the RHS of \eq{eq:Bellman_cost} for the optimal action only. Now, using $P(x)$ and $J_C(x)$, we obtain the expected long-term discounted cost $C_{\lambda}$ as:
\be
\label{eq:average_cost_P}
C_{\lambda} = \E_x[J_C(x) | \pi_{\lambda}] = \int_{x \in \X} P(x) J_C(x) \dd x . 
\ee

\item Now, we can have three cases: C1) $C_{\lambda} = \Cth$, C2) $C_{\lambda} < \Cth$ or C3) $C_{\lambda} > \Cth$. In case C1, the algorithm terminates and the optimal policy is the pure policy $\pi_{\lambda}$. Otherwise, the algorithms continues as follows. In case C2, we update $\lambda^+$ as $\lambda^+ = \lambda$ whereas, in case C3 we set $\lambda^- = \lambda$ and we initiate a new iteration, going back to step (2) above, using the new values of $\lambda^-$ and $\lambda^+$ (which represent our dynamically adapted search interval). If, instead, the difference between $C_{\lambda^-}$ and $C_{\lambda^+}$ is smaller than a small constant $\varepsilon > 0$, the algorithm stops returning $\pi_{\lambda^-}$, $\pi_{\lambda^+}$ and the value of the mixing probability $p$, which is obtained as follows:
\be
p C_{\lambda^-} + (1-p) C_{\lambda^+} = \Cth \, \Rightarrow \, p = \frac{\Cth-C_{\lambda^+}}{C_{\lambda^-}-C_{\lambda^+}} .
\ee
Hence, the optimal policy that solves \eq{eq:CMDP} is a mixed policy that, at the beginning of each stage, uses policy $\pi_{\lambda^-}$ with probability $p$ and policy $\pi_{\lambda^+}$ with probability $1-p$.
\end{enumerate}


\subsection{Optimal Policy - Complexity and Usage} 
\label{sec:polcomplex}

\cc{Let $\varepsilon$ be the desired numerical precision. The number of iterations involved in the dichotomic search for the optimal $\lambda$ is: $O(\log_2(\lambda_{\max}/\varepsilon))$, where $\lambda_{\max}$ is the upper end of the related search interval. A tight upper bound on the complexity of the {\it value iteration algorithm} (see \eq{eq:optimal_mapping}), which is executed once for each value of $\lambda$ is $O(1/((1-\alpha)^2\varepsilon)^{2n+m})$, where in our case $n=2$ and $m=1$, see~\cite{DP-Complexity-89}. The complexities associated with solving \eq{eq:recursion_steady_01} and \eq{eq:average_cost_P} are dominated by that of value iteration.}

\cc{In general, the proposed algorithm substantially reduces the complexity associated with finding the optimal policy, which would be infeasible through an exhaustive search. Although, the computational complexity is rather high, that the energy management policies neither have to be computed at runtime nor they have to be obtained by the sensor nodes.}

\cc{Instead, we propose the following. First of all, for the considered location, time of the year and type of solar module, we must derive an energy source model according to the procedure described in Section~\ref{sec:source} (see also~\cite{EnergyCon-14}). This model must then be used with the algorithm of Section~\ref{sec:polcomp} to obtain {\it online} optimal energy management policies for the considered settings. This algorithm is executed offline and only once for a given source model. The resulting policies will correspond to simple tables associating the (quantized) amount of charge in the battery with a corresponding optimal action (control $u$). Note that they can be conveniently stored in memory arrays, preloaded into the nodes' memory and looked up in $O(1)$ time. Also, note that the policies will be non-decreasing piecewise linear functions of the battery state $x_b$ and, as such, a further compression of the memory required for their storage is possible through numerical fitting.}

\cc{The sensor nodes will only have to execute at runtime the action dictated by the current policy, which corresponds to retrieving the optimal action from the policy table. }

\section{Numerical Results}
\label{sec:results}

In this section we comment numerical results about the solution of the combined optimization described in Section~\ref{sec:architecture}, which includes P1: that finds a suitable reward function $r(u)$ (throughput as a function of the energy consumption $u$) and P2: that uses $r(u)$ to obtain optimal online energy management policies that maximize the throughput, while keeping the bottleneck node (and, as a consequence, all other nodes in the network) energetically self-sufficient. \cc{In particular, Section~\ref{sec:optpol} discusses the general behavior of the optimal policies, Section~\ref{sec:performance} provides simulation results on their throughput and outage time, comparing our solution with that of proposed in~\cite{kansal2007power}, Section~\ref{sec:hetero} discusses the robustness of our solution when the amount of charge harvested by the sensor nodes differ, and Section~\ref{sec:relax} considers the relaxation of further assumptions.}\\

\noindent \textbf{Network setup:} in the following subsections, we consider a network of $N=48$ sensor nodes which transmit their data to a sink through a multi-hop topology of five hops. Problem P1 for this network has been solved in Section~\ref{sec:nodeanal}, where it is referred to as ``5-hop medium-network''. The corresponding reward function $r(u)$ is plotted in Fig.~\ref{fig:reward} and the corresponding network parameters are given in Table~\ref{tab:nets}.

For the energy inflow, we have considered a photovoltaic outdoor power source, adopting the framework of~\cite{EnergyCon-14} with two states $x_s \in \mS=\{0,1\}$; where $x_s = 0$ is the high-energy state (i.e., modeling daytime), whereas $x_s = 1$ is a state where the energy harvested is nearly zero (night). For the transition probabilities we have $p_{ij}=1$ if $i \neq j$ and $p_{ij}=0$ if $i=j$ with $i,j \in \mS$. The probability distribution functions $f_\iota(i | x_s)$ and $f_\tau(t | x_s)$ are derived using the SolarStat tool as detailed in~\cite{EnergyCon-14} using their ``night-day clustering approach''. We have selected Los Angeles as the installation location, \cc{considering a tilt of $45^{\circ}$ and an azimuthal displacement with respect to the real South of $30^{\circ}$ for the solar panels (Solarbotics SCC-3733 Monocrystalline solar technology~\cite{Solarbotics})}. Irradiance data from years 1999-2010 available at~\cite{nrel} have been employed for the calculation of $f_\iota(i | x_s)$ and $f_\tau(t | x_s)$. 

\cc{For the characterization of the optimal policies we have considered square solar panels with side going from $3$ to $12$~centimeters (in steps of one centimeter), considering the irradiance data collected for the months of August and December as these respectively corresponds to the best and worst case in terms of amount of energy harvested.}

\cc{The energy buffer size has been taken in $b_{\max} \in \{100,250,500,1000\}$~mAh, whereas the buffer threshold has been set to $b_{\rm th} = 50$~mAh imposing an outage of $1\%$, i.e., $t_{\rm out}=0.01$ (see Eq.~\eq{eq:outage}).}

\subsection{Evaluation of the Policies}
\label{sec:optpol}

As an illustrative example, in Figs.~\ref{fig:op_alpha_xs0} and \ref{fig:op_alpha_xs1} we show the optimal action $u(x) = \mu(x)$ and $P(x)$, where $\mu(x)$ is defined as $\mu(x) = p \mu_{\lambda^-}(x) + (1-p) \mu_{\lambda^+}(x)$ and $P(x)$ is the steady-state distribution induced by the optimal policy, see Section~\ref{sec:smdp}. Note that the policies shown in these figures are all {\it feasible} as they satisfy the cost constraint, while also providing the maximum possible throughput for the corresponding value of the discount factor $\alpha$. For these plots, we have considered $\alpha \in \{0.01,0.5,0.9\}$, a maximum buffer size $b_{\max}=250$~mAh, $b_{\rm th} = 0.2b_{\max} = 50$~mAh and $\tout=0.01$. 

\figresone
\figrestwo

With $\alpha=0.9$ and $x_s=0$, see Fig.~\ref{subfig:op_alpha_xs0}, the optimal policy does not transmit when the buffer state $x_b$ is below or close to $b_{\rm th}$, whereas for higher values of $x_b$ the optimal action $u(x)$ increases linearly. With our network parameters the maximum energy consumption is $u_{\max} \simeq 34$~mA, that for the considered example is never reached by the optimal policy with $\alpha=0.9$, even for a full buffer. This is due to the constraint on the minimum buffer level. To see this, we recall that the sensor node has to make its decision $u(x)$ at the beginning of each stage and the amount of energy that will actually be harvested during the stage is only known statistically through $f_\iota(i | x_s)$ and $f_\tau(t | x_s)$. In our case, for $x_s=0$ and $x_b=100\%$, picking $u(x) = u_{\max}$ would lead to a  violation of the constraint on the buffer. In fact, the optimal policy for each $x_b$ picks the maximum $u(x)$ that, on average, satisfies the constraint with equality. For this example, this value for $x_b=100\%$ is about $24$~mA (referred in the following as {\it maximum admissible energy expenditure}). This means that in the very favorable cases, where the energy inflow is abundant, the optimal policy may leave some of the harvested energy unused ({\it energy wastage}). As we discuss shortly, this is avoided by increasing the buffer size $b_{\max}$. 

By looking at the steady-state distribution for $x_s=0$ and $\alpha=0.9$, see Fig.~\ref{subfig:ss_alpha_xs0}, we observe that $P(x)$ remains low for $x_b < b_{\rm th}$ and is instead maximum for $x_b = b_{\max}$. This means that adopting the optimal policy allows the nodes to maximize the time spent with a full buffer and operating according to the maximum admissible energy expenditure.
For state $x_s=1$, we see that the optimal policy spends the minimum allowed energy consumption, $u_{\min}$ which corresponds to the energy required to keep the network operational, $\iomin$
In this way, the network saves energy during the low-energy state ($x_s=1$), resuming the transmission of data packets in the high energy state ($x_s=0$). 
To summarize, each energy management policy induces a steady-state distribution of the buffer state. The optimal policy in this case makes it so that the steady-state probability of operating with a full buffer is maximized when the system is in the high-energy state, see Fig.~\ref{subfig:op_alpha_xs0}; this is a desirable property as the sensor nodes can then maximize the time during which the maximum admissible energy expenditure is allocated. On the other hand, in the low-energy state the node will only allocate $u_{\min}$. This  leads to a modest energy consumption which, in turn, implies that the steady-state distribution of the buffer state is preserved during the low-energy state (e.g., night), so that the node at the beginning of the next high-energy state (e.g., day) has a full buffer and can transmit right away using the maximum allowed rate.

As discussed in Section~\ref{sec:smdp}, a small $\alpha$ corresponds to a greedy transmission behavior. This is evident from the policies in Figs.~\ref{subfig:op_alpha_xs0} ($x_s=0$) and~\ref{subfig:op_alpha_xs1} ($x_s=1$) for $\alpha \in \{0.01,0.5\}$. The increased greediness reshapes the steady-state distribution $P(x)$. In particular, for $\alpha \in \{0.01,0.5\}$ and $x_s=0$, $P(x)$ assumes negligible values when $x_b > 50\%$. Hence, although the optimal policies would dictate to transmit using $u_{\max}$ for these values of $x_b$, the time spent in these states is negligible. As a result, the throughput achieved by the greedier policies is smaller \cc{(the throughput reduction is as large as $23\%$ for the considered example)}.

\figresthree
\figresfour

In Figs.~\ref{fig:op_Bsize_xs0} and \ref{fig:op_Bsize_xs1}, we look at the same performance for a fixed discount factor $\alpha=0.9$ and a varying  buffer size $b_{\max} \in \{100,250,500,1000\}$~mAh. In particular, from Fig.~\ref{subfig:op_Bsize_xs0} we observe that the buffer size greatly affects the shape of the optimal policy. In fact, an increasing $b_{\max}$ also implies that the excess energy that is harvested during a stage can always be accumulated. Also, whenever the buffer is full or above $50/60\%$, with large buffers it is possible to transmit allocating the maximum energy consumption $u_{\max}$, as the buffer is sufficiently large to assure that the constraint will be satisfied at the end of the stage, irrespective of the amount of energy that will be harvested. While a big leap in performance is observed as we go from $b_{\max}=100$~mAh to $b_{\max}=500$~mAh \cc{(the throughput is about three times larger.)}, increasing $b_{\max}$ to $1000$~mAh only leads to marginal throughput improvements, \cc{smaller than $10$\%.} This is because a buffer size of $500$~mAh is already sufficient to absorb unexpected energy peaks during the day (therefore minimizing the energy wastage) and as well to allow for the consumption of the maximum current $u_{\max}$ while satisfying the buffer constraint. The fact that a buffer of $500$~mAh suffices in our scenario is also testified by the steady-state distribution in Fig.~\ref{subfig:ss_Bsize_xs0}, where we see that a buffer size of $1000$~mAh has a negligible probability of getting filled beyond $58\%$. \mr{Finally, we discuss the impact of $b_{\rm th}$. The energy buffer is allowed to decrease below this threshold to an extent controlled by $\tout$ (see \eq{eq:outage}). When $\tout \to 0$, optimal policies effectively maintain the buffer above $b_{\rm th}$ and this is equivalent to having a reduced battery capacity (of size $b_{\max}-b_{\rm th}$). This, in turn, leads to less aggressive policies (see Fig.~\ref{subfig:op_Bsize_xs0}) that result in a smaller throughput. As an example, for the considered setup and $b_{\max}=500$, when $b_{\rm th}$ goes from $100$ to $300$ we observe a throughput reduction of about $35\%$. This reduction is nonlinear in $b_{\max}-b_{\rm th}$ (a linear relation would imply a reduction of $50\%$).}

\subsection{Performance Analysis}
\label{sec:performance}


In this section, we evaluate the performance of the proposed solution focusing on a single network instance and considering the setup discussed at the beginning of Section~\ref{sec:results}. Also, we implemented the technique proposed in Kansal et al.~\cite{kansal2007power}, referred to here as ``Kansal'', comparing it against our approach for the same network topology and energy arrival trace, obtained from real data for the city of Los Angeles, see~\cite{EnergyCon-14}. For a fair comparison, we implemented Kansal's energy prediction model with parameter $\alpha=0.5$ and setting their $\rho_{\rm min}$ and $\rho_{\rm max}$ parameters as our optimal working points for the minimum and maximum drained current, i.e., $(\tU^{\rm min}, \tdcmin)$ and $(\tUlim, \tdclim)$, respectively. In addition, we implemented their dynamic duty cycle adaptation strategy not only by letting it change the duty cycle, but also letting it set the new optimal working point according to the new desired energy expenditure. Finally, for the energy buffer we set $b_{\max}=250$~mAh, $b_{\rm th}=50$~mAh, for the computation of our optimal policies we set $\alpha=0.9$,\footnote{Not to be confounded with the Kansal's $\alpha=0.5$ mentioned above.} and the same reward function $r(u)$ (defined in Section~\ref{sec:nodeanal}, see Fig.~\ref{fig:reward}) was used to compute the throughput for both techniques. \bn{For the comparison we choose $b_{\max}$ such as to evaluate which solution is preferable with different system configurations: in particular, we show that our approach is robust regardless of the amplitude of the energy variations.}

\figresfiveter

\cc{Fig.~\ref{fig:comparisons} shows the average throughput (Fig.~\ref{subfig:thr_comp}) and the outage probability (Fig.~\ref{subfig:cost_comp}) for the two schemes. Our solution is represented through solid lines, whereas dashed lines are used for Kansal's. Also, we denote the results related to August and December with square and round markers, respectively. In both figures, the $x$-axis shows the panel side length in centimeters.}

\cc{Our solution is outperformed in terms of throughput, but, on the other hand, it effectively maintains the outage probability within the prescribed threshold, while Kansal's scheme spends up to $44 \%$ of the time in outage (i.e., with a buffer charge $x_b$ smaller than $b_{\rm th}$) and up to $32 \%$ of the time with an empty battery (not shown due to space constraints). This is because our scheme delivers the maximum throughput, subject to the given buffer outage constraint. As a further evidence of the different behavior of the two techniques, in Fig.~\ref{fig:fluct_comp} we show their energy consumption and battery variations for the same energy arrival trace during a timespan of three days.}

\cc{In this figure, we show the hourly variations of the harvested current, $i$, the chosen action (or control), $u$, and the instantaneous battery state, $x_b$, for both solutions. Here, $i$ is represented for both approaches through shaded areas, while the control $u$ is indicated with as a solid line for Kansal and with a dash-dotted line for our approach. Similarly, the two battery states are represented with dashed and dotted lines for Kansal and our approach, respectively.} 


\cc{Two differences can be observed from Fig.~\ref{fig:fluct_comp}: the first is that the policy adopted in the low energy state (night) by our solution is always more conservative than Kansal's, while the same policy is adopted during the day by the two schemes. The second observation is that Kansal's more aggressive behavior leads to battery outages. In fact, while during the second day Kansal successfully maintains energy neutrality, in the first and third days its battery got depleted for about one third of the time.}

In conclusion, we can say that our approach gives priority to the network sustainability, while Kansal's privileges its throughput. This is also reflected by the fact that our control is decided based on the amount of available charge in the battery, while Kansal tries to predict the future current availability to exploit it as efficiently as possible. \bn{However, large variations in the energy availability are likely to lead to high prediction errors that, in turn, negatively affect the outage probability of Kansal}. In conclusion, the adaptability of our scheme to the battery state makes it also robust to the degradation of the battery performance.

\figresfivebis

\section{Relaxation of the Assumptions}
\label{sec:relax}

Here, we address the relaxation of the assumptions made during the analysis: namely, the homogeneity of energy sources, the transmission periodicity, the instantaneous network parameter update, and the fixed topology.

\subsection{Heterogeneous Energy Sources}
\label{sec:hetero}

\cc{The stochastic MDP analysis of Section~\ref{sec:smdp} leads to optimal online policies in the case where the energy arrival process is homogeneous, i.e., all nodes have the same energy harvesting statistics. However, as pointed out in~\cite{Culler-Harvesting-12} in actual deployments different sensor nodes may be affected by slightly differing conditions such as blockage effects due to the surrounding objects, that may partially shade the nodes, obstructing the direct sunlight.}

\cc{In this section, we adapt our analysis to the case where the energy harvesting statistics at the nodes differ. We do so following a two-step approach: 1) we extend the energy source model so that to account for the diversity in the harvested energy and we reuse the analysis of Section~\ref{sec:smdp} with the new source model to obtain a new set of energy consumption policies 2) we use these new policies  according to a simple and practical heuristic. Simulation results that prove the effectiveness of this approach are provided at the end of this section.}

\negativeskip

\cc{\paragraph{Energy source} For the energy sources we account for an additional parameter vector $\bm{p}$, which includes parameters related to the deployment of the solar modules (such as the azimuthal angle, the tilt, the presence of obstructing objects, etc.). Hence, the new statistics for a given node are redefined as $f_\iota(i | x_s,\bm{p})$ and $f_\tau(t |x_s,\bm{p})$ for the input current $i$ and the permanence time $t$ when in state $x_s \in \mathcal{S}$, respectively. Equation~\eq{eq:pdf_joint_charge} generalizes to:
\be
f_\delta(d | u, x_s,\bm{p}) = \int_{t_{\min}(x_s)}^{t_{\max}(x_s)} f_{\tau}(t | x_s,\bm{p}) f_{\iota}(d/t + u | x_s,\bm{p}) |t|^{-1} \dd t \, , \, d \in \R .
\ee
Now, referring to $\bm{\rho}$ as the random vector associated with $\bm{p}$ (its realization), we indicate with $f_{\bm{\rho}}(\bm{p})$ the pdf describing the parameter space. Hence, the new pdf of the harvested charge in state $x_s$ is obtained as:
\be
\label{eq:pdf_joint_charge_new}
f_\delta(d | u, x_s) = \int_{\mathcal{D}(\bm{\rho})} f_\delta(d | u, x_s,\bm{p}) f_{\bm{\rho}}(\bm{p}) \dd \bm{p} \, ,
\ee
where $\mathcal{D}(\bm{\rho})$ is the parameter space.}

\figressix

\cc{As a practical example, for the results that follow we consider a scalar r.v. $\rho$ describing the amount of shade received during the day by a particular sensor node. In fact, in accordance with~\cite{Culler-Harvesting-12}, we found that this is the parameter that affects the most the amount of harvested energy during the day. Here, we assume that the r.v. $\rho$ can take four distinct values, i.e., $\mathcal{D}(\rho)=\{0.4,0.6,0.8,1\}$, which indicate the fraction of sunlight that hits the sensor node. Hence, $p=1$ means that the solar module receives all the available sunlight for the considered location, whereas with $p=0.4$ only $40\%$ of the sunlight is absorbed, while the remaining $60\%$ is blocked. Moreover, we considered a mass distribution function $f_{\rho}(p)$ that assigns a probability $0.55$ to $p=1$ and $0.15$ to each of the remaining cases $p \in \{0.4,0.6,0.8\}$.}

\cc{In this case, \eq{eq:pdf_joint_charge_new} reduces to the following probability mixture:
\be
\label{eq:mixture}
f_\delta(d | u, x_s) = \sum_{p \in \mathcal{D}(\rho)} f_\delta(d | u, x_s,p) f_{\rho}(p) \, .
\ee} 

\cc{We have used the SolarStat tool to obtain $f_\delta(d | u, x_s,p)$ for all $p \in \mathcal{D}(\rho)$. Fig.~\ref{fig:pdf_mixture} shows the resulting pdfs $f_\iota(i|x_s,p)$ (Fig.~\ref{subfig:pdf_mixture_energy}) and $f_\tau(t|x_s,p)$ (Fig.~\ref{subfig:pdf_mixture_tau}) for $x_s=0$ and $p \in \mathcal{D}(\rho)$. Also, a thick solid line is used to indicate the mixture densities.}

\cc{Note that this approach makes it possible to account for heterogeneity in the solar source statistics, modeling our uncertainty on the actual amount of shade that will be received by each sensor node. This uncertainty is then embedded into the source model and the algorithm of Section~\ref{sec:polcomp} is reused with this new source model to generate new energy management policies.}

\negativeskip

\cc{\paragraph{Heuristic and results} First, we define the second bottleneck node (SBN) as the node located in the subtree originating from the bottleneck node (BN) that has the second-highest energy consumption, the node with the highest being the BN. \bn{The worst case for our control policies is when the BN has a shading coefficient equal to $1$ (the available solar radiation is absorbed in full), while the SBN has the smallest shading coefficient $0.4$. In this case, the BN is more likely to experience the most abundant energy inflow (see Fig.~\ref{fig:pdf_mixture}). Thus, its energy buffer will be likely fuller than that of the SBN and, in turn, the BN might choose too aggressive a policy than what the SBN can efficiently adopt. Although this problem may be partially mitigated by the smaller energy consumption of the SBN with respect to that of the bottleneck node, it is still possible that the SBN experiences some battery outages.}}

To make the entire network self-sustainable, an additional expedient is in order. RPL DAO messages are used to periodically report relevant data to the sink, such as the location of the nodes, etc. Thus, it is possible to leverage these messages to collect, at the sink, additional information such as the battery state of all nodes and let the sink choose the policy based on the minimum among all buffer states (instead of solely using the energy buffer state of the BN). \bn{This worst case control strategy makes the adopted policy slightly suboptimal due to the delay associated with the delivery of RPL messages, but assures that the entire network is self-sustainable.}

\figresheterofluct

\cc{Next, we show some simulation results considering that the BN has no shading, i.e., $p=1$, whereas we assume that the SBN has either $p=0.4$ or $p=1$. Moreover, we set the topology parameters of the SBN so as to reproduce the worst case scenario in terms of energy consumption, i.e., we assume that the SBN has the same number of interfering nodes ($\nin$) and packets ($\nint$) as the BN and just one node less ($\nc-1$) for the number of children. In Fig.~\ref{fig:hetero_fluct}, we show the corresponding simulation results considering real solar traces for a timespan of three days for the best ($p=1$) and worst ($p=0.4$) case in terms of energy harvested by the SBN. Note that for $p=0.4$ the energy collected by the SBN (dark shaded area) is only $40\%$ of that harvested (light shaded area) for $p=1$. In both cases, our heuristic scheme opts for a rather aggressive policy during the high energy state (solid and dash-dotted lines for the worst and best case, respectively), whereas in the worst case ($p=0.4$) it adopts much more conservative policies during the low energy state. In fact, in this case the battery level used for the selection of the policy is much lower due to the lower amount of current harvested by the SBN. Compare, for instance, the buffer state in the best case (dotted line) with that of the worst case (dashed line) at about time $1.6$~days: for $p=0.1$ the battery is completely filled up during the day, while for $p=0.4$ the battery is only filled to about half of its capacity, and should then be sparingly used to endure a full night.}

\cc{Finally, in Fig.~\ref{fig:hetero_comp} we show the average throughput (dashed line) for the network and the outage probability (solid line) for the SBN varying the shading conditions $p \in [0.4,1]$. In all the tested cases, we used $p=1$ for the bottleneck node. Note that the outage probability is always very small and almost always smaller than $0.1\%$. As expected, using our conservative approach may impact the throughput performance: this impact is negligible (less than $5\%$) for $p>0.6$, but becomes substantial (up to $30\%$) in the most unfavorable case, i.e., where the SBN has $p=0.4$.}

\figresheterocomp







\negativeskip

\cc{\subsection{Transmission Periodicity} in section~\ref{sec:node} we assumed that nodes periodically sense the environment and generate their data at a constant rate of $\fu$ packets per second. However, this is not strictly necessary, in fact, what really impacts the energy consumption is the {\it total number} of packets sent during a decision epoch. We preferred to study a periodic transmission process  because it allows for a simpler mathematical analysis, leading to a closed form solution for problem P1.}

\cc{In addition, the transmission periodicity can be enforced at the application level adopting a traffic shaping technique, i.e., by spacing out  subsequent packets, through the user of transmission timers, so that the transmission rate will be no higher than $\fu$ packets per second see, e.g.,~\cite{Castellani-IoT-CongControl-14}. This implementation trick can be useful to reduce the collision probability. In fact, reducing the traffic burstiness helps maintaining the ratio $1/\fup$ large, which translates into a low number of collisions.}

\cc{In the paper, we considered the network application to periodically sample environmental parameters. However, from the above discussion, it is easy to see that our solution can be as well applied to networks where the objective is that of communicating alarms or events to the sink. In this case, our scheme supports up to $\Delta_k/\tU$ events per epoch per node, where $\Delta_k$ is the decision epoch duration.} 

\cc{Finally, note that the latency in the communication from the nodes to the sink is not governed by $\tU$, but by $\tdc$. In fact, as soon as an event occurs, the node detecting it can send the alarm to its next hop within at most $\ttx$ seconds, which is dominated by $\toff$ in the low energy period and by $\tdata$ in the high energy period.\footnote{\cc{We recall that $\ttx = \ton + \toff + \tdata + (\fup/\fu-1) \tdc$.}} Thus, delivering an alarm or an event from a node located $h$-hops away from the sink will take about $h \max(\tdata, \toff)$ seconds, independently of $\tU$.}

\negativeskip

\cc{\subsection{Instantaneous Update} Our solution requires that all the nodes change their working point as soon as the energy source transitions to a new state. Although this is infeasible instantaneously, a simple and effective approximation can be employed. In particular, it is possible to exploit the information dissemination service provided by RPL to let the sink broadcast the new working point to all the sensor nodes. This procedure takes a finite amount of time and eventually terminates with all nodes knowing the new working point. During this lapse of time different nodes in the network may use a different working point.}

\cc{Soon after the energy source transition, as a consequence of the adoption of new parameters, two different configurations will coexist in the network: a group of nodes will have a rather high duty cycle and another group will instead have a smaller one. Many solutions have been proposed in the literature to allow the interaction of nodes with differing duty cycles. Here, we advocate the use of a very simple technique based on a {\it grace period}. During the grace period, nodes will wake up according to the highest between the two duty cycles and will send preambles using the $\toff$ associated with the smallest of the two.} 


As a drawback of this procedure, nodes will consume a higher amount of energy during the grace period. However, RPL can disseminate the new configuration to the entire network in about $h \tdc$ seconds if the longest path is at most $h$ hops long. \bn{Since the length of a grace period is related to RPL dissemination time, the worst case duration is obtained when the duty cycle is smaller (low energy state) and for bigger networks; for instance, with our settings and a duty cycle $d_c=1 \%$ the longest grace period is shorter than one second, which is negligible compared to the duration of decision epochs.} Nevertheless, to overcome this limitation more advanced techniques can be used, along the lines of~\cite{Vigorito-SECON-07}.


\negativeskip

\cc{\subsection{Fixed Topology} Our reward function, $r(u)$, inherently depends on the topology through $\nc$, $\nin$, and $\nint$. Thus, the topology must remain static in order for a policy to maintain its optimality. However, this does not mean that the topology cannot change. In fact, note that topology information is periodically reported to the sink through RPL DAO messages. Hence, the impact of a changed topology can be estimated at the sink through the calculation of new topology parameters. At this point, if the throughput degradation is deemed too high or certain nodes are likely to deplete their batteries due to their increased load, the adoption of a new energy management policy at all nodes can be triggered. In this case, the sink will send a new policy to the nodes as if a transition of the energy source were occurred. When new nodes are added to the network, we let these behave as if they were in a grace period (see our discussion above) until they receive the new policy. }

%

\section{Conclusions}
\label{sec:conclusions}

In this paper, we have provided a comprehensive mathematical framework for the design of energy scavenging wireless sensor networks. Specifically, we have investigated the general class of problems related to the long term and self-sufficient operation of wireless sensor networks powered by renewable energy sources. Our approach consisted in two nested optimization processes: the inner one (P1) characterizes the optimal operating point of the network subject to a given energy consumption figure (assumed constant), while the outer (P2) provides optimal energy management policies to make the system energetically self-sufficient, given the result of the inner problem and the statistical description of the energy source. 

As a first step, we have defined an original energy consumption model describing the behavior of the bottleneck node (i.e., the node consuming the highest amount of energy) for a given routing topology and channel access technology. Secondly, we have proven that it is sufficient to grant the self-sufficiency of the bottleneck to assure that all network nodes are also self-sufficient. Thus, we have solved P1 analytically, by deriving a closed form expression for the optimal duty cycle and the optimal information generation rate that are to be used by all nodes to guarantee their perpetual and autonomous operation. This result was derived by neglecting packet collisions at first, and it was subsequently extended through a heuristic to keep the effect of packet collisions into account. 

Hence, using the solution of P1 and a statistical description of the energy source, we have formulated P2, a discrete time constrained Markov decision problem (DT-CMDP), \cc{returning the online policies that maximize the long term average throughput of the network}, while assuring its self-sufficiency in the presence of a stochastic energy source. We have solved P2 using a Lagrangian relaxation technique, which permits a convenient exploration of the solution space. \cc{Also, we described how the obtained policies can be implemented to overcome the computational complexity of the approach at the sensor nodes.}

We have then used our framework to explore the impact of key system parameters on the design of energy harvesting sensor networks. In detail, we have assessed the impact of network topologies on the reward function, also studying the impact of battery and photovoltaic panel sizes on the optimal energy consumption strategies. Thus, the framework has been utilized to derive the long term average network performance, which includes the network throughput and the steady state probabilities of the battery charge state when the optimal policies are adopted by the nodes. \cc{Finally, we thoroughly validated our optimal policies against state of the art approaches, also proving its robustness when our main assumptions are relaxed. Our solution proved to be more conservative than the state of the art, and, although at the price of a slightly lower throughput, it assures the self-sustainability of all sensor nodes for all battery sizes and environmental conditions.}


\section*{APPENDIX}
\appendix

\section{Channel Access Modeling in the Presence of Packet Collisions}
\label{app:MAC_modeling}

To take collisions and channel transmission errors into account, we derived the following fixed point analysis. Note that our collision model is similar to the one considered in previous work, see, e.g.,~\cite{Yang-12}. The analysis that we present in what follows differs in the fact that we consider the transmission of periodic endogenous traffic, and this allows for a closed-form expression of the collision probability, which is derived next. 
We refer to the packet error probability for the transmission of the bottleneck node as $\etx$, which depends on the selected modulation and coding scheme and on the channel impairments (attenuation, noise, etc.), see, e.g., Chapter 6 of~\cite{Goldsmith-book-05}. Here, we consider $\etx$ fixed. Also, we refer to $\nin \geq 0$ as the number of interfering nodes and to $\ecoll$ as the packet collision probability. 

Given that a packet is successful when no channel errors occur (w.p. $1-\etx$) and it is not collided (w.p. $1-\ecoll$), the total packet error probability is obtained as $\ep = \ecoll + \etx - \ecoll \etx$. Now, note that when $\ep  \geq 0$, due to the increased number of packet losses and the associated retransmissions, we have that the packet transmission rate of the bottleneck node increases to $\fup \geq \fu$, where $\fu$ is the original information rate. Hence, one packet is transmitted on average every $1/\fup$ seconds, where $1/\fup$ is the new average inter-transmission time. We assume that the transmission within this time period occurs picking a transmission instant uniformly at random in $[0,1/\fup]$. Moreover, given our LPL MAC, whenever a packet is transmitted, there exists a vulnerability period\footnote{The vulnerability period is a tunable parameters reflecting the time needed for practical architectures to put the radio into the RX state and detect incoming packets.} of $\tv$ seconds and a collision event occurs whenever any of the $\nin$ interferers picks its own transmission time within this interval; the probability of this event to occur is $\pcoll = \fup \tv$ (see Fig.~\ref{fig:coll} for a graphical example). Note that $\pcoll$ corresponds to the probability that a given interferer picks its transmission time within period $\tv$, given that this transmission instant is (assumed) uniformly distributed in $[0,\tU^\prime]$, where $\tU^\prime = 1/\fup$ is the inter-packet transmission interval in the presence of retransmissions. 

\figcoll

Given this, the probability that a collision event is due to $k \in \{1,\dots,\nin\}$ interferers is given by ${\nin \choose k} \pcoll^k (1-\pcoll)^{(\nin-k)}$ and the probability that the packet sent by the bottleneck node collides is finally obtained as: $\ecoll = 1 - (1-\pcoll)^{\nin}  = 1 - (1-\fup \tv)^{\nin}$, which corresponds to the probability that at least one of the interferers transmits in the vulnerable interval. Note that the previous equation can be solved for $\fup$, expressing the latter as a function, $g_1(\cdot)$, of the other parameters:
\be
\label{eq:g1}
\fup = g_1(\ecoll, \tv, \nin) = [1-(1-\ecoll)^{(1/\nin)}] \tv^{-1} .
\ee
On the other hand, for a packet error rate $\ep$, $\fup$ can be related to $\fu$ through the following function $g_2(\cdot)$: 
\be
\label{eq:g2}
\fup = g_2(\ecoll, \etx, \fu) = \fu(1-\ep)^{-1} = \fu (1-\ecoll-\etx+\ecoll\etx)^{-1} . 
\ee 
Observing that $\fu$ is given, $\tv$ is a (hardware dependent) constant and $\etx$ and $\nin$ are also constant for a given transmission scenario (topology, modulation and channel model), we have that the only unknwown parameter is the collision probability $\ecoll$. Since, $g_1(\cdot)$ and $g_2(\cdot)$ both return the packet transmission rate $\fup$, the working point for the system is obtained by imposing $g_1(\cdot) = g_2(\cdot)$, solving for $\ecoll$ and retaining the smallest real solution to the previous equality. The steady state transmission rate $\fup$ is attained using this value of the collision probability with either $g_1(\cdot)$ or $g_2(\cdot)$. This is a practical method to obtain $\fup$ at equilibrium, in the presence of channel errors and collisions. Note, however, that a solution is not always guaranteed to exist and this occurs when the offered traffic exceeds the maximum capacity of the considered access channel. In Appendix~\ref{app:collision}, we provide an approximated formula to conveniently calculate $\ecoll$ and a stability analysis to mathematically assess when the channel access admits a solution.

\section{Collision Probability and Feasibility Condition for the Channel Access}
\label{app:collision}

\noindent \textbf{Collision probability approximation:} here we derive a closed-form approximation for the collision probability $\ecoll$ at equilibrium. As discussed in Appendix~\ref{app:MAC_modeling}, this is obtained by looking at the points where $g_1(\cdot)$ and $g_2(\cdot)$ intersect (see below for the necessary condition for this to occur). When these function do intersect, they have two real solutions in the range $\ecoll \in [0,1]$ and the $\ecoll$ at equilibrium is the smallest real solution. From the equality $g_2(\ecoll, \etx, \fu) = g_1(\ecoll, \tv, \nin)$, using $x=\ecoll$, we get:
\be
\label{eq:ec_equality}
(1 - \etx - x + \etx x ) (1 - (1-x)^{1/\nin}) - \fu \tv = 0 .
\ee
Now, we employ the Taylor expansion of $(1-x)^{1/\nin}$, around the point $x_0=0$:
\be
(1-x)^{1/\nin} = 1 - \frac{x}{\nin} + \OO{x^2} ,
\ee
that, used into \eq{eq:ec_equality} leads to:
\be
\label{eq:se_eq_ni}
x^2 (1-\etx) - x (1-\etx) + \fu \tv \nin = 0 .
\ee
The discriminant of \eq{eq:se_eq_ni} is $\Delta = 1-4(\fu \tv \nin)/(1-\etx)$, thus, the condition $\Delta \geq 0$ implies $\nin \leq \left \lfloor (1-\etx)/ (4 \fu \tv) \right \rfloor$. When the latter is verified, the solution for the collision probability is given by the smallest solution of \eq{eq:se_eq_ni}, i.e.:
\be
\label{eq:approx_ec}
\ecoll \simeq \frac{1 - \sqrt{\Delta}}{2} .
\ee
For illustrative purpose, considering $\etx \leq 0.3$ and $\fu \tv \leq 0.001$, which is largely verified in practice\footnote{As shown in Section~\ref{sec:results}, feasible values for $\fu$ are typically larger than one packet per minute that, considering $\tv \leq 0.01$~s, leads to $\fu \tv \leq 1.6\cdot 10^{-4}$.}, \eq{eq:approx_ec} is accurate up to the third decimal place for $\nin \leq 20$ and up to the second decimal place for $\nin \leq 50$. Note that these settings for $\etx$ and $\fu \tv$ are rather extreme and more accurate results are achieved for the practical network examples of this paper. For these, we have that $\fu \tv= 0.0004$, $\etx = 0.1$ and $\nin = 5$ and, with these parameters, the gap between the actual average number of retransmissions $n_{\rm retx}^\prime = \ep/(1-\ep)$ (considering the impact of packet collisions) and the approximation $n_{\rm retx} = \etx/(1-\etx)$ (considering $\ecoll=0$) leads to a relative error of $100 (n_{\rm retx}^\prime - n_{\rm retx})/n_{\rm retx}^\prime=2.18\%$.\\

\noindent \textbf{Feasibility collision for the channel access:} in what follows, we examine the condition under which the channel access problem of Appendices~\ref{app:MAC_modeling} and~\ref{app:collision}, whereby $n_i$ nodes transmit over the same medium, is feasible. Intuitively, a random access channel has a limited ``hosting capacity''. When too many users transmit over it at too high a rate, exceeding the capacity limit, the random access system becomes unstable. In this case, the collision probability tends to increase indefinitely, leading to a zero throughput for all users. Next, we mathematically derive the condition under which the channel access system of Section~\ref{sec:node} is stable as a function of the parameters $\fu$, the transmission rate of the node (of their endogenous traffic, without considering collisions), $\tv$, the vulnerability period and $\nin$, the number of nodes that transmit over the same medium (interferers).    

Mathematically, a finite solution for $\ecoll$ exists only when the two curves $g_1(\cdot)$ (see \eq{eq:g1}) and $g_2(\cdot)$ (see \eq{eq:g2}) intersect. Through a more accurate inspection of the behavior of \eq{eq:g1} and \eq{eq:g2}, it is easy to see that a solution to $g_1(\cdot) = g_2(\cdot)$ does not exist when we have that $ g_2(\ecoll, \etx, \fu) > g_1(\ecoll, \tv, \nin)$, for all values of $\ecoll \in [0,1]$. Through some algebra, it is easy to verify that this condition corresponds to:
\be
\label{eq:infeasibility_cond}
\fu \tv > (1-\etx) (1-\ecoll) (1-(1-\ecoll)^{1/\nin}) \stackrel{def}{=} g_3(\ecoll,\etx,\nin) , \, \forall \, \ecoll \in [0,1] \, . 
\ee
Now, the LHS of \eq{eq:infeasibility_cond} is a constant, whereas the RHS is a continuous function of $\ecoll$ that has a maximum in $\ecollm$, where:
\be
\ecollm = 1 - \left (  \frac{\nin}{1+\nin} \right )^{\nin} \, . 
\ee 
Note that condition \eq{eq:infeasibility_cond} is verified if the LHS is strictly greater than the RHS ($g_3(\ecoll,\etx,\nin)$) for all values of $\ecoll$ and this {\it must also hold} for $\ecoll = \ecollm$. In this case, $g_1(\cdot)$ and $g_2(\cdot)$ do not intersect and, in turn, the system does not admit a stable working point. The previous reasonings formally prove that the {\it feasibility condition} for the channel access is: 
\be
\label{eq:feasibility_cond}
\fu \tv  \leq g_3(\ecollm,\etx,\nin) = (1-\etx) \left ( \frac{1}{1+\nin} \right ) \left ( \frac{\nin}{1+\nin} \right )^{\nin} ,
\ee
as when \eq{eq:feasibility_cond} is verified $g_1(\cdot)$ and $g_2(\cdot)$ intersect in at least one point.\\

\section{Problem P1: Derivation of the Closed Form Solution}
\label{app:closed}

In what follows, we derive the closed form expression of the optimal working point $(\tUopt,\tdcopt)$ for a collision-free channel. The first step is to rewrite \eq{eq:currtx} by neglecting packet collisions, i.e., $\ecoll=0$, which implies $\fu^{\prime}/\fu-1=\etx/(1-\etx)$ and, in turn:
\bea
\label{eq:currtxnc}
\itx & = & (\ic + \itr)[\tdc/2+\ton/2+\tdata + (\etx/(1-\etx)) \tdc]\times \nonumber \\ 
  & & \times[(1+\nc)/\tU+(2+\nc)/\trpl]. 
\eea
Subsequently, we rewrite \eq{eq:currtx}-\eq{eq:curroff} isolating the terms depending on $\tU$ and $\tdc$ and introducing coefficients $\{c_1,\dots,c_5\}$ and $\{a_1,\dots,a_{11}\}$ (see Table~\ref{tab:coeff1}):
\bea
\label{eq:coeff11}
\itx & = & c_1 (\tdc a_1/\tU +a_2/\tU + \tdc a_3 + a_4) \\
\label{eq:coeff12}
\irx & = & c_2  (a_5\tU + a_6) \\
\label{eq:coeff13}
\iinte & = & c_2 (a_7\tU +a_8) \\
\label{eq:coeff14}
\icpu & = & c_3 a_9 \tU \\
\label{eq:coeff15}
\ridle & = & 1-\rtx-\rrx-\rinte-\rcpu = \nonumber \\ 
 & = & a_{10} - a_1\tdc /\tU - a_3\tdc - a_{11}/\tU\\ 
\label{eq:coeff16}
\iidle & = & \ridle (c_4 + c_5 \ton/\tdc) .
\eea
Rewriting \eq{eq:coeff11}-\eq{eq:coeff16}, and introducing coefficients $\{b_1,\dots,b_{15}\}$ (see Table~\ref{tab:coeff1}), leads to:
\bea
\label{eq:coeff17}
\itx & = & b_1\tdc /\tU +b_2/\tU + b_3 \tdc + b_4 \\
\label{eq:coeff18}
\irx & = & b_5\tU + b_6 \\
\label{eq:coeff19}
\iinte & = & b_7\tU +b_8 \\
\label{eq:coeff20}
\icpu & = & b_9 \tU \\
\label{eq:coeff21}
\iidle & = & b_{10}\tdc/\tU + b_{11}/\tU + b_{12}\tdc + b_{13} + b_{14}/\tdc + b_{15}/ (\tdc\tU) .
\eea
$\io(\tU, \tdc)$ is thus obtaned using \eq{eq:iot_tot}. For compactness, $\io(\tU, \tdc)$ is expressed using a fourth set of coefficients ($\{d_1,\dots,d_6\}$ of Table~\ref{tab:coeff2}):
\be
\label{eq:coeff22}
\io(\tU, \tdc) = d_1 \tdc/\tU + d_2/\tU + d_3\tdc + d_4 + d_5/\tdc + d_6/ (\tdc\tU) .
\ee
Now, taking the first order derivative of \eq{eq:coeff22} with respect to $\tdc$ we obtain:
\be
\label{eq:coeff23}
\frac{\partial \io(\tU, \tdc)}{\partial \tdc} = d_3+d_1/\tU - (d_5+d_6/\tU)/\tdc^2, 
\ee
which leads to the following result:
\be
\label{eq:coeff24}
\frac{\partial \io(\tU, \tdc) }{\partial \tdc} = 0 \, \Rightarrow \, \tdcopt(\tU)=\pm \sqrt{\frac{d_{6}/\tU + d_{5}}{d_{1}/\tU + d_{3}}} ,
\ee
where the wanted solution is the one with the plus sign. Note that $\tdcopt(\tU)$ is the optimal duty cycle, which minimizes the power consumption for a given inter-packet transmission time $\tU$ (endogenous traffic). At this point, we compute \eq{eq:coeff22} for $\tdcopt(\tU)$, subtracting $u$ (i.e., the target current budget) and equating to zero:
\be
\label{eq:coeff25}
\io(\tU,\tdcopt(\tU))-u = \frac{\tdcopt(\tU)^2(d_1/\tU+d_3)+(d_6/\tU + d_5)}{\tdcopt(\tU)} + d_2/\tU + d_4 - u = 0.
\ee
Now, raising \eq{eq:coeff25} to the second power and reordering leads to:
\be
\label{eq:coeff26}
(4 d_1 d_6 - d_2^2)/\tU^2 + (4d_1d_5+4d_3d_6-2d_2d_7)/\tU + 4d_3d_5-d_7^2 = e_0/\tU^2 + e_1/\tU + e_2 = 0.
\ee
Note that \eq{eq:coeff26} is solved for $\tU$, with $u \in [u_{\min},u_{\max}]$, with $\iomin=u_{\min}$ and $u_{\max}=\iolim$. It is easy to verify that the solution of problem P1, $\tU^*$, is the only positive solution of the previous equation. 

For the calculation of $\iomin$ and $\iolim$, we proceed as follows. First, for what concerns the minimum current consumption $\iomin$, we first obtain the optimal duty-cycle for the case where no data gathering operations are performed (i.e., $\tU$ goes to infinity), $\tdcmin = \lim_{\tU \to +\infty} \tdcopt(\tU) = \sqrt{d_5/d_3}$. Hence, we use this result together with \eq{eq:coeff22} to compute $\iomin$:
\be
\label{eq:coeff27}
\iomin = \lim_{\tU \to +\infty} \io(\tU,\tdcopt(\tU)) = d_3\sqrt{d_5/d_3} + d_4 + d_5 \sqrt{d_3/d_5} .
\ee
To obtain $\iolim$ we first define $\tdclim = \tdcopt(\tUlim)$, where $\tUlim$ is obtained from $\ridle(\tUlim,\tdc)=0$ (meaning that the node is always busy and maximizes its transmission activity). From the latter equality we get:
\be
\label{eq:coeff27bis}
\tUlim(\tdc)=\frac{a_1\tdc+a_{11}}{a_{10}-a_3\tdc}, 
\ee 
which, together with \eq{eq:coeff24}, leads to:
\bea
\label{eq:coeff29}
\tdclim & = & \sqrt{\frac{d_6/\tUlim+d_5}{d_1/\tUlim+d_3}}=\nonumber\\
  & = & \sqrt{\frac{d_6(a_{10}-a_3\tdclim)+d_5(a_1\tdclim+a_{11})}{d_1(a_{10}-a_3\tdclim)+d_3(a_1\tdclim+a_{11})}} ,
\eea
where the equality in the second line follows from replacing $\tUlim$ with \eq{eq:coeff27bis}. Thus, raising \eq{eq:coeff29} to the second power and solving for $\tdclim$ leads to the third order equation:
\be
\label{eq:coeff30}
f_3 (\tdclim)^3 + f_2 (\tdclim)^2 + f_1 \tdclim + f_0 = 0
\ee
and $\tdclim$ is the largest solution of \eq{eq:coeff30}. Finally, $\tUlim$ is obtained plugging $\tdclim$ into \eq{eq:coeff27bis} and $\iolim=\io(\tUlim,\tdclim)$ is finally calculated from \eq{eq:coeff22}.

\tabcoeffabc
\tabcoeffdef

\section{On the correctness of the bottleneck analysis}
\label{app:stability}

In this appendix, we analyze the network stability given that the system is tuned on the bottleneck node and all other nodes use the same operating point of the latter. To prove that when the bottleneck node is energetically self-sufficient, the same holds true for all the other nodes in the network, we will show that $\io(\nc,\nin,\nint)$ is an increasing function of $\nc$, $\nin$, and $\nint$ (all the other parameters remaining fixed). To this end, let $\nc^{\rm b}$, $\nin^{\rm b}$, $\nint^{\rm b}$ and $\io^{\rm b}$ respectively be the topology parameters and the output current of the bottleneck node. Given that for all nodes the following inequalities hold:
\be
\label{eq:stab1}
\nc \leq \nc^{\rm b} , \, \, \nin \leq \nin^{\rm b}, \, \, \nint \leq \nint^{\rm b},
\ee
for any sensor node in the network, we have that $\io \leq \io^{\rm b}$, which proves our claim.
%
In what follows, we only show that $\io(\nc,\nin,\nint)$ is an increasing function of $\nc$, as the proof for the other variables develops along the same lines.

First of all, we take the first order derivative of $\io(\nc,\nin,\nint)$ with respect to $\nc$:
\bea
\label{eq:stab3}
\frac{\partial \io}{\partial \nc} & = & \frac{\partial \itx}{\partial \nc} + \frac{\partial \irx}{\partial \nc} + \frac{\partial \iidle}{\partial \nc} = \nonumber\\
 & = & c_1\frac{\partial \rtx}{\partial \nc} + c_2\frac{\partial \rrx}{\partial \nc} + \left ( c_4 + c_5 \frac{\ton}{\tdc} \right )\frac{\partial \ridle}{\partial \nc} = \nonumber\\
 & = & c_1\frac{\partial \rtx}{\partial \nc} + c_2\frac{\partial \rrx}{\partial \nc} - \left ( c_4 + c_5 \frac{\ton}{\tdc} \right )\bigg(\frac{\partial \rtx}{\partial \nc}+\frac{\partial \rrx}{\partial \nc}\bigg)  = \nonumber\\
 & = & \frac{\partial \rtx}{\partial \nc} \left ( c_1-c_4-c_5 \frac{\ton}{\tdc} \right ) + \frac{\partial \rtx}{\partial \nc} \left (c_2-c_4-c_5 \frac{\ton}{\tdc} \right ).
\eea
Hence, we proceed showing that $\partial \rtx / \partial \nc$, $\partial \rrx / \partial \nc$, $(c_1-c_4-c_5\ton/\tdc)$ and $(c_2-c_4-c_5\ton/\tdc)$ are all positive quantities. For the two derivatives it holds:
\bea
\label{eq:stab4}
\frac{\partial \rtx}{\partial \nc} & = & \frac{\tdc/2 + \ton/2 + \tdata + (\fu^\prime/\fu-1)}{1/\tU + 1/\trpl} \nonumber\\
\frac{\partial \rrx}{\partial \nc} & = & \frac{\tdata}{1/\tU + 1/\trpl},
\eea
and it is easy to show that all the addends of the two sums are positive, because all of them are either time or frequency quantities that are positive by definition. The term $\fu^\prime/\fu-1$ is also positive since $\fu^\prime$ is the arrival rate in the presence of channel errors, which implies that $\fu^\prime \geq \fu$.

Finally, for what concerns the other two terms, they can be re-written as:
\bea
\label{eq:stab5}
c_1-c_4-c_5\ton/\tdc & = & \itr + \ic - \is - (\ir + \ic - \is)\ton/(\ton + \toff)  \geq \nonumber\\
 & \geq & (\ir + \ic - \is)\toff/(\ton + \toff) \nonumber\\
c_2-c_4-c_5\ton/\tdc & = & \ir + \ic - \is - (\ir + \ic - \is)\ton/(\ton + \toff)  = \nonumber\\
 & = & (\ir + \ic - \is)\toff/(\ton + \toff),
\eea
where the inequality in the second line holds since $\itr \geq \ir$ for all radio technologies. Also, note that $\is \ll \ir$, $\ir + \ic - \is > 0$ and $\toff/(\ton + \toff)$ is by definition positive, which prove that both terms in \eq{eq:stab5} are greater than or equal to zero. Thus, $\partial \io / \partial \nc$ is the sum of positive terms, which implies that $\io(\nc,\nin,\nint)$ is an increasing function of $\nc$ and that $\io \leq \io^{\rm b}$ for every sensor node.

\bibliographystyle{ACM-Reference-Format-Journals}

\received{received}{revised}{accepted}

\end{document}